\documentclass[a4paper,11pt]{article}
\usepackage{jcappub}
\usepackage{amsmath}
\usepackage{graphicx}
\usepackage{multirow}
\usepackage{tikz}

\usepackage{xcolor}

\makeatletter
\gdef\@fpheader{}
\makeatother

\newlength{\fullw}
\newlength{\halfw}
\newlength{\threefigw}
\newlength{\twofigw}
\newlength{\onefigw}
\newlength{\roww}
\DeclareMathOperator{\sinc}{sinc}
\DeclareMathOperator{\rect}{rect}
\DeclareMathOperator{\rectbar}{\overline{rect}}
\DeclareMathOperator{\fresnelC}{\bar{C}_2}
\DeclareMathOperator{\fresnelS}{\bar{S}_2}

\newcommand{\boldmathsymbol}[1]{{\ensuremath{\boldsymbol{#1}}}}
\newcommand{\sss}[1]{{\scriptscriptstyle{#1}}}
\newcommand{\mean}[1]{\left\langle #1 \right\rangle}

\newcommand{\heaviside}[1]{\mathrm{\Theta}\!\left( #1 \right)}

\newcommand{\invfourier}[1]{\mathcal{F}^{-1}\!\left( #1 \right)}
\newcommand{\invfourierb}[1]{\mathcal{F}^{-1}\!\left[ #1 \right]}

\newcommand{\vect}[1]{\boldmathsymbol{#1}}
\newcommand{\negleft}{\negthinspace\left}
\newcommand{\dirac}[1]{\delta\negleft(#1\right)}

\newcommand{\Mpc}{\mathrm{Mpc}}

\newcommand{\Mp}{M_\mathrm{Pl}}
\newcommand{\GN}{G_{\mathrm{\sss{N}}}}

\newcommand{\Rset}{\mathbb{R}}

\newcommand{\uTT}{\mathrm{\sss{TT}}}
\newcommand{\ud}{\mathrm{d}}
\newcommand{\uc}{\mathrm{c}}

\newcommand{\us}{\mathrm{s}}

\newcommand{\ue}{\mathrm{e}}
\newcommand{\ueq}{\mathrm{eq}}
\newcommand{\ugw}{\mathrm{gw}}

\newcommand{\upmat}{\mathrm{puremat}}
\newcommand{\umat}{\mathrm{mat}}
\newcommand{\urad}{\mathrm{rad}}
\newcommand{\umix}{\mathrm{mix}}
\newcommand{\uini}{\mathrm{ini}}

\newcommand{\uss}{\mathrm{ss}}
\newcommand{\ucc}{\mathrm{cc}}
\newcommand{\ucs}{\mathrm{cs}}
\newcommand{\usc}{\mathrm{sc}}

\newcommand{\ur}{\mathrm{r}}
\newcommand{\um}{\mathrm{m}}

\newcommand{\be}{\vect{e}}
\newcommand{\bepsilon}{\vect{\epsilon}}
\newcommand{\bx}{\vect{x}}
\newcommand{\by}{\vect{y}}
\newcommand{\bk}{\vect{k}}
\newcommand{\bkhat}{\be_k}
\newcommand{\bq}{\vect{q}}
\newcommand{\bqhat}{\be_q}

\newcommand{\rhogw}{\rho_\ugw}
\newcommand{\rhoc}{\rho_\uc}

\newcommand{\calC}{\mathcal{C}}
\newcommand{\calD}{\mathcal{D}}

\newcommand{\calE}{\mathcal{E}}

\newcommand{\calP}{\mathcal{P}}

\newcommand{\calH}{\mathcal{H}}
\newcommand{\calU}{\mathcal{U}}
\newcommand{\calUinf}{\mathcal{U}_\infty}
\newcommand{\calCFT}{\hat{\calC}}
\newcommand{\calDFT}{\hat{\calD}}

\newcommand{\calEFT}{\hat{\calE}}
\newcommand{\CFT}{\hat{C}}
\newcommand{\DFT}{\hat{D}}
\newcommand{\EFT}{\hat{E}}

\newcommand{\Mpl}{\Mp}

\newcommand{\rbu}{u}
\newcommand{\rbv}{v}
\newcommand{\rbupmat}{\rbu_{\upmat}}
\newcommand{\rbumat}{\rbu_{\umat}}
\newcommand{\rburad}{\rbu_{\urad}}
\newcommand{\rbvpmat}{\rbv_{\upmat}}
\newcommand{\rbvmat}{\rbv_{\umat}}
\newcommand{\rbvrad}{\rbv_{\urad}}
\newcommand{\sbj}[1]{j_{#1}}
\newcommand{\sby}[1]{y_{#1}}

\newcommand{\murad}{\mu_{\urad}}

\newcommand{\x}{\kappa}
\newcommand{\xbar}{\bar{\x}}
\newcommand{\xini}{x_\uini}

\newcommand{\G}[1]{G_{#1}}
\newcommand{\Gpmat}[1]{G^{\upmat}_{#1}}
\newcommand{\Gmat}[1]{G^{\umat}_{#1}}
\newcommand{\Grad}[1]{G^{\urad}_{#1}}

\newcommand{\Ks}{K_{\us}}
\newcommand{\Ksmat}{\Ks^\um}
\newcommand{\Ksrad}{\Ks^\ur}
\newcommand{\Ke}{K_{\ue}}
\newcommand{\Kerad}{\Ke^\ur}
\newcommand{\Kemat}{\Ke^\um}

\newcommand{\Imu}{I_{\mu}}
\newcommand{\Imumat}{I_{\mu}^\umat}
\newcommand{\Imurad}{I_{\mu}^\urad}
\newcommand{\Idmu}{I_{\mu'}}
\newcommand{\Idmurad}{\Idmu^\urad}
\newcommand{\Idmumat}{\Idmu^\umat}
\newcommand{\Ix}{I_{\x}}
\newcommand{\Ixrad}{\Ix^\urad}
\newcommand{\Ixmat}{\Ix^\umat}
\newcommand{\Ixbar}{I_{\xbar}}
\newcommand{\Ixbarrad}{\Ixbar^\urad}
\newcommand{\Ixbarmat}{\Ixbar^\umat}

\newcommand{\Icc}{I_{\ucc}}
\newcommand{\Iss}{I_{\uss}}
\newcommand{\Ics}{I_{\ucs}}
\newcommand{\Isc}{I_{\usc}}

\newcommand{\ahat}{\hat{a}}

\newcommand{\OmegaR}{\Omega_{\urad}}
\newcommand{\OmegaM}{\Omega_{\umat}}

\newcommand{\OmegaGW}{\Omega_{\ugw}}
\newcommand{\OmegaGWrad}{\Omega_{\ugw}^\urad}
\newcommand{\OmegaGWmat}{\Omega_{\ugw}^\umat}

\newcommand{\calPh}{\calP_h}
\newcommand{\Sh}{S_h}

\newcommand{\etaini}{\eta_\uini}
\newcommand{\etaeq}{\eta_\ueq}

\newcommand{\xeq}{x_\ueq}
\newcommand{\A}{A}
\newcommand{\B}{B}
\newcommand{\dA}{\dot{\A}}
\newcommand{\dB}{\dot{\B}}

\begin{document}

\title{Interferences in the Stochastic Gravitational Wave Background}

\author{Disrael Camargo Neves da Cunha}
\author{and Christophe Ringeval}

\affiliation{Cosmology, Universe and Relativity at Louvain,
  Institute of Mathematics and Physics, Louvain University, 2 Chemin
  du Cyclotron, 1348 Louvain-la-Neuve, Belgium}

\emailAdd{disrael.camargo@uclouvain.be}
\emailAdd{christophe.ringeval@uclouvain.be}

\date{today}

\abstract{Although the expansion of the Universe explicitly breaks
  the time-translation symmetry, cosmological predictions for the
  stochastic gravitational wave background (SGWB) are usually derived
  under the so-called stationary hypothesis. By dropping this
  assumption and keeping track of the time dependence of gravitational
  waves at all length scales, we derive the expected unequal-time (and
  equal-time) waveform of the SGWB generated by scaling sources, such
  as cosmic defects. For extinct and smooth enough sources, we show
  that all observable quantities are uniquely and analytically
  determined by the holomorphic Fourier transform of the anisotropic
  stress correlator. Both the strain power spectrum and the energy
  density parameter are shown to have an oscillatory fine structure,
  they significantly differ on large scales while running in phase
  opposition at large wavenumbers $k$. We then discuss scaling sources
  that are never extinct nor smooth and which generate a singular
  Fourier transform of the anisotropic stress correlator. For these,
  we find the appearance of interferences on top of the
  above-mentioned fine-structure as well as atypical behaviour at
  small scales. For instance, we expect the rescaled strain power
  spectrum $k^2 \calPh$ generated by long cosmic strings in the matter
  era to oscillate around a scale invariant plateau. These singular
  sources are also shown to produce orders of magnitude difference
  between the rescaled strain spectra and the energy density parameter
  suggesting that only the former should be used for making reliable
  observable predictions. Finally, we discuss how measuring such a
  fine structure in the SGWB could disambiguate the possible
  cosmological sources.}


\maketitle

\section{Introduction}
\label{sec:intro}

The statistical homogeneity and isotropy of the Universe imply that
all gravitational wave sources of natural origin must collectively
contribute to the generation of a stochastic background. All types of
merger discovered so far by the LIGO-Virgo-Kagra experiments
guarantee the existence of a background of astrophysical origin, which
is actively searched for~\cite{Desvignes:2016yex, Perera:2019sca,
  Arzoumanian:2020vkk,Kerr:2020qdo, Abbott:2021xxi}. The other
possible mechanisms to generate a SGWB are of cosmological
origin. Because all astrophysical sources fade away above some
redshift, it is very well possible that the first detection of a SGWB
could be of cosmological origin thereby providing unexpected
discoveries (see, e.g., Refs.~\cite{Allen:1996vm, Christensen:2018iqi,
  Caprini:2018mtu} for reviews).

In fact, this possibility has long been
considered~\cite{1979ApJ...234.1100D, 1980A&A....89....6C,
  Vilenkin:1981, 1983MNRAS.203..945B, Hogan:1984is, Vachaspati:1984gt,
  Accetta:1988bg, Bennett:1990ry, Caldwell:1991jj}. For instance,
measuring the effective number of relativistic degrees of freedom in
the cosmic plasma gives an upper limit to the amplitude of the
(sub-Hubble) SGWB at the time of recombination~\cite{Aghanim:2018eyx},
and also as early as during Big-Bang Nucleosynthesis
(BBN)~\cite{Fields:2019pfx, 2021MNRAS.502.2474P}. These types of
constraint are derived from the changes in the expansion rate of the
Universe induced by the overall gravitating effect of gravitational
waves. As such, they are sensitive to their integrated ``energy'' and
the constraints are given in terms of the so-called energy density
parameter $\OmegaGW$. Other detection channels in cosmology exist as
well. The $B$-mode polarisation of the Cosmic Microwave Background (CMB)
anisotropies is sensitive to spin one and spin two metric fluctuations
and it can be used to constrain gravitational
waves~\cite{Kamionkowski:1996zd, Hu:1997mn}. Observational bounds are
given on the amplitude of the primordial (strain) power spectrum
$\calPh(k)$, at a given wavenumber~\cite{Akrami:2018odb,
  Ade:2020zig}. Here $k$ is the wavenumber associated with a
\emph{spatial} Fourier transform. Interferometers, and pulsar arrays,
are also sensitive to the strain of a passing GW, and the measurable
quantity is the so-called strain power spectral density
$\Sh(\omega)$~\cite{Romano:2016dpx}. Here $\omega$ is the angular
frequency associated with a \emph{temporal} Fourier transform.  Both
approaches are rooted in what is measurable within a given
apparatus. Apart from the Earth's and Solar System's motion, direct
detection experiments assume measurements to be done at a fixed
location. The parameter against which measurements are made, and
stochasticity can be inferred, is the time. In Cosmology the
situation is similar. The proper motion of the comoving observer,
which is along the time direction, is neglected and the parameters
against which measurements are made, and stochasticity can be
inferred, are the spatial coordinates. If we are in presence of free
gravitational plane waves, propagating in the vacuum, and if we neglect the
expansion of the Universe, General Relativity tells us that
$\omega=\pm k$. This is certainly a very good assumption, today, but
only if the sources are switched off, and/or become rapidly
uncorrelated within the time/length scale of the
measurements. Otherwise, at a given location, one could be measuring
the correlated superimposition of all past emitted GW and the overall
signal can be quite complex. There are known physical situations, even
in Minkowski spacetime, for which the spatial versus temporal
extension of the sources have been shown to drastically change the
observed signal~\cite{Caprini:2006rd, Caprini:2009fx,
  Caprini:2009yp}. Moreover, the definition of stochasticity with
respect to time, or space, is not necessarily the same. Indeed, the
Friedmann-Lema\^{\i}tre-Roberton-Walker (FLRW) metric explicitly
breaks the time translation invariance and cosmological quantities do
not only depend on the time difference between two events. Even though
the expansion of the Universe can certainly be neglected during short
time intervals, cumulative effects could appear. For instance, the
breaking of the time-reversal symmetry in FLRW spacetime allows for a
cosmic strings network to generate non-Gaussianities in the CMB in the
form of a non-vanishing bispectrum~\cite{Hindmarsh:2009qk}. Would the
network evolve in a Minkowski background, the induced CMB bispectrum
would exactly vanish. It is however a common practice to not make a
distinction between $k$ and $\omega$ in the cosmological predictions
when comparison to direct detection bounds is needed. Also, simple
scaling relations are also assumed to hold between energy and strain,
such as $\OmegaGW \simeq k^2/(12\calH^2) \calPh$, where $\calH$ is
the conformal Hubble parameter. In view of the previous
discussion, one may wonder whether doing these replacements are always
justified.

In this paper, we revisit the derivation of the SGWB generated by
cosmological sources that are present for extended periods of time
during the expansion of the Universe. As a physically motivated
situation, we restrict our analysis to scaling defects. For those, the
sources remain self-similar with the Hubble radius at all times and
they can be of very small spatial extension in some directions while
being infinite in the others, depending on their topology. Our approach
assumes spatial stochasticity, which is compatible with the
statistical spatial-translation symmetry of the FLRW metric. All the
time-dependent terms, at all length scales, are kept and this allows
us to derive the corresponding strain and energy two-point
correlation functions at unequal times. Indeed, if the stochastic
sources of GW are correlated in time, one could expect to have
non-trivial unequal-time correlations as well. In this respect, our
findings extend the work of Refs.~\cite{Figueroa:2012kw,
  Figueroa:2020lvo}, in which the equal-time power spectrum of any
scaling defects have been derived. Under some conditions that we
discuss in section~\ref{sec:scalsrc}, taking the equal-time average of
our results gives back the spectra presented in these works. However,
when these conditions are not met, as it could be the case for cosmic
strings in the matter era, we find that new effects can show up at all
wavenumbers such as the appearance of interferences and violations of
the relation $\OmegaGW \simeq k^2/(12\calH^2) \calPh$.

The paper is organised as follows. In section~\ref{sec:evol} we recap
the Green's functions method to solve for the linearised tensor metric
perturbations, and their time derivative, around a FLRW metric, in
presence of sources. Compared to previous works, a special attention
has been paid to not discard the Hubble expansion terms and to
properly match the radiation and matter era solutions. In
section~\ref{sec:scalsrc}, we formally solve the evolution equations
in the case of interest, namely when the anisotropic stress is
generated by scaling defects in both the radiation and matter era. We
then prove in section~\ref{sec:convolution} that when the Fourier
transform of the anisotropic stress is holomorphic, a situation
associated with extinct and smooth sources, it is possible to exactly
evaluate the unequal-time waveform of the two-point correlation
functions associated with the strain and the energy density. These
expressions are derived in the main text and
appendix~\ref{sec:extinct}. For these well-behaved sources, averaging
the fine structure of the correlators gives back the standard
expectations. As motivated counter-examples, we discuss in
section~\ref{sec:constant} the case of ``constant'' sources and in
section~\ref{sec:singular} the case of ``singular'' sources, both
inducing a non-holomorphic Fourier transform of their correlators. We
find a very different fine structure than the one associated with
extinct sources, the most pronounced effects being induced by singular
sources in the matter era for which we consider a cosmic strings-like
correlator. All along the paper, we are keeping the time-dependence of
the observables, and this ensures that the waveform of the correlators
contains both the full spatial and temporal information. A critical
discussion and possible observable implications of our results are
finally presented in the conclusion, in section~\ref{sec:conclusion}.

\section{Evolution equations}
\label{sec:evol}

In this section, we introduce our notations and recap the basic
equations governing the evolution of tensor mode fluctuations in a
FLRW metric. From the Green's function method, we then derive the
formal solutions in presence of sources for both the matter and
radiation era, at all length scales, and through the transition
radiation to matter.

\subsection{Linearised tensor modes}

We consider $h_{ij}(\eta,\bx)$ to be the divergenceless and traceless
gauge invariant tensor fluctuations around a FLRW metric, i.e., the
line element reads
\begin{equation}
\ud s^2 = a^2(\eta)\left\{-\ud \eta^2 + \left[\delta_{ij} +
  h_{ij}(\eta,\bx) \right] \ud x^i \ud x^j  \right\},
\label{eq:pertmetric}
\end{equation}
where all scalar and vector perturbations are assumed to
vanish. Furthermore, we will be working in Fourier space and decompose
\begin{equation}
h_{ij}(\eta,\bx)= \dfrac{1}{(2 \pi)^3} \int_{-\infty}^{\infty} h_{ij}(\eta,\bk)
   e^{\imath \bk
    \bx} \ud^3\bk,
\label{eq:FTh}
\end{equation}
where $\imath^2=-1$. In the helicity basis~\cite{Zaldarriaga:1996xe,
  Alexander:2004wk}, the polarisation degrees of freedom of the
gravitational waves become manifest and one has
\begin{equation}
h_{ij}(\eta,\bk) = \sum_{r=-2,+2} h_{r}(\eta,\bk) \epsilon_{ij}^r(\bkhat),
\label{eq:hrdef}
\end{equation}
where the helicity basis tensor $\epsilon_{ij}^{r}(\bkhat)$ depends only on
the direction $\bkhat \equiv \bk/k$. In a spherical orthonormal
basis $(\bkhat,\be_1,\be_2)$, one can define the complex basis vectors
($\imath^2=-1$)
\begin{equation}
\bepsilon^{+1} \equiv \dfrac{\be_1 + \imath \be_2}{\sqrt{2}}\,,\qquad
\bepsilon^{-1} \equiv \dfrac{\be_1 - \imath \be_2}{\sqrt{2}}\,,
\end{equation}
from which the helicity basis tensor can be defined as $\bepsilon^{\pm
  2} = \bepsilon^{\pm 1} \otimes \bepsilon^{ \pm 1}$. Evaluated at the
same $(\bkhat,\be_1,\be_2)$, one has
\begin{equation}
\epsilon_{ij}^{r *} \epsilon^{ij}_s = \delta^r_s,
\end{equation}
and, since $h_{ij}(\eta,\bx)$ is a real number, one has
$\epsilon_{ij}^{\pm 2}(-\bkhat) = \epsilon_{ij}^{\pm 2
  *}(\bkhat)$. From equation~\eqref{eq:pertmetric}, the linearised
Einstein equations give, in the absence of spatial curvature,
\begin{equation}
h_{ij}'' + 2 \calH h_{ij}' - \Delta h_{ij} = \dfrac{2}{\Mp^2} a^2 \varPi_{ij},
\label{eq:einstein}
\end{equation}
where a prime denotes derivative with respect to the conformal
time. The reduced Planck mass is defined as $\Mp^2=
1/(8\pi\GN)$, $\calH= a'/a$ is the conformal Hubble parameter and the
anisotropic stress $a ^2 \varPi_{ij}(\eta,\bx)=\delta
T_{ij}^{\uTT}(\eta,\bx)$ is the divergenceless and traceless part of
the source stress tensor. After decomposing the anisotropic stress in
the helicity basis and defining the mode function
\begin{equation}
\mu_r \equiv a(\eta)h_r,
\label{eq:mudef}
\end{equation}
equation~\eqref{eq:einstein}, in Fourier space, simplifies to the
well-known equation of a sourced parametric
oscillator~\cite{Mukhanov:1990me}
\begin{equation}
\mu_r''(\eta,\bk) + \left(k^2 - \dfrac{a''}{a} \right) \mu_r(\eta,\bk)
= \dfrac{2}{\Mp^2} a^3 \varPi_r(\eta,\bk).
\label{eq:muevol}
\end{equation}
This equation shows that both helicy states propagate identically and,
at linear order, in an isotropic way. Exact solutions to this equation
can be derived assuming that the background expansion of the universe
is driven by a gravitating fluid having a constant equation of state
parameter.

\subsection{Green's functions}

For a dominating background cosmological fluid having $P=w \rho$, with
constant $w$, one has $\rho(\eta) \propto a^{-3(1+w)}$ and the first
Friedmann-Lema\^{\i}tre equation implies $a(\eta) \propto
\eta^{2/(1+3w)}$. The tensor modes verify
\begin{equation}
\mu_r'' + \left[k^2 - \dfrac{n(n+1)}{\eta^2} \right] \mu_r = \dfrac{2}{\Mp^2} a^3 \varPi_r,
\label{eq:riccatibessel}
\end{equation}
where we have introduced the constant
\begin{equation}
n \equiv \dfrac{1-3w}{1+3w}\,.
\end{equation}
In the radiation era $n=0$, in the matter era $n=1$ and for
cosmological constant domination $n=-2$. Under this form, the
homogeneous part of equation~\eqref{eq:riccatibessel} is a
Riccati-Bessel equation which admits analytical solutions for all
positive and negative integer values of $n$\footnote{See
  Ref.~\cite{Abramovitz:1970aa}, Eq.~(10.3.1).}. At fixed $k$, the two
linearly independent solutions are Riccati-Bessel functions
\begin{equation}
\rbu(\eta) = k\eta \, \sbj{n}(k\eta), \qquad \rbv(\eta) = k \eta \, \sby{n}(k\eta),
\label{eq:rbfuncs}
\end{equation}
where $\sbj{n}(x)$ and $\sby{n}(x)$ are the spherical Bessel functions
of order $n$. From these homogeneous solutions, one can immediately
construct the retarded Green's function $\G{\xi}(\eta,\bk)$ associated
with equation~\eqref{eq:riccatibessel}, in which the source term is
replaced by the distribution $\delta(\eta-\xi)$. It reads
\begin{equation}
\begin{aligned}
  \G{\xi}(\eta,k)& = \dfrac{\rbu(\xi) \rbv(\eta) - \rbv(\xi) \rbu(\eta)}{W(\xi)}
  \heaviside{\eta - \xi} \\
  & = \dfrac{(k \xi) (k \eta)}{k}\left[\sbj{n}(k\xi) \sby{n}(k\eta) -
    \sby{n}(k\xi) \sbj{n}(k\eta)\right]\heaviside{\eta - \xi},
\end{aligned}
\label{eq:green}
\end{equation}
where $\heaviside{x}$ is the Heaviside function and the Wronskian has
been simplified as~\cite{Abramovitz:1970aa}
\begin{equation}
W(\xi) \equiv \rbu(\xi) \rbv'(\xi) - \rbu'(\xi) \rbv(\xi) = k.
\end{equation}
Assuming that the source vanishes for $\eta < \etaini$, the solution
of equation~\eqref{eq:riccatibessel} finally reads
\begin{equation}
\mu_r(\eta,\bk) = \dfrac{2}{\Mp^2 k} \int_{\etaini}^{\eta} k
\G{\xi}(\eta,k) a^3(\xi) \varPi_r(\xi,\bk) \ud \xi.
\label{eq:musol}
\end{equation}
From the explicit expression \eqref{eq:green}, one can show that the
solution for $\mu_r'(\eta,\bk)$ takes the simple form
\begin{equation}
\mu_r'(\eta,\bk) = \dfrac{2}{\Mp^2} \int_{\etaini}^{\eta}
\G{\xi}'(\eta,k) a^3(\xi) \varPi_r(\xi,\bk) \ud \xi,
\label{eq:dmusol}
\end{equation}
where $\G{\xi}'$ stands for $\partial \G{\xi}(\eta,k)/\partial
\eta$. Equations~\eqref{eq:musol} and \eqref{eq:dmusol} are valid as
long as the expansion of the universe is associated with a constant
$n$ value. Therefore, they can be used if $\eta$ and $\etaini$, or the
support of the anisotropic stress, are confined within the radiation
era. In this case, one has
\begin{equation}
\rburad(\eta) = \sin(k\eta), \qquad \rbvrad(\eta) = -\cos(k\eta),
\qquad k
\Grad{\xi}(\eta,k) = \sin\left[k(\eta - \xi)\right] \heaviside{\eta-\xi}.
\label{eq:greenrad}
\end{equation}
Interestingly, the Green's function only depends on time differences due
to the conformal symmetry of the radiation era, and this ensures the
validity of the stationary assumption for \emph{free} gravitational
waves at all length scales. However, this is not necessarily the case
in presence of sources.

If $\eta$ and $\etaini$ belong to the matter era, one can use again
equations~\eqref{eq:musol} and \eqref{eq:dmusol} with
\begin{equation}
\begin{aligned}
\rbupmat(\eta) & = -\cos(k \eta) + \dfrac{\sin(k \eta)}{k \eta}\,, \qquad
\rbvpmat(\eta) = -\sin(k \eta) - \dfrac{\cos(k \eta)}{k \eta}\,, \\
k \Gpmat{\xi}(\eta,k) &= \left\{ \sin\left[k(\eta - \xi)\right] -
\dfrac{k(\eta-\xi)}{k\eta \, k\xi} \cos\left[k(\eta-\xi)\right] +
\dfrac{1}{k\eta \, k\xi} \sin\left[k(\eta-\xi)\right] \right\} \heaviside{\eta-\xi}.
\end{aligned}  
\label{eq:greenpmat}
\end{equation}

For $\etaini$ in the radiation era and $\eta$ in the matter era, it is
still possible to find a solution provided one assumes an
instantaneous transition between the two eras. In that case, one can
split the time support of the anisotropic stress into radiation and
matter era, and consider that all gravitational waves sourced in the
radiation era freely propagate into the matter era on top of the ones
sourced during the matter era (the equations are linear). Notice that
there is no analytical solution of equation~\eqref{eq:muevol} for a
mixture of matter and radiation. In the following, the perturbation
modes will be approximated as evolving through an instantaneous
transition.

\subsection{Transition radiation-matter}
\label{sec:transition}

In order to implement the transition radiation to matter for the
perturbation modes, one has to determine how radiation generated
gravitational waves propagate into the matter era. The usual approach
to this problem is to ignore the transition and extend the solution of
$\mu_r$ into the matter era. Although this is justified on small
scales, because both Green's functions asymptote to the same functional
form [see Eqs.~\eqref{eq:greenrad} and \eqref{eq:greenpmat}], they do
significantly differ on large scales and the matching requires to
properly patch the radiation and matter era manifolds together, up to
order one in the metric perturbations. The matching conditions for
cosmological perturbations, and thus gravitational waves, require
continuity of both the background metric, namely $a(\eta)$ and
$a'(\eta)$, as well as the continuity of $h_{ij}$ and
$h'_{ij}$~\cite{Deruelle:1995kd, Martin:1997zd, Watanabe:2006qe}.
Therefore, $\mu_r$ and $\mu_r'$ are also continuous at the transition
radiation to matter.

Let us first ensure continuity of the background metric. Solving the
matter era Friemann-Lema\^{\i}tre equations and ensuring continuity of
the scale factor and the Hubble parameter gives the unique solution
\begin{equation}
a(\eta \ge \etaeq) = \dfrac{\calH^2(\eta_0)}{4} \OmegaM \left(\eta+\etaeq \right)^2,
\label{eq:amat}
\end{equation}
with $a(\eta_0)=1$,
\begin{equation}
\etaeq \equiv \dfrac{\sqrt{\OmegaR}}{\calH(\eta_0) \OmegaM}\,,
\end{equation}
and where $\OmegaR$ and $\OmegaM$ are the density parameters of
radiation and matter, today. Equation~\eqref{eq:amat} shows that, in a
matter era preceded by a radiation era, one does not have $a \propto
\eta^2$ but $a \propto (\eta + \etaeq)^2$. The modifications induced
on the mode functions, and on the Green's function, are however
trivial and obtained by shifting the time variable accordingly. One
gets
\begin{equation}
\begin{aligned}
  \rbumat(\eta) & = \rbupmat(\eta+\etaeq) =
-\cos\left[k(\eta+\etaeq)\right] +
\dfrac{\sin\left[k(\eta+\etaeq)\right]}{k\left(\eta+\etaeq\right)}\,,
\\
\rbvmat(\eta) & = \rbvpmat(\eta+\etaeq) =
-\sin\left[k(\eta+\etaeq)\right] - \dfrac{\cos\left[k(\eta+\etaeq)\right]}{k(\eta+\etaeq)}\,,
\label{eq:uvmat}
\end{aligned}
\end{equation}
and
\begin{equation}
\begin{aligned}
k \Gmat{\xi}(\eta,k) &= \left\{ \sin\left[k(\eta - \xi)\right] -
\dfrac{k(\eta-\xi)}{k(\eta+\etaeq) \, k(\xi+\etaeq)}
\cos\left[k(\eta-\xi)\right] \right. \\ & \left. +
\dfrac{1}{k(\eta+\etaeq) \, k(\xi+\etaeq)} \sin\left[k(\eta-\xi)\right] \right\} \heaviside{\eta-\xi}.
\end{aligned}  
\label{eq:greenmat}
\end{equation}

Let us now consider the matching of the tensor perturbation modes. Dropping the
explicit dependence in the helicity state and assuming the
radiation-era solution of the mode function $\murad(\eta,k)$ to be known,
it propagates during the matter era as
\begin{equation}
\mu(\eta \ge\etaeq,k) = c_1(k) \rbumat(\eta,k) + c_2(k) \rbvmat(\eta,k),
\end{equation}
where the continuity conditions at $\eta=\etaeq$ read
\begin{equation}
  \begin{aligned}
    \murad(\etaeq,k) & = c_1(k) \rbumat(\etaeq,k) + c_2(k) \rbvmat(\etaeq,k),\\
    \murad'(\etaeq,k) & =c_1(k) \rbumat'(\etaeq,k) + c_2(k) \rbvmat'(\etaeq,k).
  \end{aligned}
\end{equation}
The matter-era mode functions entering these equations are given by
equation~\eqref{eq:uvmat} and \emph{not} by
equation~\eqref{eq:greenpmat}.  These equations uniquely determine the
functions $c_1(k)$ and $c_2(k)$, and, after some algebra, one gets for
both helicity states
\begin{equation}
  \mu(\eta \ge \etaeq,k) = \A(x,\xeq) \murad(\etaeq,k) + \dfrac{\B(x,\xeq)}{k} \murad'(\etaeq,k),
\label{eq:muradtomat}
\end{equation}
where we have defined
\begin{equation}
  x \equiv k \eta
\end{equation}
and
\begin{equation}
  \begin{aligned}
    \A(x,\xeq) &\equiv \dfrac{\xeq - x +4 x \xeq ^2 + 4
      \xeq^3}{4\xeq^2(x+\xeq)} \cos(x-\xeq) + \dfrac{1+2 x \xeq - 2\xeq^2}{4\xeq^2(x+\xeq)} \sin(x-\xeq),
    \\ \B(x,\xeq) & \equiv \dfrac{1 + 2 x \xeq +
      2\xeq^2}{2\xeq(x+\xeq)} \sin(x-\xeq) + \dfrac{\xeq -x}{2\xeq(x+\xeq)}\cos(x-\xeq).
  \end{aligned}
\end{equation}
The evolution of $\mu'(\eta \ge \etaeq,k)$ is also uniquely determined
from equation~\eqref{eq:muradtomat} and reads
\begin{equation}
  \mu'(\eta \ge \etaeq,k) = k \dA(x,\xeq) \murad(\etaeq,k) + \dB(x,\xeq) \murad'(\etaeq,k),
\label{eq:dmuradtomat}
\end{equation}
with $\dA \equiv \partial \A/\partial x$ and $\dB \equiv \partial \B/\partial x$, or explicitly
\begin{equation}
  \begin{aligned}
    \dA(x,\xeq) &= \dfrac{x-\xeq + 2 x^2 \xeq - 2 \xeq^3}{4
      \xeq^2(x+\xeq)^2} \cos(x-\xeq) \\ & + \dfrac{-1 + x^2 +3\xeq^3
      -8 x \xeq^3 - 4x^2 \xeq^2 - 4 \xeq^4}{4\xeq^2(x+\xeq)^2}
    \sin(x-\xeq),\\ \dB(x,\xeq) &= \dfrac{x-\xeq +4 x \xeq^2 + 2x^2\xeq 
      + 2\xeq^3}{2\xeq(x+\xeq)^2} \cos(x-\xeq) +
    \dfrac{-1+x^2-\xeq^2}{2\xeq(x+\xeq)^2} \sin(x-\xeq).
  \end{aligned}
\end{equation}

\subsection{Unequal-time power spectra}

\label{sec:averaging}

Among the simplest statistical properties that one can measure over a
gravitational wave background is the unpolarised spatial two-point
correlation function of the strain, i.e.,
\begin{equation}
\mean{h_{ij}(\eta_1,\bx) h^{ij}(\eta_2,\bx+\by)}_{V} \equiv \dfrac{1}{V}
\int h_{ij}(\eta_1,\bx) h^{ij}(\eta_2,\bx + \by) \ud^3\bx,
\end{equation}
where $V$ is the (infinite) volume over which the averaging is
performed. By construction, this function depends on $\by$ only. Using
the Fourier and helicity state decomposition of
equations~\eqref{eq:FTh} and \eqref{eq:hrdef}, over $V$, one gets
\begin{equation}
\begin{aligned}
  \mean{h_{ij}(\eta_1,\bx) h^{ij}(\eta_2,\bx+\by)}_{V} & = \dfrac{V}{(2
  \pi)^3} \int \sum_{r,s} \epsilon_{ij}^r(\bkhat)
\epsilon^{ij}_s(\bqhat) h_r(\eta_1,\bk) h^s(\eta_2,\bq) e^{\imath \bq
  \by} \dirac{\bk+\bq}\ud^3 \bk \ud^3 \bq \\
& = \dfrac{V}{(2\pi)^3} \int \ud^3 \bq \left[\sum_r h_r^*(\eta_1,\bq)
  h_r(\eta_2,\bq)\right] e^{\imath \bq\by},
\end{aligned}
\label{eq:twopoint}
\end{equation}
which is the inverse Fourier transform (over the volume $V$) of the
total strain power spectrum $P_h = \sum_r P_{h_r}$ with
\begin{equation}
P_{h_r}(\eta_1,\eta_2,\bk) \equiv h_r^*(\eta_1,\bk) h_r(\eta_2,\bk).
\end{equation}
Without any additional assumption, the spatial averaging could, in
principle, depends on the volume location. However, in a FLRW
space-time, statistical invariance by translation ensures that this is
not the case and that the result cannot depend on $\bx$ nor its domain
$V$. For this reason, it is equally possible to \emph{define} an
ensemble average by immediately enforcing statistical invariance by
translation. This amounts to define the ensemble average by
\begin{equation}
\mean{h_r^*(\eta_1,\bk) h_r(\eta_2,\bq)} \equiv \dfrac{(2\pi)^3}{V}
\dirac{\bk-\bq} P_{h_r}(\eta_1,\eta_2,\bk),
\end{equation}
which ensures the absence of correlations between different wave
vectors. As this derivation shows, there is no reason, a priori, to
have correlations depending only on the time difference $\eta_2 -
\eta_1$.

Further simplifications can however be made using the expected
statistical isotropy of the cosmological sources. This symmetry of the
FLRW metric implies that $P_{h_r}(\eta_1,\eta_2,\bk)$ depends on $k$
only, and not on $\bkhat$, such that equation~\eqref{eq:twopoint}
becomes also isotropic and reads
\begin{equation}
\mean{h_{ij}(\eta_1,\bx) h^{ij}(\eta_2,\bx+\by)} = \int_0^\infty \dfrac{\ud q}{q} 
\calP_h(\eta_1,\eta_2,q) \sinc(q y).
\end{equation}
Here $\sinc(x) \equiv \sin(x)/x$ is the sine cardinal function and we
have introduced the (spherical) strain power spectrum $\calP_h=  \sum_r
\calP_{h_r}$ with
\begin{equation}
\calP_{h_r}(\eta_1,\eta_2,k) \equiv \dfrac{k^3 V}{2\pi^2} P_{h_r}(\eta_1,\eta_2,k).
\label{eq:calPdef}
\end{equation}
This is the quantity constrained by CMB
measurements~\cite{Kuroyanagi:2013ns}. In the following we also
consider the power spectra constructed on $h_r'$, $\mu_r$ and
$\mu_r'$.

\subsection{Generalised energy density parameter}

As mentioned in the introduction, a few cosmological constraints are
associated with the overall gravitating effects of gravitational
waves. For this reason, one can also define the following two-point
correlation function
\begin{equation}
\rhogw(\eta_1,\eta_2,\by) \equiv \dfrac{\Mp^2}{4 a(\eta_1) a(\eta_2)}
\mean{h_{ij}'(\eta_1,\bx) {h^{ij}}'(\eta_2,\bx + \by)}.
\label{eq:rhogwdef}
\end{equation}
At equal times, and vanishing spatial separation $y=0$, this
expression gives back the energy density of gravitational waves given
by the leading term of Landau-Lifchitz pseudo stress
tensor~\cite{Landau:1989}. Let us notice that the ensemble average is
on space, which is precisely accounting for the cumulative
gravitational effects of all gravitational waves at a given time. At
unequal times and non-vanishing spatial separation,
equation~\eqref{eq:rhogwdef} gives how the derivatives of
$h_{ij}(\eta,\bx)$ are correlated in space. Exactly as detailed in
section~\ref{sec:averaging}, in FLRW, one can decompose $\rhogw$ in
Fourier space as
\begin{equation}
\begin{aligned}
  \rhogw(\eta_1,\eta_2,y) &= \dfrac{V \Mp^2}{4 a(\eta_1) a(\eta_2)}
  \int_0^\infty \dfrac{\ud k}{k} \sum_r \dfrac{k^3}{2\pi^2}
  P_{h'_r}(\eta_1,\eta_2,k) \sinc(k y) \\
  & \equiv \int_0^{\infty} \dfrac{\ud k}{k} \dfrac{\ud
    \rhogw}{\ud \ln k}(\eta_1,\eta_2,k) \sinc(ky),
\end{aligned}
\label{eq:rhogw}
\end{equation}
where the last line \emph{defines} the energy density per logarithmic
wavenumber. The density parameter in real space is defined as
$\OmegaGW = \rhogw(\eta,\eta,0)/\rhoc$, where $\rhoc(\eta) \equiv 3
\Mpl^2 H^2(\eta)$ is the critical density, $H$ being the Hubble
parameter. From equation~\eqref{eq:rhogw}, we can generalise this
definition to unequal times and distinct spatial locations, the  Fourier
transform of which being
\begin{equation}
\OmegaGW(\eta_1,\eta_2,k) \equiv \dfrac{1}{\sqrt{\rhoc(\eta_1)}
  \sqrt{\rhoc(\eta_2)}} \dfrac{\ud \rhogw(\eta_1,\eta_2,k)}{\ud \ln k}
= \dfrac{\sum_r \calP_{h_r'}(\eta_1,\eta_2,k)}{12 \calH(\eta_1)
  \calH(\eta_2)}\,,
\label{eq:OmegaGW}
\end{equation}
where the last equality comes from equation~\eqref{eq:calPdef}. It is
a dimensionless quantity whose expression gives back the usual
definition when considered at equal times~\cite{Caprini:2018mtu}. In
the next section, we use the Green's functions method to determine the
actual values of these correlators in presence of sources.

\subsection{Matching power spectra}

The observable quantities we are interested in are
$\calP_h(\eta_1,\eta_2,k)$ and $\OmegaGW(\eta_1,\eta_2,k)$, or,
equivalently, the unequal times power spectra $P_{h_r}$ and
$P_{h_r'}$. From the definition~\eqref{eq:mudef} one has
\begin{equation}
\begin{aligned}
  P_{h_r}(\eta_1,\eta_2,k) & =
  \dfrac{P_{\mu_r}(\eta_1,\eta_2,k)}{a(\eta_1) a(\eta_2)}\,, \\
  P_{h_r'} (\eta_1,\eta_2,k) & = H(\eta_1) H(\eta_2) \left[
    \dfrac{P_{\mu_r'}(\eta_1,\eta_2,k)}{\calH(\eta_1) \calH(\eta_2)} + P_{\mu_r}(\eta_1,\eta_2,k) -
    \dfrac{P_{\x_r}(\eta_1,\eta_2,k)}{\calH(\eta_1)} \right. \\ &
    \left. - 
    \dfrac{P_{\xbar_r}(\eta_1,\eta_2,k)}{\calH(\eta_2)} \right],
\end{aligned}
\label{eq:Phs}
\end{equation}
where $P_{\x_r}$ and $P_{\xbar_r}$ are the cross power spectra
\begin{equation}
P_{\x_r}(\eta_1,\eta_2,k) = {\mu_r'}^*(\eta_1,k)
  \mu_r(\eta_2,k), \qquad P_{\xbar_r}(\eta_1,\eta_2,k) =
\mu_r^*(\eta_1,k) \mu_r'(\eta_2,k).
\end{equation}
They are not independent as one has $P_{\xbar_r}(\eta_1,\eta_2,k) =
P_{\x_r}^*(\eta_2,\eta_1,k)$. In presence of sources, using the mode
evolution equations~\eqref{eq:musol} and \eqref{eq:dmusol}, these
power spectra are given by
\begin{equation}
\begin{aligned}
  P_{\mu_r}(\eta_1,\eta_2,\bk) & = \dfrac{4}{k^2 \Mp^4}
  \int_{\etaini}^{\eta_1} \ud \xi \int_{\etaini}^{\eta_2} \ud \xi'
  k \G{\xi}^*(\eta_1,k) \, k \G{\xi'}(\eta_2,k)
  a^3(\xi) a^3(\xi') \varPi_r^*(\xi,\bk) \varPi_r(\xi',\bk), \\
  P_{\mu_r'}(\eta_1,\eta_2,\bk) & = \dfrac{4}{\Mp^4} \int_{\etaini}^{\eta_1} \ud \xi \int_{\etaini}^{\eta_2} \ud \xi'
  {\G{\xi}'}^*(\eta_1,k) \G{\xi'}'(\eta_2,k)
  a^3(\xi) a^3(\xi') \varPi_r^*(\xi,\bk) \varPi_r(\xi',\bk), \\
  P_{\x_r}(\eta_1,\eta_2,\bk) & = \dfrac{4}{k \Mp^4}
  \int_{\etaini}^{\eta_1} \ud \xi \int_{\etaini}^{\eta_2} \ud \xi'
  {\G{\xi}'}^*(\eta_1,k) \, k \G{\xi'}(\eta_2,k)
  a^3(\xi) a^3(\xi') \varPi_r^*(\xi,\bk) \varPi_r(\xi',\bk).
\end{aligned}
\label{eq:Pmus}
\end{equation}
It is important to recall that these expressions are valid only for
$\etaini$, $\eta_1$ and $\eta_2$ belonging to the same expansion era
within each of the integration domain. However, using the matching
condition of section~\ref{sec:transition}, it is possible to freely
propagate the radiation-era modes into the matter era. In that
situation, and assuming for the time being that $\varPi_r(\eta \ge
\etaeq,\bk)=\varPi_r^\urad(\eta,\bk) \heaviside{\etaeq - \eta}$, one
has the following relation
\begin{equation}
\begin{aligned}
  P_{\mu_r}^\urad(\eta_1>\etaeq,\eta_2>\etaeq,\bk) & = \A_1 \A_2 P_{\mu_r}^\urad(\etaeq,\etaeq,\bk) +
\dfrac{\B_1 \B_2}{k^2} P_{\mu_r'}^\urad(\etaeq,\etaeq,\bk) \\ & + \dfrac{\A_1
  \B_2 + \B_1 \A_2}{k} P_{\x_r}^\urad(\etaeq,\etaeq,\bk).
\label{eq:Pmuprop}
\end{aligned}
\end{equation}
In this expression, we have used the shortcut notations $\A_i=\A(k
\eta_i,k\etaeq)$ and $\B_i = \B(k \eta_i,k\etaeq)$, these functions
being given in section~\ref{sec:transition}. Similarly, the other
unequal times spectra are given by
\begin{equation}
\begin{aligned}
  P_{\mu_r'}^\urad(\eta_1> \etaeq,\eta_2>\etaeq,\bk) & = \dB_1 \dB_2
P_{\mu_r'}^\urad(\etaeq,\etaeq,\bk) + k^2 \dA_1 \dA_2
P_{\mu_r}^\urad(\etaeq,\etaeq,\bk) \\ & + k\left(\dA_1 \dB_2 + \dA_2 \dB_1\right) P_{\x_r}^\urad(\etaeq,\etaeq,\bk),
\end{aligned}
\label{eq:Pdmuprop}
\end{equation}
and
\begin{equation}
  \begin{aligned}
    P_{\x_r}^\urad(\eta_1>\etaeq,\eta_2>\etaeq,\bk) & = \left(\dA_1
    \B_2 + \dB_1 \A_2 \right) P_{\x_r}^\urad(\etaeq,\etaeq,\bk) + k
    \dA_1 \A_2 P_{\mu_r}^\urad(\etaeq,\etaeq,\bk) \\ & + \dfrac{\dB_1
      \B_2}{k} P_{\mu_r'}^\urad(\etaeq,\etaeq,\bk),
  \end{aligned}
\label{eq:Pxprop}
\end{equation}
again with $\dA_i=\dA(k \eta_i,k\etaeq)$ and $\dB_i=\dB(k
\eta_i,k\etaeq)$. Equations~\eqref{eq:Pmuprop} to \eqref{eq:Pxprop}
contain all the correlations induced by sources confined within the
radiation era and measured in the matter era. One should add to these
terms the contribution of sources confined in the matter era and the total
anisotropic stress is of the form
\begin{equation}
\varPi_r(\eta,\bk) = \varPi_r^\urad(\eta,\bk)
\heaviside{\etaeq - \eta} + \varPi_r^\umat(\eta,\bk)
\heaviside{\eta-\etaeq}.
\label{eq:radmatvarpi}
\end{equation}
Plugging this expression into equation~\eqref{eq:Pmus} gives three
contributions
\begin{equation}
\begin{aligned}
&   P_{\mu_r}(\eta_1>\etaeq,\eta_2>\etaeq,\bk)  =
P_{\mu_r}^\urad(\eta_1,\eta_2,\bk) +
P_{\mu_r}^\umat(\eta_1,\eta_2,\bk)  \\ & +
\left[ \A_1 P_{\mu_r}^\umix(\etaeq,\eta_2,\bk) + \dfrac{\B_1}{k}
P_{\x_r}^\umix(\etaeq,\eta_2,\bk) + \A_2
P_{\mu_r}^\umix(\eta_1,\etaeq,\bk) + \dfrac{\B_2}{k}
P_{\xbar_r}^\umix(\eta_1,\etaeq,\bk) \right],
\label{eq:Pmuall}
\end{aligned}
\end{equation}
where the radiation contribution, first term, is given by
equation~\eqref{eq:Pdmuprop}. The second term is the matter era part
given by equation~\eqref{eq:Pmus} with the replacement $\etaini
\rightarrow \etaeq$ and $\varPi_r \rightarrow \varPi_r^\umat$. The
last term, in brackets, encodes the possible cross-correlations
between modes generated in the radiation era, propagated into the
matter era, and the modes generated in the matter era. Explicitly,
one has
\begin{equation}
\begin{aligned}
  P_{\mu_r}^\umix(\etaeq,\eta_2>\etaeq,\bk) & =  \dfrac{4}{k^2 \Mp^4}
  \int_{\etaini}^{\etaeq} \ud \xi \int_{\etaeq}^{\eta_2} \ud \xi'
  k {\G{\xi}^\urad}^*(\etaeq,k) \, k \G{\xi'}^\umat(\eta_2,k)
  a^3(\xi) a^3(\xi') \\ & \times {\varPi_r^\urad}^*(\xi,\bk)\varPi_r^\umat(\xi',\bk),
\label{eq:Pmumix}
\end{aligned}
\end{equation}
and equivalent expressions for $P_{\mu_r'}^\umix$ and
$P_{\kappa_r}^\umix$, see equation~\eqref{eq:Pmus}. The other power
spectra can be derived in the same way and they read
\begin{equation}
  \begin{aligned}
& P_{\mu_r'}(\eta_1> \etaeq,\eta_2 > \etaeq,\bk) =
    P_{\mu_r'}^\urad(\eta_1,\eta_2,\bk) +
    P_{\mu_r'}^\umat(\eta_1,\eta_2,\bk) \\ &+ \left[k \dA_1
      P_{\xbar_r}^\umix(\etaeq,\eta_2,\bk) + \dB_1
      P_{\mu_r'}^\umix(\etaeq,\eta_2,\bk) + k\dA_1
      P_{\x_r}^\umix(\eta_1,\etaeq,\bk) + \dB_2
      P_{\mu_r'}^\umix(\eta_1,\etaeq,\bk) \right],
    \label{eq:Pdmuall}
  \end{aligned}
\end{equation}
and
\begin{equation}
  \begin{aligned}
    & P_{\x_r}(\eta_1 > \etaeq,\eta_2 > \etaeq,\bk) =
    P_{\x_r}^\urad(\eta_1,\eta_2,\bk) +
    P_{\x_r}^\umat(\eta_1,\eta_2,\bk) \\ & + \left[\A_2
      P_{\x_r}^\umix(\eta_1,\etaeq,\bk) + \dfrac{\B_2}{k}
      P_{\mu_r'}^\umix(\eta_1,\etaeq,\bk) + k \dA_1
      P_{\mu_r}^\umix(\etaeq,\eta_2,\bk) + \dB_1
      P_{\x_r}^\umix(\etaeq,\eta_2,\bk) \right].      
\label{eq:Pxall}
  \end{aligned}
\end{equation}
As can be seen in equation~\eqref{eq:Pmumix}, the domains of the
integrals associated with $\xi$ and $\xi'$ do not overlap, and,
because physical sources decorrelate at large unequal times, these
terms are expected to be small. Moreover, our assumption of an
instantaneous transition between radiation and matter would not allow
us to estimate accurately their (small) value. Indeed, most of the
contribution to $P_{\mu_r}^\umix$ comes from $\xi \lesssim \etaeq$ and
$\xi' \gtrsim \etaeq$, i.e., close to equality between radiation and
matter, precisely when the exact Green's function of
equation~\eqref{eq:muevol} could be significantly different than the
ones associated with pure radiation and matter eras. As a result, a
numerical integration of the Green's function during the transition
radiation to matter would be required to accurately determine these
terms~\cite{inprogress}. For these reasons, the cross-correlations
between radiation and matter era will be neglected in the following.

In order to get some insight into the behaviour of these solutions, we
now focus our discussion to the case of scaling sources.

\section{Cosmological solutions for scaling sources}
\label{sec:scalsrc}

\subsection{Isotropic scaling sources}

We define as isotropic scaling sources any cosmological objects
having an unequal-time correlator for the anisotropic stress verifying
\begin{equation}
\mean{\varPi_r^*(\xi,\bk) \varPi_r(\xi',\bq)} = \dfrac{(2 \pi)^3}{V}
\dirac{\bk - \bq}
\dfrac{M^4 \calU_r(k\xi,k\xi')}{a^2(\xi) \sqrt{\xi} \, a^2(\xi') \sqrt{\xi'}}\,.
\label{eq:scaling}
\end{equation}
Such an expression is motivated by the universal attractor reached by
the stress-tensor of cosmic defects in an expanding
universe~\cite{Durrer:1997ep, Durrer:2002, Bevis:2006mj,
  Lazanu:2014xxa}. The dimensionless function $\calU(x,x')$ is
peculiar to each type of defects, but causality requires that it
should be analytic at small $x$~\cite{Turok:1996ud}. Also, it is
expected to vanish for large values of $x$ and $x'$, but the precise
asymptotic behaviour is very much dependent on the defect
topology~\cite{Wu:2002}. Equation~\eqref{eq:scaling} is expected to be
violated only during the transition radiation to matter as the scaling
solutions in both era usually differ. In the following, this effect is
ignored as we deal with an instantaneous transition and we introduce
the corresponding scaling functions $\calU_r^\urad(x,x')$ and
$\calU_r^\umat(x,x')$. This assumption also implies that the
correlations between radiation and matter eras are also neglected,
which consists in ignoring all ``mix'' terms appearing in
equations~\eqref{eq:Pmuall}, \eqref{eq:Pdmuall} and \eqref{eq:Pxall}.

The interest in focusing on scaling sources is that
equation~\eqref{eq:scaling} greatly simplifies the integrals appearing
in the power spectra~\eqref{eq:Pmus} and allows us to derive a closed
form expression.

\subsection{Strain spectrum today}

\label{sec:straintoday}

The gravitational wave power spectrum at unequal times that is an
observable for direct detection is $\calP_{h}(\eta_1,\eta_2,k)$, where
it is understood that both times are within the matter era (probably
close to the current conformal time $\eta_0$). From the previous
discussion, it can be split into two contributions
\begin{equation}
\calPh(\eta_1,\eta_2,k) = \calPh^{\umat}(\eta_1,\eta_2,k)+
\calPh^{\urad}(\eta_1,\eta_2,k).
\label{eq:straintot}
\end{equation}
From equations~\eqref{eq:calPdef}, \eqref{eq:Phs}, \eqref{eq:Pmus} and
\eqref{eq:scaling}, the first term can be expressed as
\begin{equation}
  \calPh^{\umat}(\eta_1,\eta_2,k) = 128 \left(\GN M^2
  \right)^2 \Imumat(x_1,x_2,k),
\label{eq:strainmat}
\end{equation}
with
\begin{equation}
\Imumat(x_1,x_2,k) = \dfrac{1}{a(\eta_1)a(\eta_2)} \int_{\xeq}^{x_1}\ud x
\int_{\xeq}^{x_2} \ud x' \Ksmat(x_1,x) \Ksmat(x_2,x')
\dfrac{a\left(\frac{x}{k}\right) a\left(\frac{x'}{k}\right)}{\sqrt{x
    x'}} \calU^\umat(x,x'),
\label{eq:Imumat}
\end{equation}
where $x_1 = k \eta_1$ and $x_2 = k \eta_2$. The correlator stands for
$\calU^{\umat} \equiv \sum_r \calU_r^\umat$ and we have defined the
convolutional kernel of the strain in the matter era as\footnote{The
  matching function $\B$ being a Wronskian, it is related to the
  strain kernel by $\B = \Ksmat(x,\xeq)$. The functions $A$ are not
  given by a Wronskian and there is not similar relation from
  them. For clarity, we keep both notation distinct.}
\begin{equation}
\begin{aligned}
  \Ksmat(x_i,x) & \equiv k
  \G{\frac{x}{k}}^\umat\left(\frac{x_i}{k}\right)
   = \left[1 + \dfrac{1}{(x_i+\xeq)(x+\xeq)} \right] \sin(x_i-x) \\ & + \left(\dfrac{1}{x_i+\xeq} -
   \dfrac{1}{x+\xeq}\right) \cos(x_i-x).
   \end{aligned}
\label{eq:Ksmat}
\end{equation}

The second term of equation~\eqref{eq:straintot} requires more
attention. If $\eta_1$ and $\eta_2$ were in the radiation era, one
would get
\begin{equation}
  \calPh^{\urad}(\eta_1<\etaeq,\eta_2<\etaeq,k) = 128 \left(\GN M^2
  \right)^2 \Imurad(x_1,x_2,k),
\label{eq:strainradinrad}
\end{equation}
with
\begin{equation}
\Imurad(x_1,x_2,k)  = \dfrac{1}{a(\eta_1)a(\eta_2)} \int_{\xini}^{x_1}\ud x
\int_{\xini}^{x_2} \ud x' \Ksrad(x_1,x) \Ksrad(x_2,x')
\dfrac{a\left(\frac{x}{k}\right) a\left(\frac{x'}{k}\right)}{\sqrt{x
    x'}} \calU^\urad(x,x').
\label{eq:Imurad}
\end{equation}
For the case where it needs to be evaluated in the matter era, this
solution is maximally extended to $\eta_1=\etaeq$ and $\eta_2=\etaeq$,
matched and freely propagated into the matter era. From
equations~\eqref{eq:calPdef}, \eqref{eq:Pmus}, \eqref{eq:Pmuprop} and
\eqref{eq:Imurad} one gets
\begin{equation}
\begin{aligned}
  \calPh^{\urad}(\eta_1> \etaeq,\eta_2>\etaeq,k) & = 128 \left(\GN M^2\right)^2
\dfrac{a^2(\etaeq)}{a(\eta_1) a(\eta_2)} \left[\A_1 \A_2
  \Imurad(\xeq,\xeq,k)  \right. \\  & \left. + \B_1 \B_2 \Idmurad(\xeq,\xeq,k)
   +\left(\A_1 \B_2 + \A_2 \B_1\right) \Ixrad(\xeq,\xeq,k) \right],
\end{aligned}
\label{eq:strainradinmat}
\end{equation}
with $\xeq = k \etaeq$ and
\begin{equation}
  \begin{aligned}    
\Idmurad(x_1,x_2,k) & = \dfrac{1}{a(\eta_1)a(\eta_2)} \int_{\xini}^{x_1}\ud x
\int_{\xini}^{x_2} \ud x' \Kerad(x_1,x) \Kerad(x_2,x')
\dfrac{a\left(\frac{x}{k}\right) a\left(\frac{x'}{k}\right)}{\sqrt{x
    x'}} \calU^\urad(x,x'),\\
\Ixrad(x_1,x_2,k) & = \dfrac{1}{a(\eta_1)a(\eta_2)} \int_{\xini}^{x_1}\ud x
\int_{\xini}^{x_2} \ud x' \Kerad(x_1,x) \Ksrad(x_2,x')
\dfrac{a\left(\frac{x}{k}\right) a\left(\frac{x'}{k}\right)}{\sqrt{x
    x'}} \calU^\urad(x,x').
  \end{aligned}
\label{eq:Irad}
\end{equation}
In equations~\eqref{eq:Imurad} and \eqref{eq:Irad}, $\calU^\urad =
\sum_r \calU_r^\urad$ and two other convolution kernels have been
defined in the radiation era, one for the strain and one for the
energy:
\begin{equation}
\Ksrad(x_i,x) \equiv k
\G{\frac{x}{k}}^\urad\left(\dfrac{x_i}{k}\right) = \sin(x_i-x), \qquad
\Kerad(x_i,x) \equiv
      {\G{\frac{x}{k}}^\urad}'\left(\dfrac{x_i}{k}\right) =
      \cos(x_i-x).
\label{eq:KsKerad}
\end{equation}
Let us first notice that the time dependence of $\calPh^\urad$ in
$\eta_1$ and $\eta_2$ is explicit and completely given by the
functions $\A_i$ and $\B_j$ appearing in
equation~\eqref{eq:strainradinmat}.  All the integrals are evaluated
at equal times $\etaeq$, and do not depend on $\eta_1$ and
$\eta_2$. The wavenumber dependence is not so simple. An explicit part
is coming from the functions $\A_i$ and $\B_j$, another quasi-explicit
part is coming from the scale factor, which is evaluated at $a(x/k)$,
and, the boundaries of the integrals are $k$-dependent. As a result,
depending on where $\calU^\urad(x,x')$ is non-vanishing, one should
expect different $k$-behaviour.

\subsection{Energy density parameter today}

\label{sec:energytoday}

Up to some Hubble terms, this is the power spectrum $\calP_{h'}$
evaluated at unequal times $\eta_1$ and $\eta_2$ in the matter
era. Exactly as for the strain power spectrum, we can split it in two
contributions
\begin{equation}
\OmegaGW(\eta_1,\eta_2,k) = \OmegaGWmat(\eta_1,\eta_2,k) + \OmegaGWrad(\eta_1,\eta_2,k).
\label{eq:omegatot}
\end{equation}
From equations~\eqref{eq:calPdef}, \eqref{eq:OmegaGW}, \eqref{eq:Phs}
and \eqref{eq:Pmus}, the first term reads
\begin{equation}
\begin{aligned}
  \OmegaGWmat(\eta_1,\eta_2,k) &= \dfrac{32}{3} \left(\GN M^2\right)^2
\left[\dfrac{k^2}{\calH(\eta_1) \calH(\eta_2)}
  \Idmumat(x_1,x_2,k) + \Imumat(x_1,x_2,k) \right. \\
  & \left.
  -\dfrac{k}{\calH(\eta_1)} \Ixmat(x_1,x_2,k) -
  \dfrac{k}{\calH(\eta_2)} \Ixbarmat(x_1,x_2,k)  \right].
\end{aligned}
\label{eq:omegamat}
\end{equation}
The integral $\Imumat$ is given in equation~\eqref{eq:Imumat} while
$\Idmumat$ and $\Ixmat$ are the analogues, for the matter era,
of those appearing in equation~\eqref{eq:Irad}. They read
\begin{equation}
  \begin{aligned}
\Idmumat(x_1,x_2,k) & = \dfrac{1}{a(\eta_1)a(\eta_2)} \int_{\xeq}^{x_1}\ud x
\int_{\xeq}^{x_2} \ud x' \Kemat(x_1,x) \Kemat(x_2,x')
\dfrac{a\left(\frac{x}{k}\right) a\left(\frac{x'}{k}\right)}{\sqrt{x
    x'}} \calU^\umat(x,x'),\\
\Ixmat(x_1,x_2,k) & = \dfrac{1}{a(\eta_1)a(\eta_2)} \int_{\xeq}^{x_1}\ud x
\int_{\xeq}^{x_2} \ud x' \Kemat(x_1,x) \Ksmat(x_2,x')
\dfrac{a\left(\frac{x}{k}\right) a\left(\frac{x'}{k}\right)}{\sqrt{x
    x'}} \calU^\umat(x,x'),
  \end{aligned}
\label{eq:Imat}
\end{equation}
where we have introduced the energy convolution kernel in the matter era
\begin{equation}
\begin{aligned}
  \Kemat(x_i,x) & \equiv {\G{\frac{x}{k}}^\umat}'\left(\dfrac{x_i}{k}\right) = \left[1 +
      \dfrac{1}{(x_i+\xeq)(x+\xeq)} - \dfrac{1}{(x_i+\xeq)^2} \right]
  \cos(x_i-x) \\ & +
      \left[\dfrac{1}{x+\xeq} - \dfrac{1}{x_i+\xeq} - \dfrac{1}{(x_i+\xeq)^2 (x+\xeq)} \right]
      \sin(x_i-x).
\label{eq:Kemat}
\end{aligned}
\end{equation}
Another integral
\begin{equation}
\Ixbar(x_1,x_2,k) \equiv \Ix^*(x_2,x_1,k),
\label{eq:Ixbardef}
\end{equation}
has been defined, but it is exactly equal to $\Ix(\eta_1,\eta_2,k)$
for real symmetric correlators $\calU$. Comparing
equation~\eqref{eq:strainmat} and \eqref{eq:omegamat} immediately shows
that the relation $\OmegaGW \simeq k^2/(12 \calH^2) \calPh$ does not
hold at large scales. At small scales, provided the term in
$k^2/\calH^2$ dominates, one still has to verify that $\Idmumat \simeq
\Imumat$, which requires some assumptions on the function
$\calU^\umat$.

The second term in equation~\eqref{eq:omegatot} is first integrated in
the radiation era. For $\eta_1< \etaeq$ and $\eta_2 < \etaeq$, it
takes a functional form identical to equation~\eqref{eq:omegamat}, namely
\begin{equation}
\begin{aligned}
  \OmegaGWrad(\eta_1<\etaeq,\eta_2<\etaeq,k) &= \dfrac{32}{3} \left(\GN M^2\right)^2
\left[\dfrac{k^2}{\calH(\eta_1) \calH(\eta_2)}
  \Idmurad(x_1,x_2,k) + \Imurad(x_1,x_2,k) \right. \\
  & \left.
  -\dfrac{k}{\calH(\eta_1)} \Ixrad(x_1,x_2,k) -
  \dfrac{k}{\calH(\eta_2)} \Ixbarrad(x_1,x_2,k)  \right].
\end{aligned}
\label{eq:omegaradinrad}
\end{equation}
In order to determine its value in the matter era, we first evaluate
it at $\eta_1=\eta_2=\etaeq$, match and freely propagate the solution
into the matter era. After some algebra, one gets the expression
\begingroup \allowdisplaybreaks
\begin{align}
  \OmegaGWrad(\eta_1>\etaeq &,\eta_2>\etaeq,k) = \dfrac{32}{3} \left(\GN M^2
  \right)^2 \dfrac{a^2(\etaeq)}{a(\eta_1) a(\eta_2)} \times \\
  \Bigg\{& 
  \left[\dfrac{k^2}{\calH(\eta_1) \calH(\eta_2)}\dB_1 \dB_2 + \B_1
    \B_2 - \dfrac{k}{\calH(\eta_1)} \dB_1 \B_2 -
    \dfrac{k}{\calH(\eta_2)} \B_1 \dB_2 \right] \Idmurad(\xeq,\xeq,k)
  \nonumber \\
  + & \left[\dfrac{k^2}{\calH(\eta_1)\calH(\eta_2)} \dA_1 \dA_2 + \A_1
    \A_2 - \dfrac{k}{\calH(\eta_1)} \dA_1 \A_2 -
    \dfrac{k}{\calH(\eta_2)} \A_1 \dA_2 \right] \Imurad(\xeq,\xeq,k)
  \nonumber \\
  + & \left[\dfrac{k^2}{\calH(\eta_1)\calH(\eta_2)} \left(\dA_1 \dB_2
    + \dB_1 \dA_2\right) + \A_1 \B_2 + \B_1 \A_2 -
    \dfrac{k}{\calH(\eta_1)}\left(\dA_1 \B_2 + \dB_1 \A_2\right)
    \right. \nonumber \\  & \left. -
    \dfrac{k}{\calH(\eta_2)} \left(\A_1 \dB_2 + \B_1 \dA_2 \right)
    \right] \Ixrad(\xeq,\xeq,k) \Bigg\}.
\label{eq:omegaradinmat}
\end{align}
\endgroup Equations~\eqref{eq:strainmat}, \eqref{eq:strainradinrad},
\eqref{eq:strainradinmat}, \eqref{eq:omegamat},
\eqref{eq:omegaradinrad} and \eqref{eq:omegaradinmat} are new. They
give the unequal-time correlators of the strain, and energy density,
of gravitational waves in the matter and radiation era, at all length
scales. However, in order to determine their complete time and
wavenumber dependence it is necessary to evaluate all the convolution
integrals appearing in these formulas, which is the subject of the
next section.

\subsection{Convolution integrals for extinct sources}
\label{sec:convolution}
The integrals $\Imu$, $\Idmu$ and $\Ix$ are not independent. From the
definition of the convolution kernels, one can check that
\begin{equation}
\Ke(x_i,x) = \dfrac{\partial \Ks(x_i,x)}{\partial x_i}\,,
\end{equation}
in both the radiation and matter era. Moreover, these kernels being
proportional to the Green's functions, this implies some relations
between the integrals. One has
\begin{equation}
\Idmu(x_1,x_2,k) = \dfrac{\partial^2\Imu(x_1,x_2,k)}{\partial x_1
  \partial x_2}\,,\qquad \Ix(x_1,x_2,k) = \dfrac{\partial\Imu(x_1,x_2,k)}{\partial x_1}\,,
\label{eq:partialx1x2}
\end{equation}
where, in equations~\eqref{eq:Imumat} and \eqref{eq:Imurad}, one
should pay attention that the factor $1/[a(\eta_1) a(\eta_2)]$ must be out of the
derivation. As a result, only $\Imu(x_1,x_2,k)$, with its dependence
in $x_1$ and $x_2$ has to be known.

\subsubsection{Radiation era}
\label{sec:radholo}
For presenting the method, let us first focus on the simplest of all
these integrals, which is $\Imurad$. Using the radiation strain kernel
of equation~\eqref{eq:KsKerad}, it reads
\begin{equation}
\begin{aligned}
  \Imurad(x_1,x_2,k) & = \int_{\xini}^{x_1} \ud x \int_{\xini}^{x_2} \ud x'
\sin(x_1-x) \sin(x_2-x')\,  \dfrac{\ahat_1(x,k) \ahat_2(x',k)}{\sqrt{x x'}}
\calU^\urad(x,x'),
\end{aligned}
\end{equation}
where we have defined the functions
\begin{equation}
\ahat_i(x,k) \equiv \dfrac{a\left(\frac{x}{k}\right)}{a(\eta_i)}\,.
\end{equation}
This convolution integral is quite close to a sine Fourier transform,
but the domain of integration is not infinite and we would like to
keep track of $x_1$ and $x_2$ in both the boundaries and the sine arguments, they
are precisely the terms we are interested in. We can pursue this route
by defining the new variables
\begin{equation}
y \equiv x - \xini, \qquad y'\equiv x'-\xini,
\end{equation}
from which one has
\begin{equation}
\Imurad(y_1,y_2,k) = \sin(y_1) \sin(y_2) \Icc - \sin(y_1)
\cos(y_2) \Ics - \cos(y_1) \sin(y_2) \Isc +
\cos(y_1) \cos(y_2) \Iss.
\label{eq:ImuradfromIccss}
\end{equation}
Four new simpler integrals have been defined
\begin{equation}
\begin{aligned}
  \Icc(y_1,y_2,k) & \equiv \int_0^{y_1}\ud y \int_0^{y_2} \ud y' \cos(y)
  \cos(y') \calC_k(y,y'), \\
  \Iss(y_1,y_2,k) & \equiv \int_0^{y_1}\ud y \int_0^{y_2} \ud y' \sin(y)
  \sin(y') \calC_k(y,y'), \\
  \Ics(y_1,y_2,k) & \equiv \int_0^{y_1}\ud y \int_0^{y_2} \ud y' \cos(y)
  \sin(y') \calC_k(y,y'), \\
 \Isc(y_1,y_2,k) & \equiv \int_0^{y_1}\ud y \int_0^{y_2} \ud y' \sin(y)
  \cos(y') \calC_k(y,y'),
\end{aligned}
\label{eq:Iccss}
\end{equation}
where the function $\calC_k$ stands for
\begin{equation}
\calC_k(y,y') \equiv \dfrac{\ahat_1(\xini+|y|,k)
  \ahat_2(\xini+|y'|,k)}{\sqrt{\xini+|y|} \sqrt{\xini+|y'|}}
\calU(\xini+|y|,\xini + |y'|).
\label{eq:calCdef}
\end{equation}
The index $k$ is a reminder that this function is an explicit function
of the wavenumber $k$ due to its dependence in $\xini$ and in the
function $\ahat_i$. The reason of having introduced $|y|$ and $|y'|$
is that we can now extend its domain to the whole $\Rset^2$. Doing so,
one can rewrite all integrals of equation~\eqref{eq:Iccss} in terms of
complex exponentials.

Once more, for simplicity, let us focus first on the $\Icc$ integral. It can be
rewritten as
\begin{equation}
\Icc(y_1,y_2,k) = \dfrac{1}{4} \iint_{-\infty}^{+\infty} \ud y
 \ud y' e^{-\imath y} e^{-\imath y'}
\rect\left(\dfrac{y}{2 y_1}\right) \rect\left(\dfrac{y'}{2 y_2}\right)
\calC_k(y,y'),
\label{eq:Iccexp}
\end{equation}
where the $\rect(x)$ function is unity for $-0.5<x<0.5$ and vanishes
elsewhere. Written under this form, we have made explicit that all
these integrals are Fourier transforms of $\calC_k$ multiplied by some
sharp window functions, and evaluated at unit frequencies. Equally, we
can use the convolution theorem and re-expressed $\Icc$ in another
form. Defining the Fourier transform
\begin{equation}
\calCFT_k(\gamma,\gamma') \equiv \iint_{-\infty}^{+\infty} \ud y
 \ud y' e^{-\imath(\gamma y + \gamma' y')} \calC_k(y,y'),
\end{equation}
one gets
\begin{equation}
\Icc(y_1,y_2,k) = \dfrac{y_1 y_2}{4\pi^2} \iint_{-\infty}^{\infty}
  \ud \gamma \ud \gamma' \sinc\left[(1-\gamma)y_1\right]
  \sinc\left[(1-\gamma') y_2\right] \calCFT_k(\gamma,\gamma'),
\label{eq:Iccsinc}
\end{equation}
where the sine cardinal functions arise from the Fourier transform of
the rectangular window functions.  One can rapidly check what is
going on for $y_1$ and $y_2$ becoming large. The functions
\begin{equation}
  y \sinc[y(1-\gamma)] \underset{\infty}{\to} \pi
  \delta(1-\gamma),
\end{equation}
 and, if $\calCFT_k(\gamma,\gamma')$ is a smooth function, the
 integral approaches the ($k$-dependent) value
\begin{equation}
  \Icc(y_1\gg 1,y_2\gg 1,k) = \dfrac{1}{4} \calCFT_k(1,1).
\label{eq:Icclimit}
\end{equation}
In fact, as we show in the appendix~\ref{sec:holomorphic}, if
$\calCFT_k(\gamma,\gamma')$ is a holomorphic function, this limit is
actually the exact value of the integral and does not depend on $y_1$
and $y_2$! This could appear surprising when considering how $y_1$ and
$y_2$ enter equation~\eqref{eq:Iccss}, but the Paley-Weiner
theorem states that if $\calC_k(y,y')$ is at compact support within
the domain of integration (and square integrable), then its Fourier
transform is holomorphic. Therefore, $\calCFT_k(\gamma,\gamma')$ is
holomorphic when the source is actually ``switched off'' at the times
$y_1$ and $y_2$ of the measurements. When this is the case,
equation~\eqref{eq:Iccss} shows that the integral can not depend on
$y_1$ and $y_2$. These scaling sources will be referred to as
``extinct'' in the following.

The other integrals of equation~\eqref{eq:Iccss} can be dealt in a
similar manner. We get
\begin{equation}
\begin{aligned}
\Iss(y_1,y_2,k) = -\dfrac{1}{4} \iint_{-\infty}^{+\infty} \ud y \ud y'
e^{-\imath y} e^{-\imath y'} \rectbar\left(\dfrac{y}{2 y_1}\right)
\rectbar\left(\dfrac{y'}{2 y_2}\right) \calC_k(y,y'),
\end{aligned}
\label{eq:Issexp}
\end{equation}
where $\rectbar(0<x<0.5)=1$, $\rectbar(-0.5<x<0)=-1$ and it vanishes
elsewhere. From this expression, one obtains
\begin{equation}
\Iss(y_1,y_2,k) = \dfrac{y_1^2 y_2^2}{16 \pi^2}
\iint_{-\infty}^{+\infty} \ud \gamma \ud \gamma' (1-\gamma)
(1-\gamma') \sinc^2\left(\dfrac{1-\gamma}{2}y_1\right)
\sinc^2\left(\dfrac{1-\gamma'}{2}y_2\right) \calCFT_k(\gamma,\gamma').
\label{eq:Isssinc}
\end{equation}
If $\calCFT_k$ is smooth, the limit of large $y_1$ and $y_2$ can be
determined. Using
\begin{equation}
\dfrac{y}{2} \sinc^2\left(\dfrac{1-\gamma}{2} y \right)
\underset{\infty}{\to} \pi \delta(1-\gamma),
\end{equation}
one gets
\begin{equation}
\Iss(y_1 \gg 1,y_2 \gg 1,k) =  \dfrac{y_1 y_2}{4} \lim_{\gamma,\gamma' \to
  1} (1-\gamma) (1-\gamma') \calCFT_k(\gamma,\gamma') = 0.
\end{equation}
Notice that this limit is non trivial as we have used $y_1\gg 1$ and
$y_2\gg 1$ to replace the sine cardinal functions by Dirac
distributions. The correct derivation, again for holomorphic functions
$\calCFT_k(\gamma,\gamma')$, can be found in the
appendix~\ref{sec:holomorphic} and, for them, this result holds for
all $y_1$ and $y_2$.

The cross integrals $\Isc$ and $\Ics$ can be expressed in a similar way
as equations~\eqref{eq:Iccexp} and \eqref{eq:Issexp}, but with a
product of $\rect(x)$ and $\rectbar(x)$. Following the same method, we
get
\begin{equation}
\begin{aligned}
  \Ics(y_1,y_2,k) &= \dfrac{y_1 y_2^2}{8\pi^2}\iint_{-\infty}^{+\infty}
\ud \gamma \ud \gamma' (1-\gamma') \sinc[(1-\gamma)y_1]
\sinc^2\left(\dfrac{1-\gamma'}{2}y_2\right) \calCFT_k(\gamma,\gamma'),\\
 \Isc(y_1,y_2,k) &= \dfrac{y_1^2 y_2}{8\pi^2}\iint_{-\infty}^{+\infty}
\ud \gamma \ud \gamma' (1-\gamma) 
\sinc^2\left(\dfrac{1-\gamma}{2}y_1\right)\sinc[(1-\gamma')y_2] \, \calCFT_k(\gamma,\gamma').
\label{eq:Icssinc}
\end{aligned}
\end{equation}
The large $(y_1,y_2)$ limits for smooth $\calCFT_k(\gamma,\gamma')$ are
\begin{equation}
\begin{aligned}
  \Ics(y_1\gg 1,y_2\gg 1,k) & =  \dfrac{y_2}{4} \lim_{\gamma'\to 1}
  (1-\gamma')\calCFT_k(1,\gamma') = 0, \\
  \Isc(y_1\gg 1,y_2\gg 1,k) & =  \dfrac{y_1}{4} \lim_{\gamma \to 1}
  (1-\gamma)\calCFT_k(\gamma,1) = 0,
\label{eq:Iscsinc}
\end{aligned}
\end{equation}
again exact for holomorphic Fourier transforms $\calC_k(y,y')$.

All in all, for the case of extinct sources, we have the very
simple and quite elegant result
\begin{equation}
\begin{aligned}
  \Imurad(x_1,x_2,k) & =\dfrac{\calCFT_k^\urad(1,1)}{4}\sin(x_1-\xini) \sin(x_2 - \xini)
   \\
  & = \dfrac{\calCFT_k^\urad(1,1)}{8} \left[\cos(x_1-x_2) -
    \cos(x_1+x_2-2\xini) \right],
\end{aligned}
\label{eq:Imuradholo}
\end{equation}
from which we immediately get $\Idmurad$ and $\Ixrad$ by
equation~\eqref{eq:partialx1x2},
\begin{equation}
\begin{aligned}
  \Idmurad(x_1,x_2,k) &= \dfrac{\calCFT_k^\urad(1,1)}{4} \cos(x_1-\xini) \cos(x_2 - \xini)
   \\
& = \dfrac{\calCFT_k^\urad(1,1)}{8} \left[\cos(x_1-x_2) +
    \cos(x_1+x_2-2\xini) \right],  \\
\Ixrad(x_1,x_2,k) &= \dfrac{\calCFT_k^\urad(1,1)}{4}\cos(x_1-\xini) \sin(x_2 - \xini)
 \\& = \dfrac{\calCFT_k^\urad(1,1)}{8} \left[-\sin(x_1-x_2) +
    \sin(x_1+x_2-2\xini) \right].
\end{aligned}
\label{eq:Idmuradholo}
\end{equation}
They determine completely $\calPh^\urad$ and $\OmegaGWrad$. Confined
in the radiation era, one gets
\begin{equation}
\calPh^\urad(\eta_1<\etaeq,\eta_2<\etaeq,k) = 16 \left(\GN M^2
\right)^2 \calCFT_k^\urad(1,1) \left[\cos(x_1-x_2) -
    \cos(x_1+x_2-2\xini) \right],
\label{eq:strainradinradholo}
\end{equation}
and
\begin{equation}
\begin{aligned}
  \OmegaGWrad(\eta_1<\etaeq,\eta_2<\etaeq,k)  & = \dfrac{4}{3}
\left(\GN M^2 \right)^2 \calCFT_k^\urad(1,1) \left[
  \left(1+\dfrac{k^2}{\calH_1 \calH_2}\right) \cos(x_1-x_2) \right. \\
  & \left. +  \left(\dfrac{k}{\calH_1} -
    \dfrac{k}{\calH_2}\right)\sin(x_1-x_2) -
    \left(1-\dfrac{k^2}{\calH_1\calH_2} \right)\cos(x_1+x_2
    - 2\xini) \right. \\
   & \left. - \left(\dfrac{k}{\calH_1} + \dfrac{k}{\calH_2} \right)
    \sin(x_1+x_2-2\xini) \right].
\label{eq:omegaradinradholo}
\end{aligned}
\end{equation}
These two expressions generally differ. At equal times, for $x_1=x_2$,
they oscillate, but not in phase, with an angular frequency given by
$\omega= 2k$. The standard approximation $\OmegaGWrad \simeq
k^2/(12\calH^2) \calPh$ is recovered by not only considering the large
wavenumber limit $k \gg \calH$ but also by postulating a zero average
of these oscillations. Let us notice that the amplitude of these
oscillations is maximal for $k\gg\calH$, which implies that, at a
given time $\eta_1=\eta_2$, and scale $k$, if $\calPh^\urad$ is
maximal, $\OmegaGWrad$ vanishes.

\subsubsection{Radiation era solutions propagated into the matter era}

From the previous section, we can evaluate all the convolution
integrals at $x_1=x_2=\xeq$ and they simplify to
\begin{equation}
  \Imurad(\xeq,\xeq,k)  = \dfrac{\calCFT_k(1,1)}{8}\left[1 -
  \cos\left(2\xeq-2\xini\right) \right],
\label{eq:Imuradeq}
\end{equation}
while
\begin{equation}
\Idmurad(\xeq,\xeq,k)
 = \dfrac{\calCFT_k(1,1)}{8}\left[1 + \cos\left(2\xeq-2\xini\right)
   \right],
 \label{eq:Idmuradeq}
\end{equation}
and
\begin{equation}
  \Ixrad(\xeq,\xeq,k)  =\dfrac{\calCFT_k(1,1)}{8} \sin\left(2\xeq-2\xini\right).
  \label{eq:Ixradeq}
\end{equation}
Plugging these expressions into equations~\eqref{eq:strainradinmat}
gives the full time dependence of $\calPh^\urad(\eta_1,\eta_2,k)$ at
any times in the matter era. One gets
\begin{equation}
  \begin{aligned}
    & \calPh^\urad(\eta_1>\etaeq,\eta_2>\etaeq,k) = \dfrac{\left(\GN M^2
    \right)^2}{2} \calCFT_k^\urad(1,1) \dfrac{a^2(\etaeq)}{a(\eta_1) a(\eta_2)} \\
   & \times  \bigg[ \dfrac{-1+4\xeq(\xeq-x_1)}{\xeq^2(x_1+\xeq)} \cos(x_1-\xini)
      + \dfrac{-x_1 +3\xeq + 8 x_1 \xeq^2 + 8\xeq^3}{\xeq^2(x_1 + \xeq)} \sin(x_1-\xini)  \\ &  + \dfrac{\cos(x_1-2\xeq +
      \xini)}{(x_1+\xeq) \xeq^2} + \dfrac{\sin(x_1 - 2 \xeq
        +\xini)}{\xeq^2}\bigg] \\ & \times \bigg[ \dfrac{-1+4\xeq(\xeq-x_2)}{\xeq^2(x_2+\xeq)} \cos(x_2-\xini)
      + \dfrac{-x_2 +3\xeq + 8 x_2 \xeq^2 + 8\xeq^3}{\xeq^2(x_2 + \xeq)} \sin(x_2-\xini)  \\ &  + \dfrac{\cos(x_2-2\xeq +
      \xini)}{(x_2+\xeq) \xeq^2} + \dfrac{\sin(x_2 - 2 \xeq
        +\xini)}{\xeq^2}\bigg].
  \end{aligned}
\label{eq:strainradinmatholo}
\end{equation}
This expression is factorized into two symmetric terms, in $x_1$ and
$x_2$, but expanding all sine and cosine functions would give four
time-dependent terms in $\cos(x_1-x_2)$, $\sin(x_1-x_2)$,
$\cos(x_1+x_2 - 2\xeq)$ and $\sin(x_1+x_2 - 2\xeq)$, modulated by
oscillatory functions depending only on the wavenumbers, such as
$\cos(2\xeq-2\xini)$. Such an expansion being quite long, it is not
reported here.

In the same manner, plugging equations~\eqref{eq:Imuradeq} to
\eqref{eq:Ixradeq} into the general expression of the energy density
parameter given in equation~\eqref{eq:omegaradinmat}, one gets
\begingroup \allowdisplaybreaks
\begin{align}
& \OmegaGWrad(\eta_1>\etaeq,\eta_2>\etaeq,k) = \dfrac{\left(\GN
M^2 \right)^2}{24} \calCFT_k^\urad(1,1) \dfrac{a^2(\etaeq)}{a(\eta_1)a(\eta_2)}
 \nonumber \\ &  \times \Bigg\{\left[\dfrac{1+4 x_1 \xeq -
     4\xeq^2}{\xeq^2(x_1+\xeq)} + \dfrac{k}{\calH_1} \dfrac{1-5\xeq^2
     +8\xeq^4 + 2x_1 \xeq (1+8\xeq^2) +
     x_1^2(-1+8\xeq^2)}{\xeq^2(x_1+\xeq)^2} \right] \nonumber \\ &
 \times \cos(x_1-\xini) + \left[\dfrac{x_1 - 3\xeq - 8x_1 \xeq^2 - 8
     \xeq^3}{\xeq^2(x_1+\xeq)} +\dfrac{k}{\calH_1} \dfrac{x_1 - 3\xeq
     + 4 x_1^2 \xeq - 4\xeq^3}{\xeq^2(x_1+\xeq)^2} \right] \nonumber
 \\ & \times \sin(x_1 - \xini) - \left[\dfrac{1}{\xeq^2(x_1+\xeq)} +
   \dfrac{k}{\calH_1} \dfrac{1-x_1^2 - 2x_1 \xeq -\xeq^2}{\xeq^2(x_1+\xeq)^2} \right]
 \cos(x_1-2\xeq + \xini) \nonumber \\ & - \left[\dfrac{1}{\xeq^2} +
 \dfrac{k}{\calH_1}\dfrac{1}{\xeq^2(x_1+\xeq)} \right]\sin(x_1-2\xeq+\xini) \Bigg\} \nonumber \\
 & \times \Bigg\{\left[\dfrac{1+4 x_2 \xeq -
     4\xeq^2}{\xeq^2(x_2+\xeq)} + \dfrac{k}{\calH_1} \dfrac{1-5\xeq^2
     +8\xeq^4 + 2x_2 \xeq (1+8\xeq^2) +
     x_2^2(-1+8\xeq^2)}{\xeq^2(x_2+\xeq)^2} \right] \nonumber \\ &
 \times \cos(x_2-\xini) + \left[\dfrac{x_2 - 3\xeq - 8x_2 \xeq^2 - 8
     \xeq^3}{\xeq^2(x_2+\xeq)} +\dfrac{k}{\calH_1} \dfrac{x_2 - 3\xeq
     + 4 x_2^2 \xeq - 4\xeq^3}{\xeq^2(x_2+\xeq)^2} \right] \nonumber
 \\ & \times \sin(x_2 - \xini) - \left[\dfrac{1}{\xeq^2(x_2+\xeq)} +
   \dfrac{k}{\calH_1} \dfrac{1-x_2^2 - 2x_2 \xeq -\xeq^2}{\xeq^2(x_2+\xeq)^2} \right]
 \cos(x_2-2\xeq + \xini) \nonumber \\ & - \left[\dfrac{1}{\xeq^2} +
 \dfrac{k}{\calH_1}\dfrac{1}{\xeq^2(x_2+\xeq)} \right]\sin(x_2-2\xeq+\xini) \Bigg\}.
\label{eq:omegaradinmatholo}
\end{align}
\endgroup
For readability, this equation is again given under a factorized form.
As for the strain spectrum, isolating the time dependence by expanding
all sine and cosine functions would give back the four
oscillatory terms in $\cos(x_1-x_2)$, $\sin(x_1-x_2)$, $\cos(x_1+x_2 -
2\xeq)$ and $\sin(x_1+x_2 - 2\xeq)$.

\subsubsection{Matter era}

\label{sec:matera}

For calculating $\Imumat$ in the case of extinct sources, the
only difference with respect to the previous section comes from the
more complicated strain convolution kernel which is written in
equation~\eqref{eq:Ksmat}. Plugging its expression into
equation~\eqref{eq:Imumat} gives an integral over various products of
sine and cosine functions mixed with terms in $1/(x+\xeq)$,
$1/(x'+\xeq)$ and their product. As we have shown in the previous
section, after defining $y=x-\xeq$ and $y'=x'-\xeq$, one ends up
having a complicated combination of integrals of the forms given in
equation~\eqref{eq:Iccss}. For extinct sources, we have just proven
that only $\Icc$ is non-zero and most of these integrals are
vanishing. The calculation is straightforward, but lengthy, and the
expression of the integrals $\Imumat$, $\Idmumat$ and $\Ixmat$ can be
found in the appendix~\ref{sec:extinct}. Here, we simply quote the
result. Defining the new functions
\begin{equation}
\calC_k^\umat(y,y') \equiv \dfrac{\ahat_1(\xeq+|y|,k)
  \ahat_2(\xeq+|y'|,k)}{\sqrt{\xeq+|y|} \sqrt{\xeq+|y'|}}
\calU^\umat(\xeq+|y|,\xeq + |y'|).
\label{eq:calCmatdef}
\end{equation}
and
\begin{equation}
  \calD_k^\umat(y,y')  \equiv \dfrac{\calC_k^\umat(y,y')}{2\xeq +
    |y|}\,, \qquad \calE_k^\umat(y,y')  \equiv  \dfrac{\calC_k^\umat(y,y')}{\left(2\xeq +
    |y|\right)\left(2\xeq + |y'|\right)}\,,
\label{eq:calDEdef}
\end{equation}
the waveform of the unequal-time strain power spectrum for
extinct sources in the matter era reads
\begin{equation}
  \begin{aligned}
    \calPh^\umat(\eta_1,\eta_2,k) & = \dfrac{16\left(\GN M^2
      \right)^2}{(x_1+\xeq)(x_2+\xeq)} \bigg\{
    (\calCFT_k^\umat+\calEFT_k^\umat)
    \left[1+(x_1+\xeq)(x_2+\xeq) \right]
    \cos(x_1-x_2) \\ & + (\calCFT_k^\umat+\calEFT_k^\umat)
    (x_1-x_2) \sin(x_1-x_2) + \big\{    
      - 2 \calDFT_k^\umat(x_1+x_2+2\xeq) \\ & +
      (\calCFT_k^\umat-\calEFT_k^\umat)\left[1-(x_1+\xeq)(x_2+\xeq)\right]
      \big\} \cos(x_1+x_2-2\xeq) \\ & + \big\{ 2 \calDFT_k^\umat
      \left[1-(x_1+\xeq)(x_2+\xeq)\right] + (\calCFT_k^\umat
      -\calEFT_k^\umat)(x_1+x_2+2\xeq) \big\} \\ & \times
      \sin(x_1+x_2-2\xeq) \bigg\},
  \end{aligned}
  \label{eq:strainmatholo}
\end{equation}
where we have used the shortcut notation $\calCFT_k = \calCFT_k(1,1)$,
$\calDFT_k = \calDFT_k(1,1)$ and $\calEFT_k = \calEFT_k(1,1)$.
Similarly, using equations~\eqref{eq:Imumatholo},
\eqref{eq:Idmumatholo} and \eqref{eq:Ixmatholo} into
equation~\eqref{eq:omegamat} gives the unequal-time energy density
parameter $\OmegaGWmat(\eta_1,\eta_2,k)$. As for the strain power
spectrum above, its expression is made of four oscillatory terms, two
encoding the coherence of the signal, varying as $\cos(x_1-x_2)$ and
$\sin(x_1-x_2)$, and two others describing oscillations as
$\cos(x_1+x_2 - 2\xeq)$ and $\sin(x_1+x_2 - 2\xeq)$. The prefactors of
these terms are functions of the wavenumbers and the Fourier transform
of the correlators. Their expression being quite long, they have been
reported in the appendix~\ref{sec:extinct}, see
equation~\eqref{eq:omegamatholo}.

\subsection{Equal-time spectra for extinct sources}

In order to understand the behaviour of the spectra derived in the
previous section, let us discuss their shape at equal times by setting
$\eta_1 = \eta_2 = \eta_0$, with $\eta_0$ either in the radiation era
or matter era.

The contribution to the strain coming from the radiation era is given
in equations~\eqref{eq:strainradinradholo} and
\eqref{eq:strainradinmatholo} and one needs to evaluate
$\calCFT_k^\urad(1,1)$ which is a function of $k$. Taking the scale
factor as for a pure radiation era, one has
\begin{equation}
\ahat_i(x,k) = \dfrac{x}{k \eta_i}\,,
\label{eq:ahatrad}
\end{equation}
and
\begin{equation}
  \calCFT_k^\urad(1,1) = \dfrac{1}{k^2 \eta_1 \eta_2}\CFT_\uini^\urad(1,1) ,
\label{eq:calCtoCrad}
\end{equation}
where $\eta_i \le \etaeq$ and $\CFT_\uini^\urad(\gamma,\gamma')$ stands for the two-dimensional
Fourier transform of the function
\begin{equation}
C_\uini^\urad(y,y') = \sqrt{\xini + |y|} \sqrt{\xini + |y'|} \, \calU^\urad(\xini
+ |y|,\xini + |y'|).
\label{eq:Crad}
\end{equation}
The index ``ini'' is a reminder that we cannot pull out the complete
$k$-dependence of this function. However, if the scaling sources have
appeared very early in the history of the Universe, one has $\etaini
\to 0$ and for all wavenumbers $k \ll 1/\etaini$, the functional shape
of $C_\uini^\urad(y,y')$ is essentially independent of
$\xini$. Therefore, $\CFT_\uini^\urad(1,1)$ is just a number and
does not depend on $k$. Obviously, the conclusion is reversed if one
considers modes $k > 1/\etaini$ for which one has $\xini > 1$. For
these modes, the function $C_\uini(y,y')$ becomes strongly dependent
on the shift $\xini$ in the correlator
$\calU^\urad(\xini+|y|,\xini+|y'|)$ and so does
$\CFT_\uini^\urad(1,1)$. In particular, if $\calU^\urad(x,x')$ decays
at large $(x,x')$, as it should, the Fourier transform will only pick
the tail of the correlator and this ensures that
$\CFT_\uini^\urad(1,1) \to 0$ for $k \gg 1/\etaini$. 
\begin{figure}
\begin{center}
  \includegraphics[width=\twofigw]{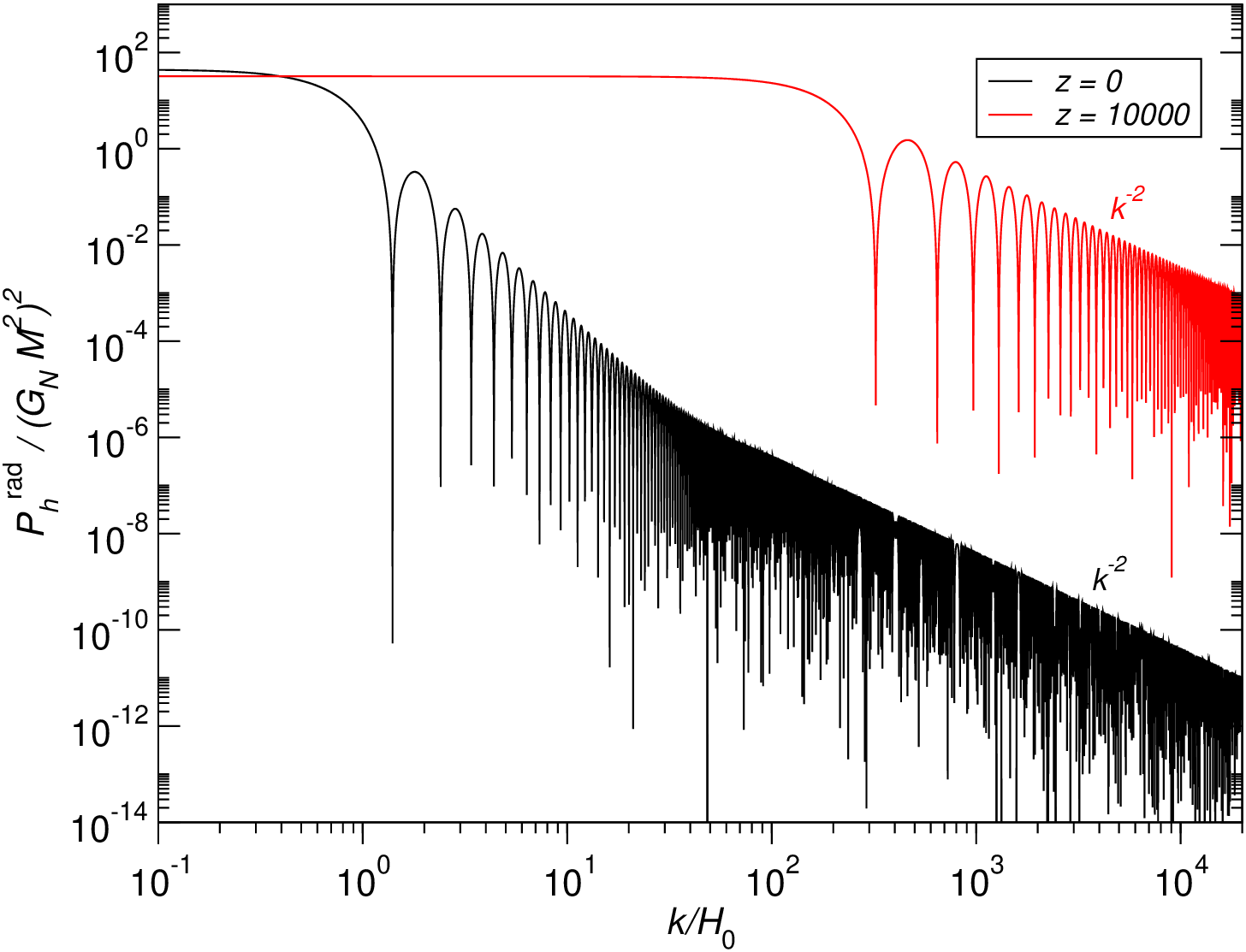}
  \includegraphics[width=\twofigw]{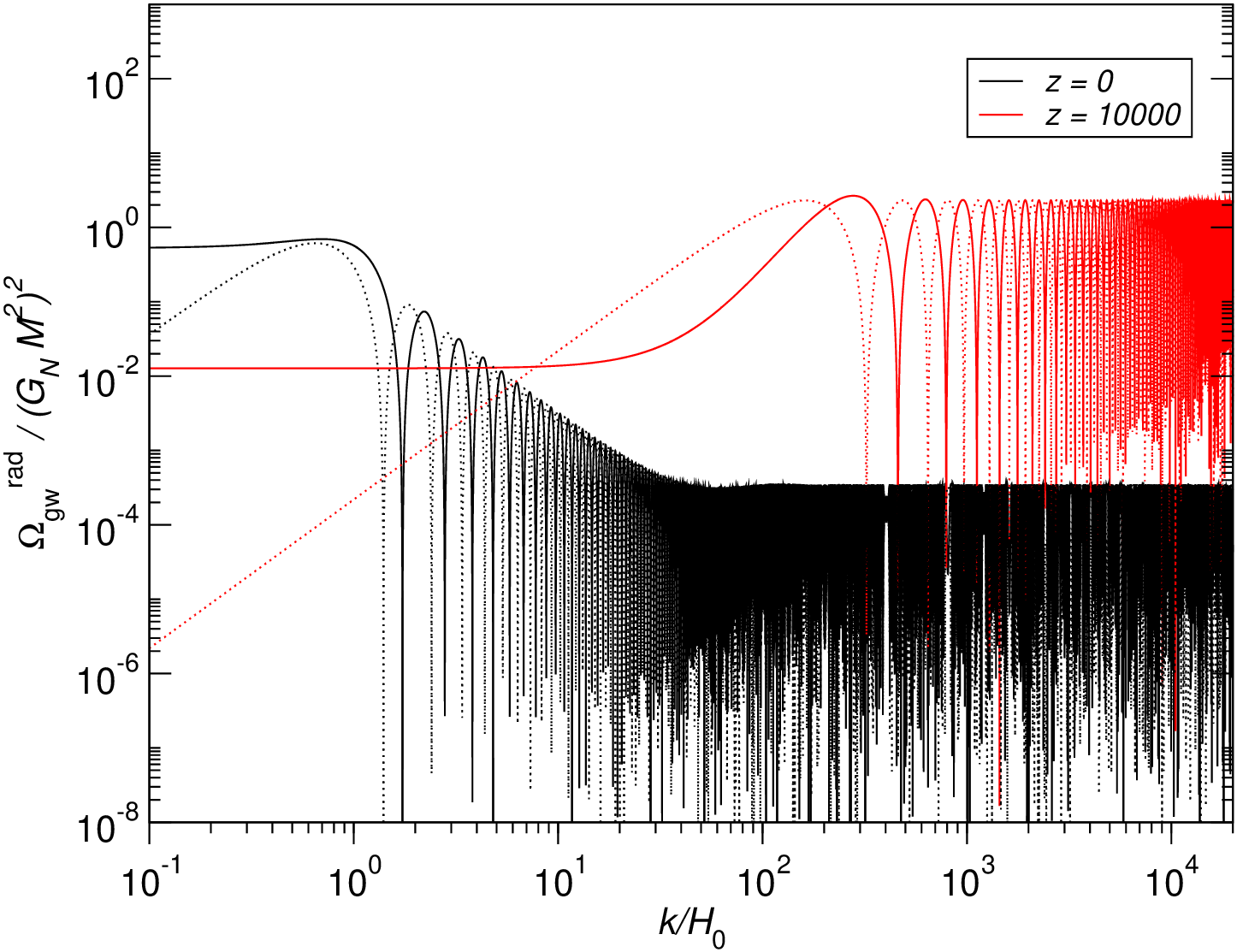}
\caption{The normalised strain power spectrum $\calPh^\urad/(\GN
  M^2)^2$ (left panel) and energy density parameter $\OmegaGWrad/(\GN
  M^2)^2$ (right panel) coming from extinct sources in the radiation
  era, measured either at $z=10^4$ (red curve) or in the matter era,
  today, at $z=0$ (black curve). For illustration purposes, the
  Fourier transform of the source correlator is set to
  $\CFT_\uini^\urad(1,1)=1$. Notice the change of slope for
  $k>1/\etaeq$ for $z=0$. The dotted curves on the right panel shows
  $k^2/(12 \calH_0)^2 \calPh^\urad$ which deviates from
  $\OmegaGWrad$ on large scales and oscillates in phase opposition at
  small scales.}
\label{fig:spectraradinmatholo}
\end{center}
\end{figure}
In figure~\ref{fig:spectraradinmatholo} we have represented the
normalised strain power spectrum (left panel) and the energy density
parameter (right panel), at equal times, as a function of
$k/\calH_0$. Two measurement redshifts have been represented, one in
the radiation era at $z=10^4$, and one today at $z=0$. For the latter,
we see that the spectrum dependence with respect to the wavenumbers
changes at scales matching equality $k=1/\etaeq$. In the right panel
of this figure, we have compared the usual approximation $k^2/(12
\calH_0^2) \calPh^\urad$, plotted as dotted curves, to the actual value of
$\OmegaGWrad$. The envelope of both matches well inside the Hubble
radius, but they do oscillate in phase opposition at large
wavenumbers. In fact, a better approximation can be obtained from
equation~\eqref{eq:strainradinrad}, \eqref{eq:omegaradinrad} and
\eqref{eq:partialx1x2}, assuming the integral to have all the same
typical amplitude, one has
\begin{equation}
\OmegaGWrad(\eta_1,\eta_2,k \gg \calH) \simeq
\dfrac{k^2}{12 \calH(\eta_1)\calH(\eta_2)} \dfrac{\partial^2 \calPh^\urad}{\partial
  x_1 \partial x_2}\,.
\label{eq:omegaradapprox}
\end{equation}
Notice that the envelope of the oscillations plotted in
figure~\ref{fig:spectraradinmatholo} matches the typical behaviour
derived in Refs.~\cite{Figueroa:2012kw, Figueroa:2020lvo}, within the
level of their approximation. On the very large scales, both
$\calPh^\urad$ and $\OmegaGWrad$ are scale invariant for extinct
sources and the approximation of equation~\eqref{eq:omegaradapprox} is
also violated. Let us mention that, as discussed in more details in
section~\ref{sec:constant}, requiring the source to be extinct for $k
\to 0$ is very contriving as the lifetime, or spatial extension, of
the sources should be irrealistically small.

\begin{figure}
\begin{center}
  \includegraphics[width=\twofigw]{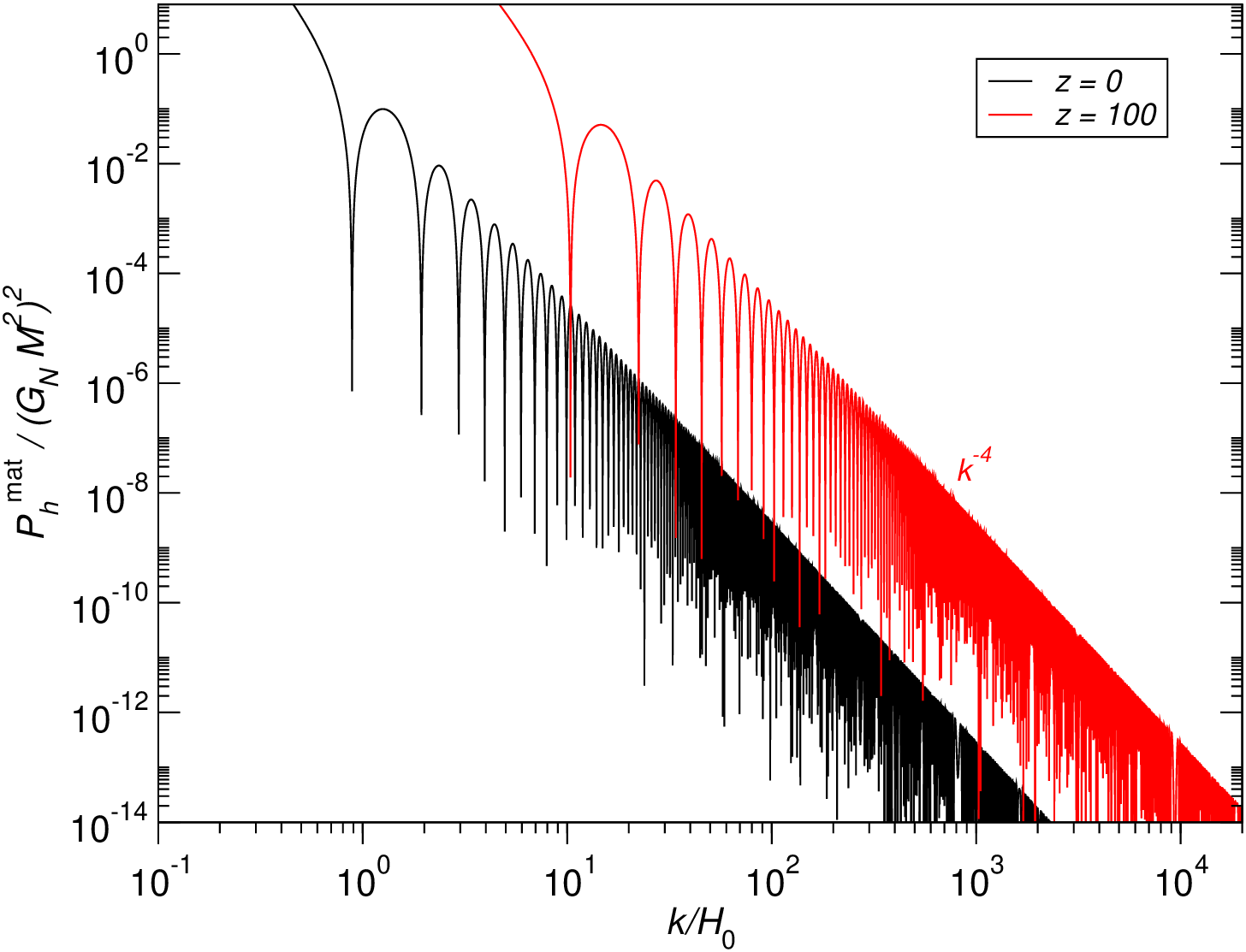}
  \includegraphics[width=\twofigw]{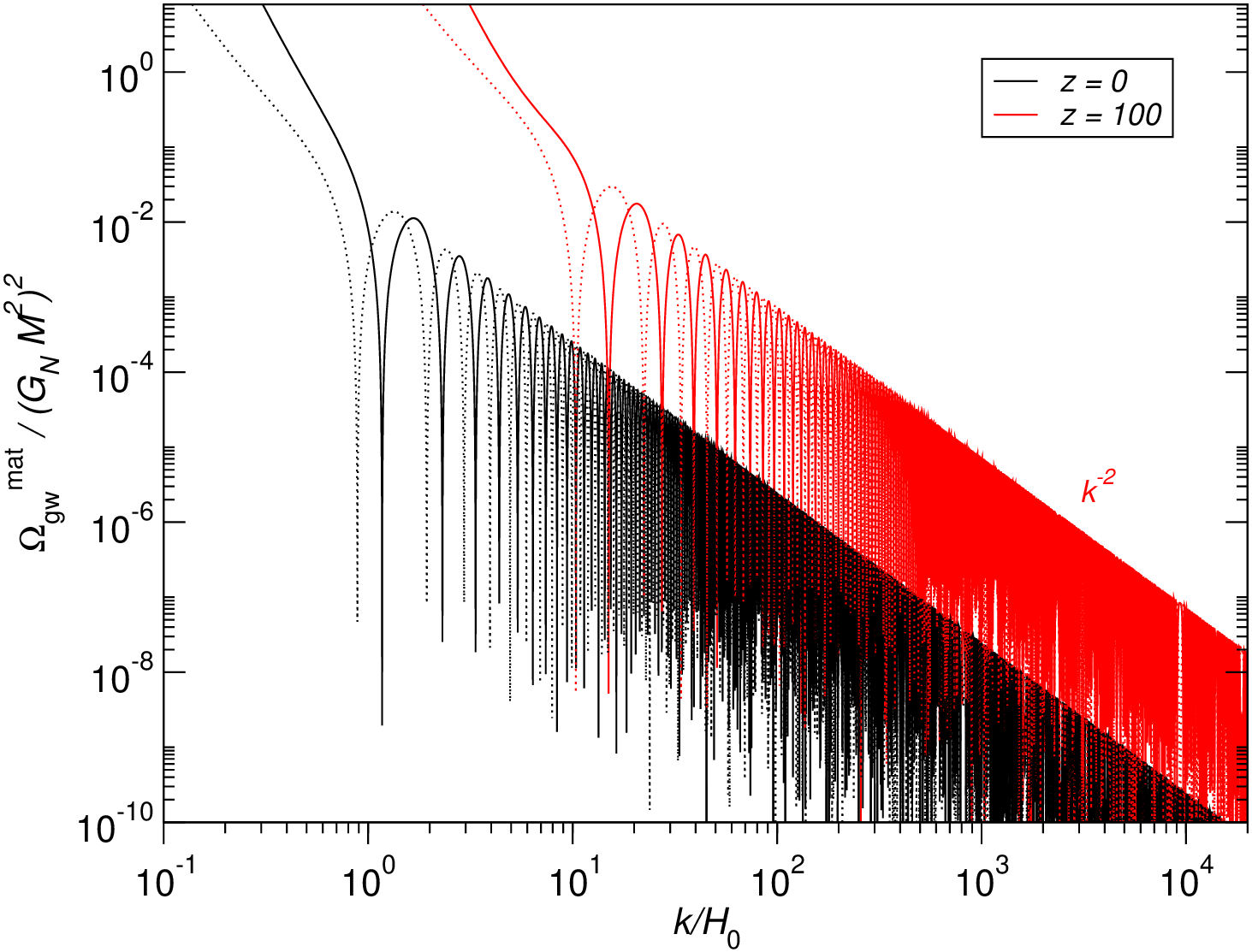}
\caption{The normalized strain power spectrum $\calPh^\umat/(\GN
  M^2)^2$ (left panel) and energy density parameter $\OmegaGWmat/(\GN
  M^2)^2$ (right panel) coming from extinct sources in the matter era,
  measured either at $z=10^2$ (red curve) or today at $z=0$ (black
  curve). For illustration purposes, the Fourier transform of the
  source correlators is set to $\CFT_\ueq^\umat(1,1)=1$,
  $\DFT_\ueq^\umat(1,1)=0$ and $\EFT_\ueq^{\umat}(1,1)=0$, which is
  the dominant term at wavenumbers $1/\eta_0 \ll k \ll 1/\etaeq$. The dotted
  curves on the right panel shows $k^2/(12 \calH_0)^2
  \calPh^\umat$ which oscillates in phase opposition of $\OmegaGWmat$
  at small scales. The envelope of both curves decays as expected in
  $1/k^2$.}
\label{fig:spectramatinmatholo}
\end{center}
\end{figure}

The derivation of the equal-time contribution coming from the matter
era extinct sources can be performed in a similar way, paying attention that the
functions $\ahat(x,k)$ are different. In the matter era, one has
\begin{equation}
\ahat_i(x,k) = \dfrac{(x+\xeq)^2}{\left[k\left(\eta_i + \etaeq\right)\right]^2}\,,
\label{eq:ahatmat}
\end{equation}
and
\begin{equation}
\calCFT_k^\umat(1,1) = \dfrac{1}{k^4 (\eta_1+\etaeq)^2 (\eta_2+\etaeq)^2} \CFT_\ueq^\umat(1,1),
\label{eq:calCtoCmat}
\end{equation}
with $\CFT_\ueq^\umat(\gamma,\gamma')$ the Fourier transform of
\begin{equation}
  C_\ueq^{\umat}(y,y') = \dfrac{\left(2\xeq + |y|\right)^{2}\left(2\xeq +
  |y'|\right)^{2}}{\sqrt{\xeq+|y|} \, \sqrt{\xeq+|y'|}} \calU^\umat(\xeq+|y|,\xeq+|y'|).
\label{eq:Cmat}
\end{equation}
The other correlators are $D_\ueq^{\umat}=C_\ueq^{\umat}/(2\xeq+|y|)$
and $E_\ueq^{\umat}=C_\ueq^{\umat}/[(2\xeq+|y|)(2\xeq+|y'|)]$ and one
has to perform three Fourier transforms to determine the matter era
power spectrum. However, $D_\ueq^{\umat}$ and $E_\ueq^{\umat}$ are
smaller than $C_\ueq^{\umat}(y,y')$ when $2\xeq+|y|>1$ but could
dominate otherwise. The index ``eq'' here is a reminder that there is
an implicit dependence in $k$ through the parameter
$\xeq=k\etaeq$. For $k \ll 1/\etaeq$, this dependence is negligible
and $\CFT_\ueq^\umat(1,1)$ should be roughly constant. In
figure~\ref{fig:spectramatinmatholo}, we have represented the
resulting $\calPh^\umat$ as a function of $k/\calH_0$ by setting
$\CFT_\ueq^\umat(1,1)=1$ with $\DFT_\ueq^\umat(1,1)=0$ and
$\EFT_\ueq^\umat(1,1)=0$. These plots should be typical of matter era
extinct sources but only in the intermediate range $1/\eta_0 \ll k \ll
1/\etaeq$. Indeed, as soon as $k > 1/\etaeq$, one does not expect
$\CFT_\ueq^\umat(1,1)$ to be constant any more and only a precise
knowledge of the function $\calU^\umat(x,x')$ would allow us to
determine how it varies with $k$. For instance, if $\calU^\umat(x,x')$
rapidly decays for $x>\xeq$ (and $x'>\xeq$), faster than
$(xx')^{-3/2}$, equation~\eqref{eq:Cmat} shows that
$\CFT_\ueq^\umat(1,1) \to 0$ and the spectra represented in
figure~\ref{fig:spectramatinmatholo} could decay faster than the
represented $\calPh^\umat \propto k^{-4}$ at $k\gg
1/\etaeq$. Nonetheless, in the regime represented, the
small scale approximation
\begin{equation}
\OmegaGWmat(\eta_1,\eta_2,k \gg \calH) \simeq
\dfrac{k^2}{12 \calH(\eta_1)\calH(\eta_2)} \dfrac{\partial^2 \calPh^\umat}{\partial
  x_1 \partial x_2}\,,
\label{eq:omegamatapprox}
\end{equation}
also holds. On the large scales, for $k<1$, one cannot neglect any more
the other Fourier transforms $\DFT_\ueq^\umat(1,1)$ and
$\EFT_\ueq^\umat(1,1)$. Moreover, as already mentioned, the assumption
of extinct sources on the largest scales is very contriving.

\subsection{Large scales and constant sources}
\label{sec:constant}

The waveforms obtained in equations~\eqref{eq:strainradinradholo},
\eqref{eq:omegaradinradholo}, \eqref{eq:strainradinmatholo},
\eqref{eq:omegaradinmatholo}, \eqref{eq:strainmatholo} and
\eqref{eq:omegamatholo} are exact provided the function
$\calC_k(y,y')$ is compactly supported in addition to be square
integrable. This ensures that its Fourier transform is holomorphic
(see appendix~\ref{sec:holomorphic}). From the definition of
$\calC_k(y,y')$ given in equation~\eqref{eq:calCdef}, this will be the
case if $\calU(x,x')$ has compact support, i.e., there should exist a
domain in the plane $(x,x')$ outside of which the correlator is
vanishing. As an example, let us assume that we require
$\calU_r(x>x_0,x'>x_0)=0$ with $x_0<x_1$ and $x_0<x_2$. From the
definition of the scaling correlator in equation~\eqref{eq:scaling},
this implies that the anisotropic stress $\varPi_r(\xi,k)$ can only be
non-vanishing in a domain of the plane $(\xi,k)$ verifying
$\xi<x_0/k$, which is very restrictive if $x_0$ is small. Conversely,
if the anisotropic stress $\varPi(\xi,k)$ vanishes for $\xi>\xi_0$,
$\calU(x,x')$ will only be compactly supported if there exists a
wavenumber $k_0$ above which $\varPi(\xi,k>k_0)=0$ and one gets $x_0 =
k_0 \xi_0 $. Here again, we see that small values of $x_0$ would be
very contriving, either on the time during which the source can be
active, or on its spatial structure which should not excite high
wavenumbers. It may be possible to relax somehow these constraints
by requiring the correlators to belong the Schwartz space but the
physical requirements for smoothness and time-limited sources will
certainly remain.

Even though the formulas obtained for extinct sources are not
approximation, we thus expect the regime for which they have been
derived to break down at large scale for any realistic anisotropic
stresses. This is illustrated by the infrared divergence of
$\calPh^\umat$ in figure~\ref{fig:spectramatinmatholo}. Interestingly,
for scaling sources such as cosmic defects, the correlators
$\calU(x,x')$ are usually trivial at small $x$ and $x'$ as they become
constant.

Let us assume that $\calU(x,x')= \calU_0$, a constant, for all $x
\le x_1$ and $x'\le x_2$. This condition implies that it is not
compactly supported within the domain of integration and the
expression obtained from extinct sources are no longer
applicable. However, the integral $\Imu$ can again be derived
exactly. In the radiation era, using equations~\eqref{eq:Imurad},
\eqref{eq:KsKerad} and \eqref{eq:ahatrad}, one finds
\begin{equation}
\begin{aligned}
  \Imurad(x_1,x_2,k) &= \dfrac{\calU_0^\urad}{k^2 \eta_1 \eta_2} \bigg\{ \sqrt{x_1} -
  \cos(x_1-\xini) \sqrt{\xini} -\sin(x_1) 
  \left[\fresnelS(x_1) - \fresnelS(\xini) \right] \\ & - \cos(x_1)
   \left[\fresnelC(x_1) - \fresnelC(\xini)
    \right] \bigg\} \times
  \bigg\{ \sqrt{x_2} - \cos(x_2-\xini) \sqrt{\xini} \\ & -\sin(x_2)  \left[\fresnelS(x_2) -
    \fresnelS(\xini) \right]  - \cos(x_2)
  \left[\fresnelC(x_2) - \fresnelC(\xini) \right] \bigg\},
\label{eq:ImuradUconst}
\end{aligned}
\end{equation}
where we have introduced the unnormalised Fresnel
integrals~\cite{Abramovitz:1970aa}
\begin{equation}
\fresnelC(x) = \dfrac{1}{2} \int_0^x
  \dfrac{\cos(t)}{\sqrt{t}} \ud t, \qquad \fresnelS(x) = \dfrac{1}{2} \int_0^x
  \dfrac{\sin(t)}{\sqrt{t}} \ud t.
\end{equation}
In the large scale limit $x_1 \to 0$ and $x_2 \to 0$, still assuming
$\xini \ll x_1$ and $\xini \ll x_2$, one gets
\begin{equation}
\Imurad(x_1,x_2,k) \simeq \dfrac{16 }{225} \calU_0^\urad \left(x_1 x_2 \right)^{3/2},
\end{equation}
which shows that the strain power spectrum varies as $\calPh^\urad
\propto k^3$, at large scales. The $\Imumat$ integral stemming from a
constant correlator $\calU(x,x') = \calU_0^\umat$ in the matter era
can also be analytically derived. From equations~\eqref{eq:Imumat},
\eqref{eq:Ksmat} and \eqref{eq:ahatmat}, one obtains
\begingroup
\allowdisplaybreaks
\begin{align}
    & \Imumat(x_1,x_2,k) = \dfrac{\calU_0^\umat}{4 k^4
    (\eta_1+\etaeq)^2 (\eta_2+\etaeq)^2 (x_1+\xeq)
      (x_2+\xeq)}  \nonumber \\ & \times \bigg\{ \sqrt{x_1} 
    \left[5+2(x_1+\xeq) (x_1+2\xeq) \right] \nonumber \\ & -\sqrt{\xeq}
    \left[5+6\xeq(x_1+\xeq)\right]\cos(x_1-\xeq) +
    \sqrt{\xeq}\left(\xeq-5 x_1\right) \sin(x_1-\xeq)  \nonumber \\ &
    + \left[ \left(5+8 \xeq x_1+ 4\xeq^2 \right) \cos(x_1)+ \left(5
      x_1 - 3 \xeq - 4x_1 \xeq^2 - 4\xeq^3\right)\sin(x_1) \right]   
    \left[ \fresnelC(\xeq) - \fresnelC(x_1) \right]  \nonumber \\
    &+ \left[\left(5+8\xeq x_1 + 4\xeq^2\right)\sin(x_1) - \left(5 x_1
      -3\xeq -4 x_1 \xeq^2 - 4\xeq^3 \right) \cos(x_1) \right]
    \left[\fresnelS(\xeq) - \fresnelS(x_1)\right]\bigg\}  \nonumber \\ & \times
    \bigg\{ \sqrt{x_2} \left[5+2(x_2+\xeq)(x_2+2\xeq) \right] \nonumber \\ & -\sqrt{\xeq}
    \left[5+6\xeq(x_2+\xeq)\right]\cos(x_2-\xeq)  +
    \sqrt{\xeq}\left(\xeq-5 x_2\right) \sin(x_2-\xeq)  \nonumber \\ & 
    + \left[ \left(5+8 \xeq x_2+ 4\xeq^2 \right) \cos(x_2)+ \left(5
      x_2 - 3 \xeq - 4x_2 \xeq^2 - 4\xeq^3\right)\sin(x_2) \right]   
    \left[ \fresnelC(\xeq) - \fresnelC(x_2) \right]  \nonumber \\
    &+ \left[\left(5+8\xeq x_2 + 4\xeq^2\right)\sin(x_2) - \left(5 x_2
      -3\xeq -4 x_2 \xeq^2 - 4\xeq^3 \right) \cos(x_2) \right]
    \left[\fresnelS(\xeq) - \fresnelS(x_2)\right]\bigg\}.
\label{eq:ImumatUconst}
\end{align}
\endgroup
In the large scale limits $x_1\to 0$ and $x_2 \to 0$ with $\xeq \ll
x_1$ and $\xeq \ll x_2$, it simplifies to
\begin{equation}
\Imumat(x_1,x_2,k) \simeq \dfrac{16 }{729} \calU_0^\umat  \left(x_1 x_2 \right)^{3/2},
\end{equation}
which again implies that $\calPh^\umat \propto k^3$ on the largest
scales.

\begin{figure}
\begin{center}
  \includegraphics[width=\twofigw]{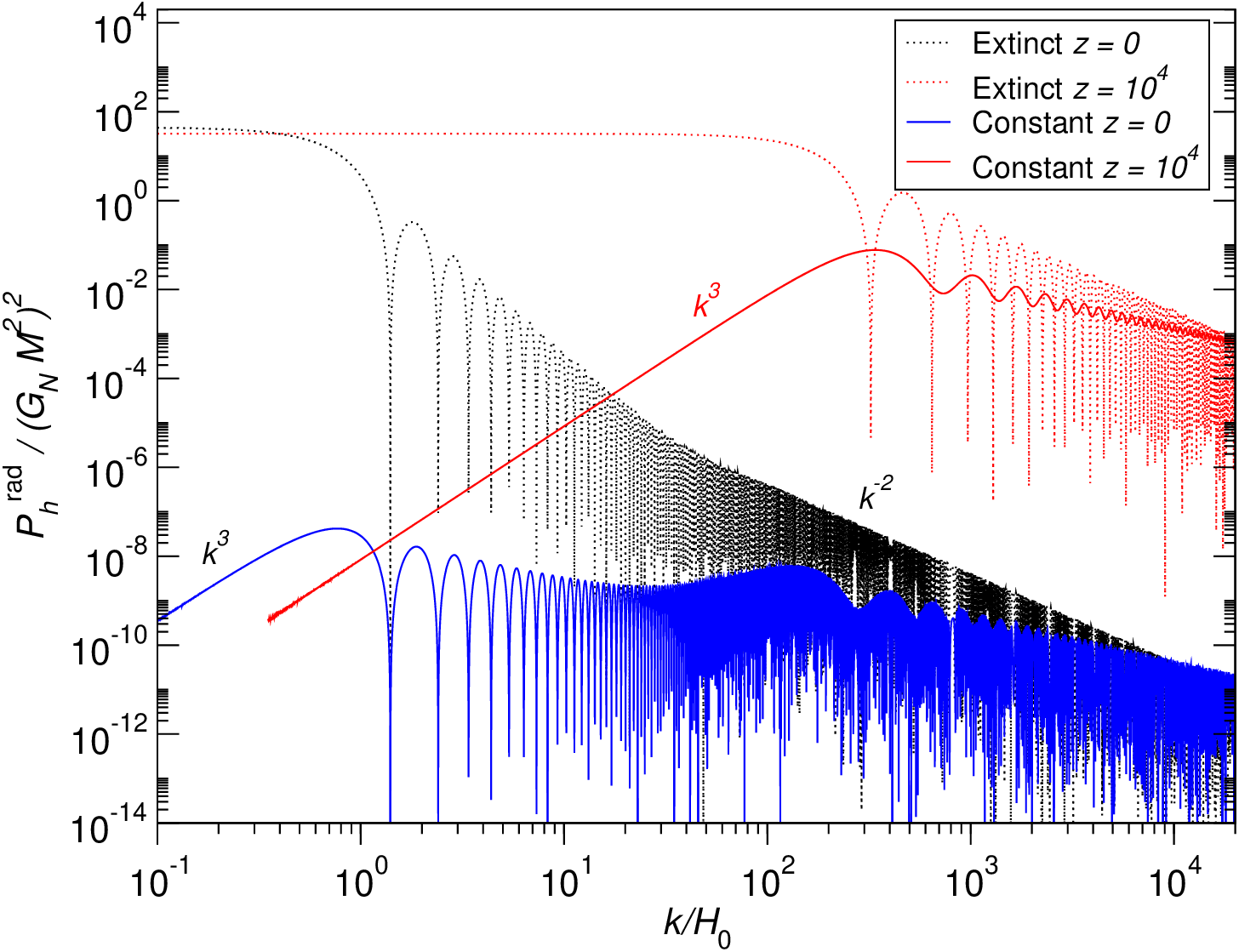}
  \includegraphics[width=\twofigw]{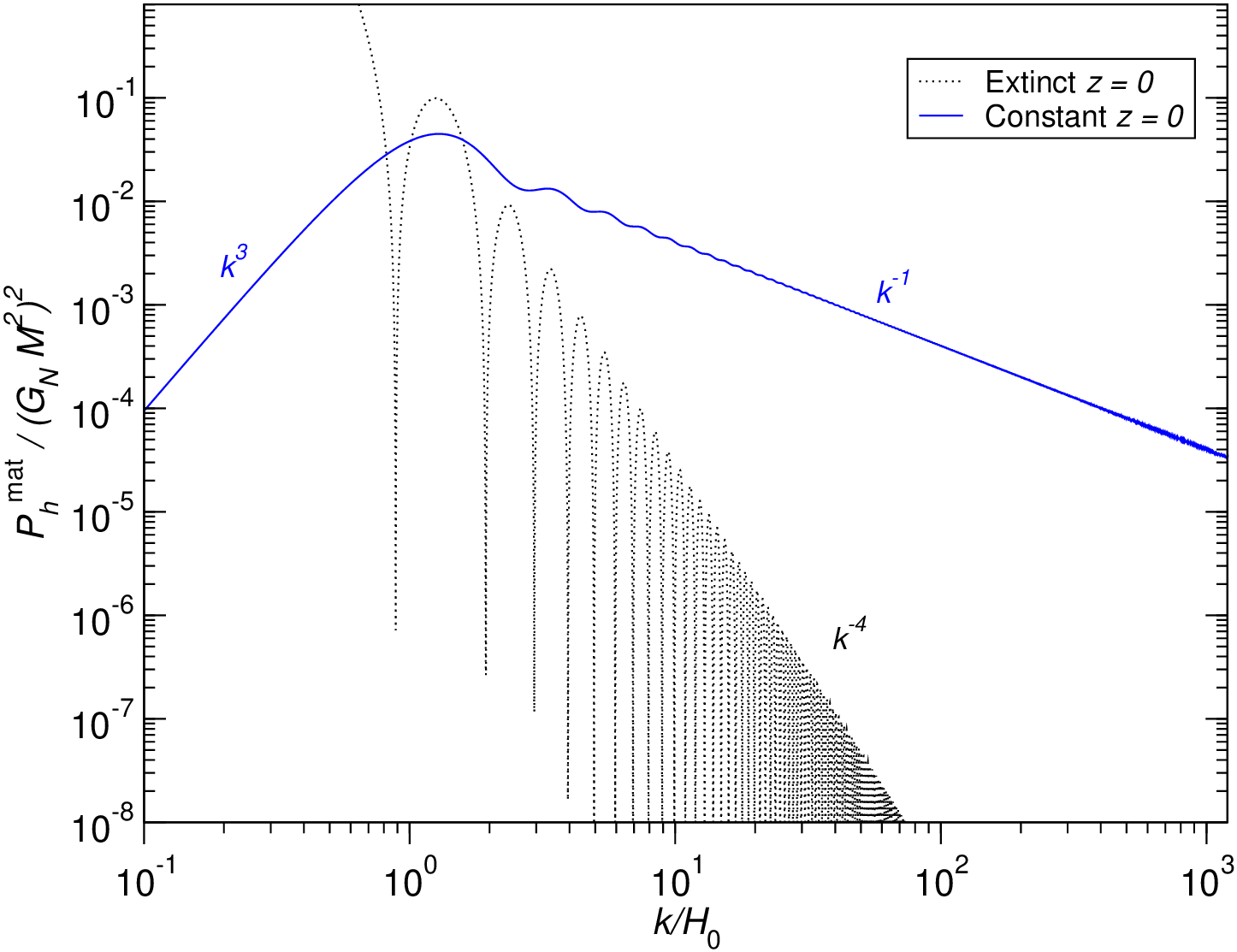}
\caption{The normalised strain power spectrum $\calPh^\urad/(\GN
  M^2)^2$ (left panel) and $\calPh^\umat/(\GN M^2)^2$ (right panel)
  generated by constant sources, i.e., having a strictly constant
  correlator $\calU(x,x')=\calU_0$ over the whole integration
  domain. The power spectrum generated during the radiation era is
  represented at $z=10^4$, but also propagated down to the matter era
  at $z=0$. The dotted curves are the strain spectra associated with
  extinct sources (see figures~\ref{fig:spectraradinmatholo} and
  \ref{fig:spectramatinmatholo}). For illustration purposes, we have
  arbitrarily set $\calU_0^\umat = \calU_0^\urad = 10^{-3}$. We
  expect realistic, but non-singular, scaling sources to be
  ``constant'' at small wavenumbers (as depicted in this figure) while
  behaving as ``extinct'' above some given wavenumber.}
\label{fig:spectraconstcorr}
\end{center}
\end{figure}

From equations~\eqref{eq:ImuradUconst} and \eqref{eq:ImumatUconst},
one could derive exact analytical formulas for the waveforms of all
the observable quantities, $\calPh^\urad$, $\calPh^\umat$,
$\OmegaGWmat$ and $\OmegaGWrad$. Obviously, these expressions would
only be valid in the $(x_1,x_2)$ domains for which $\calU(x,x')$
remains strictly constant. We do not report these calculations here,
but, as an illustration, we have plotted in
figure~\ref{fig:spectraconstcorr}, the shape of the radiation and
matter era strain power spectrum stemming from
equations~\eqref{eq:ImuradUconst} and \eqref{eq:ImumatUconst},
respectively. For comparison, we have reported the spectra associated
with the extinct sources of
figures~\ref{fig:spectraradinmatholo} and
\ref{fig:spectramatinmatholo}. Firstly, one can notice that, for
constant correlators, the oscillations are not maximal and only induce
small modulations with respect to the overall amplitude. This can be
understood from equations~\eqref{eq:ImuradUconst} and
\eqref{eq:ImumatUconst}. Oscillatory terms have a prefactor scaling as
a positive power of $\xeq$ in the matter era (and $\xini$ in the radiation era), which is always smaller than $x_1$ and
$x_2$. Moreover, all the Fresnel integrals appear as differences,
as in $\fresnelC(\xeq)-\fresnelC(x_1)$, and they vanish
asymptotically. As such, they can never drive the overall shape of the
spectra. This contrasts with the spectra associated with extinct
scaling sources. Because we do not expect $\calU(x,x')$ to remain
constant for $x_1$ and $x_2$ large, the strain power spectra
associated with realistic scaling sources should be matching the
constant correlator behaviour at small wavenumbers, up to some
wavenumber at which it turns into an extinct sources
spectra. Let us notice that, focusing only on the spectra envelope, and
omitting the differences between $\OmegaGW$ and $\calPh$, such a
behaviour is the one discussed in Refs.~\cite{Figueroa:2012kw,
  Figueroa:2020lvo}. As such, it could be considered as the standard
lore for scaling sources and we have now added its complete time and
wavenumber dependence.

However, as we discuss in the next section, it is possible for certain
scaling sources to produce a singular Fourier transform
$\calCFT(\gamma,\gamma')$ from a non-square integrable function
$\calC(y,y')$ while having a regular correlator $\calU(x,x')$.

\subsection{Small scales and singular sources}
\label{sec:singular}

In view of the previous discussion, it is instructive to discuss
scaling sources that could possibly break the assumption of being
``extinct'' on all length scales. A sufficient condition for this to
happen is that the Fourier transform $\calCFT(\gamma,\gamma')$ should
be singular for some value of $\gamma$ and $\gamma'$.

As an example, let us consider a perfectly coherent correlator that
behaves at large $x$ and $x'$ as
\begin{equation}
\calU(x\gg 1,x'\gg 1) = \dfrac{\calUinf}{\sqrt{x x'}}\,,
\label{eq:toycoherent}
\end{equation}
where $\calUinf$ is a constant. For $x=x'$, the correlator slowly
decays with the wavenumber as $1/k$ and such a behaviour is reminiscent
with the small scales behaviour of the two-point correlation functions
associated with a random distribution of line-like objects such as
long cosmic strings~\cite{PhysRevD.45.1898, Wu:2002,
  Fraisse:2007nu,Ringeval:2012tk}.

\begin{figure}
\begin{center}
  \includegraphics[width=\twofigw]{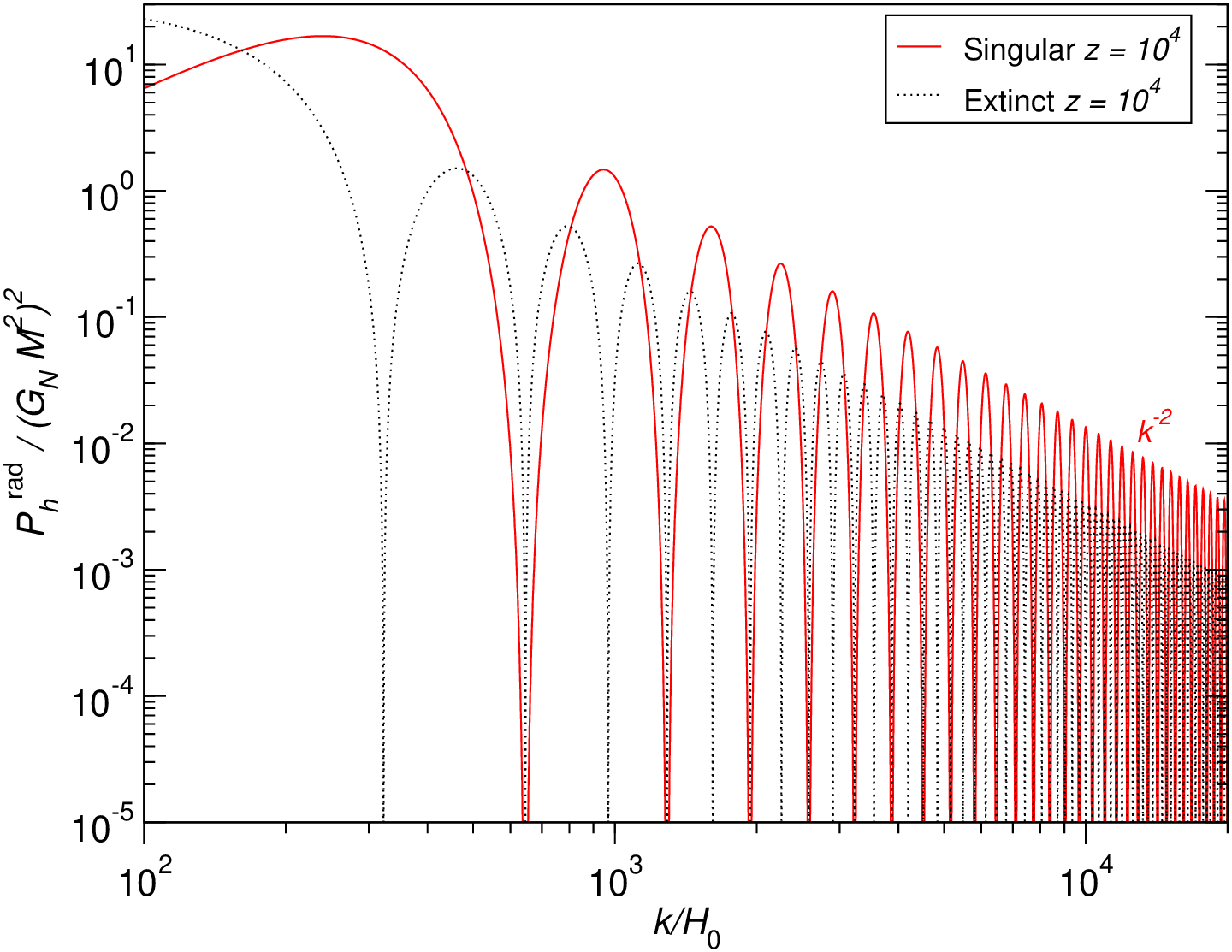}
  \includegraphics[width=\twofigw]{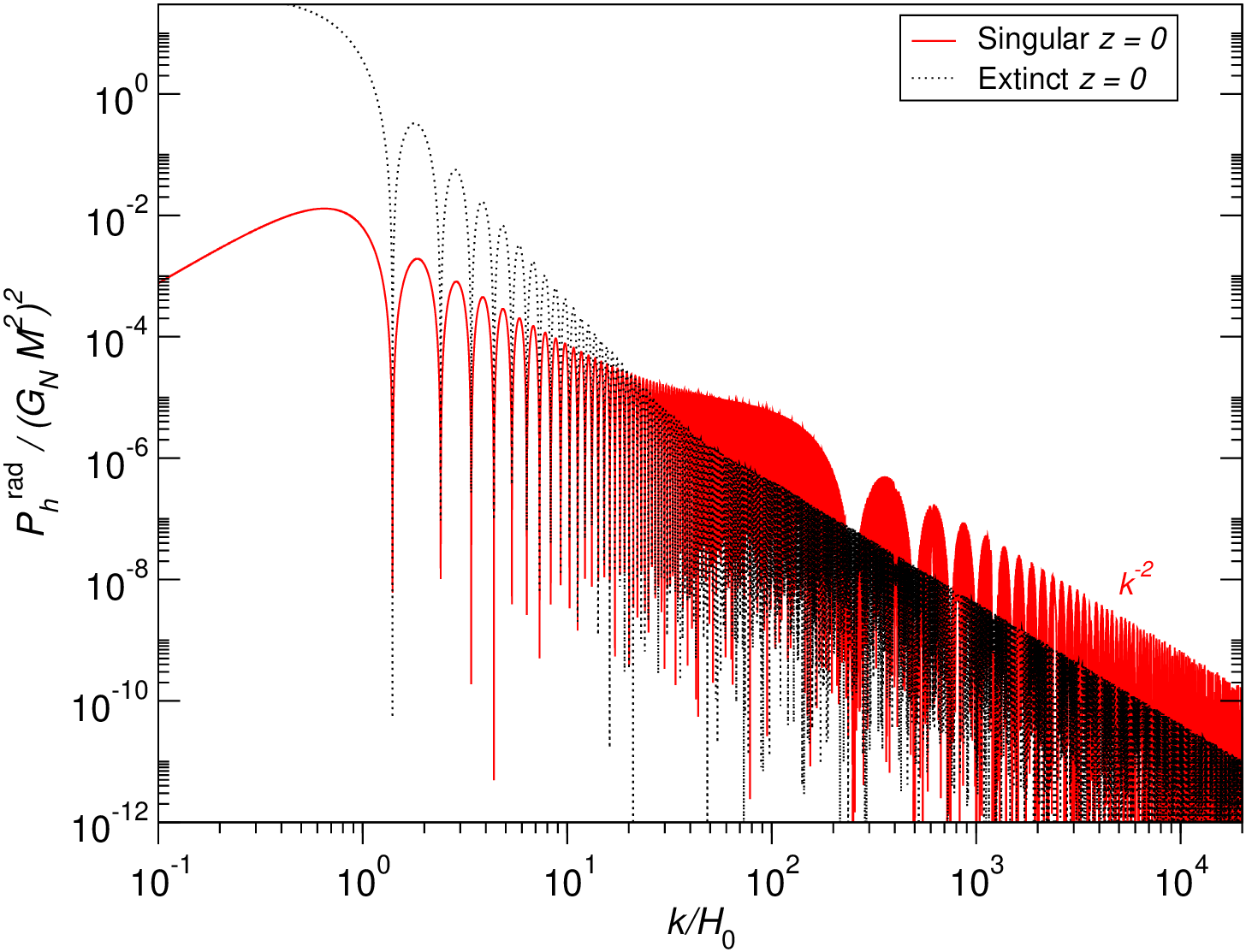}
\caption{The normalised strain power spectrum $\calPh^\urad/(\GN
  M^2)^2$ generated in the radiation era by a singular source having a
  correlator $\calU(x,x')=\calUinf/\sqrt{x x'}$. The left panel shows
  the strain spectrum generated and evaluated in the radiation era
  ($z=10^4$) whereas the right panel shows the same spectrum
  propagated and evaluated in the matter era ($z=0$). The strain
  spectrum from extinct sources is represented as dotted curves. Both
  have an envelope decreasing as $k^{-2}$ at large wavenumbers but
  exhibit completely different oscillatory patterns, as if the
  singular sources were generating interferences. For illustration
  purposes, we have set $\calUinf=1$.}
\label{fig:strainsingrad}
\end{center}
\end{figure}

In the radiation era, at large enough $y$ and $y'$, from equations~\eqref{eq:Crad}, one gets
\begin{equation}
C_\uini^\urad(y,y') = \calUinf^\urad \implies
\CFT_\uini^\urad(\gamma,\gamma') = (2\pi)^2 \calUinf^\urad \delta(\gamma) \delta(\gamma'),
\end{equation}
and the Fourier transform $\calCFT_k^\urad$ is a distribution, singular at the origin of
the plane $(\gamma,\gamma')$. It reads
\begin{equation}
\calCFT_k^\urad(\gamma,\gamma') = (2\pi)^2
\dfrac{\calUinf^\urad}{k^2 \eta_1 \eta_2}\,\delta(\gamma)\delta(\gamma').
\end{equation}
Plugging this expression into the integrals $\Icc$, $\Ics$, $\Isc$ and
$\Iss$ given by equations~\eqref{eq:Iccsinc}, \eqref{eq:Isssinc},
\eqref{eq:Icssinc} and \eqref{eq:Iscsinc} gives a completely different
result than the extinct case. In particular, $\Iss$, $\Ics$ and $\Isc$
are now non-vanishing and read\footnote{These integrals can be more
straightforwardly calculated in the $(y,y')$ space from
equations~\eqref{eq:Iccss}. We do it in Fourier space for illustrating
how the singular behaviour of $\calCFT_\uini^\urad(\gamma,\gamma')$
breaks the extinct source hypothesis.}
\begin{equation}
\begin{aligned}
  \Icc(y_1,y_2,k) &= \dfrac{\calUinf^\urad}{k^2 \eta_1 \eta_2} \sin(y_1) \sin(y_2), \qquad
  & \Iss(y_1,y_2,k) & = \dfrac{4\calUinf^\urad}{k^2 \eta_1 \eta_2} \sin^2\left(\dfrac{y_1}{2}\right)
  \sin^2\left(\dfrac{y_2}{2}\right), \\
  \Ics(y_1,y_2,k) & = \dfrac{2\calUinf^\urad}{k^2 \eta_1\eta_2} \sin(y_1)
  \sin^2\left(\dfrac{y_2}{2}\right),\qquad &  \Isc(y_1,y_2,k) & =
 \dfrac{2 \calUinf^\urad}{k^2 \eta_1 \eta_2}
 \sin^2\left(\dfrac{y_1}{2}\right) \sin(y_2) .
\end{aligned}
\end{equation}
From equation~\eqref{eq:ImuradfromIccss}, one obtains
\begin{equation}
\Imurad(x_1,x_2,k) = \dfrac{4 \calUinf^\urad}{k^2 \eta_1 \eta_2}
\sin^2\left(\dfrac{x_1-\xini}{2} \right) \sin^2\left(\dfrac{x_2-\xini}{2} \right),
\label{eq:Imuradsing}
\end{equation}
from which the strain power spectrum reads
\begin{equation}
\begin{aligned}
  \calPh^\urad(\eta_1<\etaeq,\eta_2<\etaeq,k) & = 128 \left(\GN M^2 \right)^2
\dfrac{\calUinf^\urad}{k^2 \eta_1 \eta_2} \bigg\{1  -
\cos(x_1 - \xini) - \cos(x_2 - \xini) \\ & +
\dfrac{1}{2}\cos(x_1-x_2) + \dfrac{1}{2} \cos(x_1+x_2 - 2\xini) \bigg\}.
\label{eq:strainradinradsing}
\end{aligned}
\end{equation}
This expression has to be compared to the one for extinct sources in
equation~\eqref{eq:strainradinradholo}, together with
equation~\eqref{eq:calCtoCrad} (see also
figure~\ref{fig:spectraradinmatholo}). The amplitude in front of each
oscillatory function is different and so are their associated
waveform. For instance, focusing on the equal-time spectra with
$x_1=x_2=x_0$, we see that the strain spectrum for extinct sources
varies as $[\sin(x_0-\xini)]^2/k^2$ whereas the singular one goes as
$\{\sin[(x_0-\xini)/2]\}^4/k^2$. Some oscillations have disappeared as
if interferences were appearing. Notice that focusing only on their
envelope, both spectra decay as $1/k^2$ and would be undistinguishable
without looking at their fine structure. Deriving
equation~\eqref{eq:Imuradsing} with respect to $x_1$ and $x_2$ gives
the other integrals
\begin{equation}
\Idmurad(x_1,x_2,k)  = \dfrac{\calUinf^\urad}{k^2 \eta_1 \eta_2}
\sin(x_1-\xini) \sin(x_1-\xini),
\end{equation}
and
\begin{equation}
\Ixrad(x_1,x_2,k)  = \dfrac{2 \calUinf^\urad}{k^2 \eta_1 \eta_2}
\sin(x_1-\xini) \sin^2\left(\dfrac{x_2-\xini}{2}\right).
\end{equation}
These integrals evaluated at $x_1=x_2=\xeq$ allow us to derive the
radiation strain spectrum for the singular source propagated in the
matter era by using equations~\eqref{eq:strainradinmat} and
\eqref{eq:omegaradinmat}. They inherit the $1/k^2$ behaviour in their
envelope while the waveform measured at any $\eta_i > \etaeq$ is driven
by the $A_i$ and $B_i$ functions. However, there is also an additional
modulation which is induced by the functions $\Imurad(\xeq,\xeq,k)$,
$\Idmurad(\xeq,\xeq,k)$ and $\Ixrad(\xeq,\xeq,k)$. This modulation is
visible, and compared to the extinct spectrum, in the right panel of
figure~\ref{fig:strainsingrad}. Here as well, only the presence of
these interferences would signal a singular source at large
wavenumbers. Let us also notice that these modulations involve terms
in $\xeq - \xini \simeq k \etaeq$ for $\etaini \ll \etaeq$. Therefore,
in real space, the typical length associated with these interferences
is about the Hubble radius at equality, i.e., of the order of
$100\,\Mpc$. There are also differences at smaller wavenumbers but, as
mentioned before, the correlator should behave as constant on these
scales and equation~\eqref{eq:toycoherent} may not be applicable (see
section~\ref{sec:constant}).

We can also derive the energy density spectrum within the radiation
era, as given by equation~\eqref{eq:omegaradinrad}. Since all
integrals are of the same typical amplitude, at small scales, one has
\begin{equation}
\OmegaGWrad(\eta_1<\etaeq,\eta_2<\etaeq,k\gg\calH) \simeq
\dfrac{32}{3} \left(\GN M^2 \right)^2 \dfrac{\calUinf^\urad}{\eta_1
  \calH_1 \eta_2 \calH_2} \sin(x_1-\xini) \sin(x_2-\xini),
\end{equation}
which is, as expected, proportional to the double derivative of
$\partial^2 \calPh^\urad/\partial x_1 \partial x_2$. Interestingly,
the ``interferences'' are no longer present for $\OmegaGWrad$ and it
behaves almost exactly as the one associated with extinct sources, see
equations~\eqref{eq:Idmuradholo} and \eqref{eq:omegaradinradholo}. It
oscillates with a product of sine functions instead of a product of cosine functions and both
end-up only differing by a phase shift of $\pi/2$ at large wavenumbers.

The singular source of equation~\eqref{eq:toycoherent} induces more
pronounced effects in the matter era. From equation~\eqref{eq:Cmat}
one gets
\begin{equation}
C_\ueq^\umat(y,y') = \dfrac{(2\xeq + |y|)^2 (2\xeq +
  |y'|)^2}{(\xeq+|y|)(\xeq+|y'|)} \calUinf^\umat,
\end{equation}
which is a growing function of $(y,y')$. It can be separated into
three terms according to their behaviour at large $(y,y')$
\begin{equation}
\begin{aligned}
  C_\ueq^\umat(y,y') & = \calUinf^\umat (\xeq +|y|)(\xeq+|y'|) \\ & + 
\calUinf^\umat\xeq  \left[\dfrac{(3\xeq+2|y|)(\xeq+|y'|)}{\xeq+|y|} +
  \dfrac{(3\xeq+2|y'|)(\xeq+|y|)}{\xeq+|y'|}\right] \\
& + \calUinf^\umat \xeq^2 \dfrac{(3 \xeq+2 |y|)(3 \xeq +
  2|y'|)}{(\xeq+ |y|)(\xeq + |y'|)}\,.
\end{aligned}
\end{equation}
The first term is the one that dominates asymptotically and, to
simplify the discussion, we focus only on this one\footnote{This term
would be the only one present if we were not considering the matter
era to be preceded by a radiation era.}. Moreover, doing so is
consistent with neglecting the other functions $D_\ueq^\umat(y,y')$
and $E_\ueq^\umat(y,y')$ at large wavenumbers. We have for the
Fourier transform
\begin{equation}
\calCFT_k^\umat(\gamma,\gamma') \simeq \dfrac{\calUinf^\umat}{k^4
  (\eta_1+\etaeq)^2(\eta_2+\etaeq)^2}\left[2 \pi \xeq \delta(\gamma) -
  \dfrac{2}{\gamma^2} \right]\left[2 \pi \xeq \delta(\gamma') -
  \dfrac{2}{\gamma'^2}\right].
\end{equation}
Clearly not holomorphic as it contains Dirac distributions as well as
power law terms in $1/\gamma^2$ and $1/\gamma'^2$, all singular at the
origin $\gamma=\gamma'=0$. These terms explicitly break the extinct
sources calculations. Ignoring the other functions
$\calDFT_\ueq^\umat$ and $\calEFT_\ueq^\umat$, and considering only
the asymptotic form of the matter era Green's functions, one obtains,
for the four basic integrals, the following approximations
\begin{equation}
\begin{aligned}
  \Icc(y_1,y_2,k\gg\calH) &\simeq  \dfrac{\calUinf^\umat}{k^4
    (\eta_1+\etaeq)^2(\eta_2+\etaeq)^2}
  \left[(\xeq+y_1)\sin(y_1)+\cos(y_1)-1\right] \\ & \times
  \left[(\xeq+y_2)\sin(y_2)+\cos(y_2)-1\right], \\
  \Iss(y_1,y_2,k\gg\calH) & \simeq \dfrac{\calUinf^\umat}{k^4
    (\eta_1+\etaeq)^2(\eta_2+\etaeq)^2}
  \left\{\xeq[1-\cos(y_1)]-y_1\cos(y_1)+\sin(y_1)\right\} \\ & \times \left\{\xeq[1-\cos(y_2)]-y_2\cos(y_2)+\sin(y_2)\right\},
\end{aligned}
\end{equation}
and
\begin{equation}
\begin{aligned}
 \Ics(y_1,y_2,k\gg\calH) &=\dfrac{\calUinf^\umat}{k^4
    (\eta_1+\etaeq)^2(\eta_2+\etaeq)^2}
 \left[(\xeq+y_1)\sin(y_1)+\cos(y_1)-1\right]  \\ & \times \left\{\xeq[1-\cos(y_2)]-y_2\cos(y_2)+\sin(y_2)\right\},
\end{aligned}
\end{equation}
with $\Isc(y_1,y_2,k)=\Ics(y_2,y_1,k)$. Finally, one gets
\begin{equation}
\begin{aligned}
  \Imumat(x_1,x_2,k\gg\calH) & \simeq \dfrac{\calUinf^\umat}{k^4
    (\eta_1+\etaeq)^2(\eta_2+\etaeq)^2}\left[x_1 - \xeq \cos(x_1-\xeq)
  - \sin(x_1-\xeq)\right] \\ &\times \left[x_2 - \xeq \cos(x_2-\xeq)
  - \sin(x_2-\xeq)\right], 
\label{eq:Imumatsing}
\end{aligned}
\end{equation}
and, with $\eta_i\gg \etaeq$, keeping only the leading terms, this implies
\begin{equation}
\begin{aligned}
  \calPh^\umat(\eta_1,\eta_2,k\gg\calH) & \simeq \dfrac{128 \left(\GN
    M^2 \right) \calUinf^\umat}{k^2
    \eta_1 \eta_2}\left[1 - \dfrac{\etaeq}{\eta_1} \cos(x_1-\xeq)\right]\left[1 - \dfrac{\etaeq}{\eta_2}
    \cos(x_2-\xeq) \right].
\label{eq:strainmatsing}
\end{aligned}
\end{equation}
At large wavenumbers this spectrum behaves as $1/k^2$, modulated by
small oscillations, a very different result than the expected decay
in $1/k^4$. The integrals $\Idmumat$ and $\Ixmat$ are obtained by
deriving equation~\eqref{eq:Imumatsing} with respect to $x_1$ and
$x_2$. From equations~\eqref{eq:partialx1x2} and \eqref{eq:Imumatsing}, they read
\begin{equation}
\begin{aligned}
  \Idmumat(x_1,x_2,k\gg\calH) & \simeq \dfrac{\calUinf^\umat}{k^4
    (\eta_1+\etaeq)^2(\eta_2+\etaeq)^2}\left[1 + \xeq \sin(x_1-\xeq)
  - \cos(x_1-\xeq)\right] \\ &\times \left[1 + \xeq \sin(x_2-\xeq)
  - \cos(x_2-\xeq)\right], 
\label{eq:Idmumatsing}
\end{aligned}
\end{equation}
and
\begin{equation}
\begin{aligned}
  \Ixmat(x_1,x_2,k\gg\calH) & \simeq \dfrac{\calUinf^\umat}{k^4
    (\eta_1+\etaeq)^2(\eta_2+\etaeq)^2}\left[1 + \xeq \sin(x_1-\xeq)
  - \cos(x_1-\xeq)\right] \\ &\times \left[x_2 - \xeq \cos(x_2-\xeq)
  - \sin(x_2-\xeq)\right].
\label{eq:Ixmatsing}
\end{aligned}
\end{equation}
They allow us to determine $\OmegaGWmat$ from
equation~\eqref{eq:omegamat} and one gets
\begin{equation}
\begin{aligned}
  \OmegaGWmat(\eta_1,\eta_2,k\gg\calH) & \simeq \dfrac{32}{3} \left(\GN M^2
  \right)^2 \dfrac{\calUinf^\umat}{k^4 \left(\eta_1 + \etaeq \right)^2
    \left(\eta_2+\etaeq \right)^2} \\ & \times
  \left[\dfrac{k}{\calH_1} - x_1 + \left(\dfrac{k \xeq}{\calH_1} +
    1\right)\sin(x_1-\xeq) - \left(\dfrac{k}{\calH_1} - \xeq \right)
    \cos(x_1-\xeq) \right] \\ & \times \left[\dfrac{k}{\calH_2} - x_2 +\left(\dfrac{k \xeq}{\calH_2} +
    1\right)\sin(x_2-\xeq) - \left(\dfrac{k}{\calH_2} - \xeq \right)
    \cos(x_2-\xeq) \right].
\end{aligned}
\end{equation}
It decreases as $1/k^2$ when the terms in $k/\calH$ dominate and until
the terms in $k \xeq/\calH$ take over. When they do, the
leading terms read
\begin{equation}
\OmegaGWmat(\eta_1,\eta_2,k\gg\calH) \simeq \dfrac{32}{3} \left(\GN
M^2 \right)^2  \calUinf^\umat \dfrac{\etaeq^2}{\eta_1\eta_2 (\eta_1
  \calH_1)(\eta_2 \calH_2)} \sin(x_1-\xeq) \sin(x_2-\xeq),
\end{equation}
and $\OmegaGWmat$ maximally oscillates with a scale-invariant
envelope. Compared to equation~\eqref{eq:strainmatsing}, we see that
its amplitude strongly violates the relation $\OmegaGW
\simeq k^2/(12 \calH^2) \calPh$. Instead we have
\begin{equation}
\max\left(\OmegaGWmat \right) \simeq \dfrac{\etaeq^2}{\eta_1 \eta_2}
\max\left(\dfrac{k^2}{12 \calH_1 \calH_2} \calPh \right),
\end{equation}
and the maximal amplitude reached by the energy density is typically
four orders of magnitude smaller than the typical strain power
spectrum amplitude.

\begin{figure}
\begin{center}
  \includegraphics[width=\twofigw]{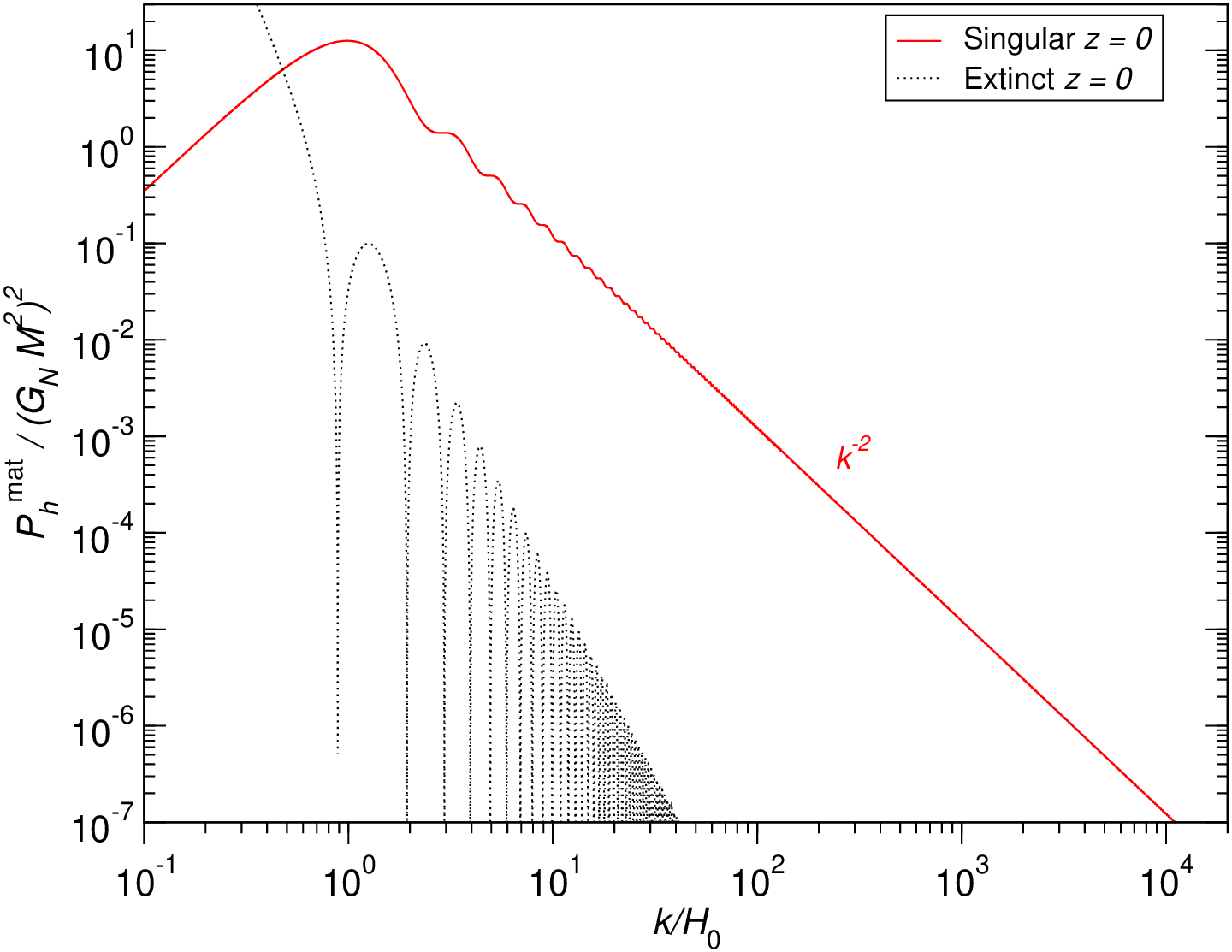}
  \includegraphics[width=\twofigw]{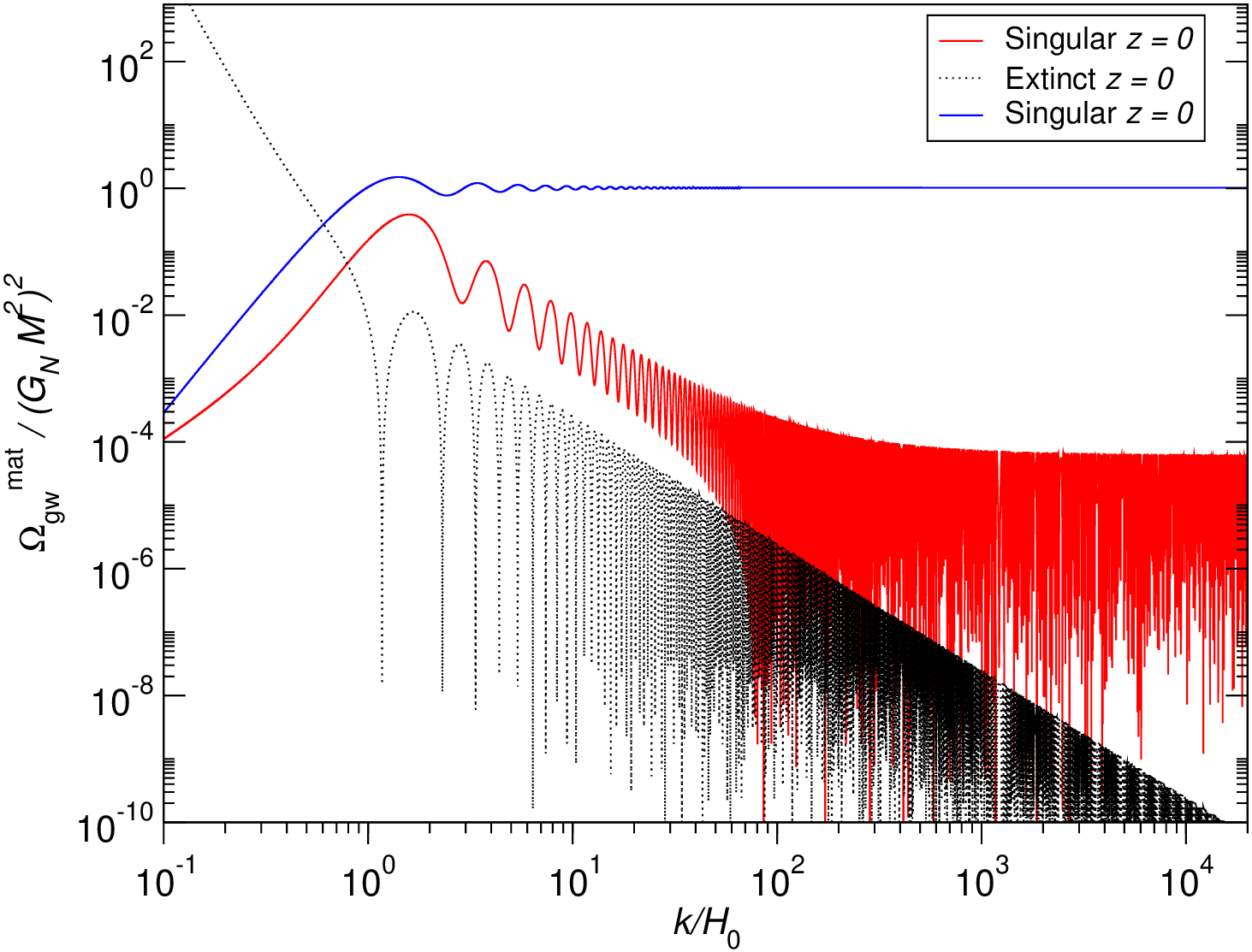}
\caption{The left panel shows the normalised strain power spectrum
  $\calPh^\umat/(\GN M^2)^2$ generated in the matter era by a singular
  source having a correlator $\calU(x,x')=\calUinf/\sqrt{x x'}$ (red
  curve). It is compared to the one associated with extinct sources
  (black dotted curve). Notice the unusual behaviour of $k^2
  \calPh^\umat$ becoming scale invariant at large wavenumbers. The
  right panel shows the corresponding energy density parameter
  $\OmegaGWmat/(\GN M^2)^2$, coming from the singular source in red,
  associated with extinct sources in black (dotted curve) and compared
  to the singular $k^2/(12 \calH_0^2) \calPh^\umat$ (blue horizontal
  curve). The usual relation $\OmegaGWmat \simeq k^2/(12 \calH_0^2)
  \calPh$ is violated on all length scales. For illustration purposes,
  we have set $\calUinf=1$.}
\label{fig:strainsingmat}
\end{center}
\end{figure}

\begin{figure}
\begin{center}
  \includegraphics[width=\twofigw]{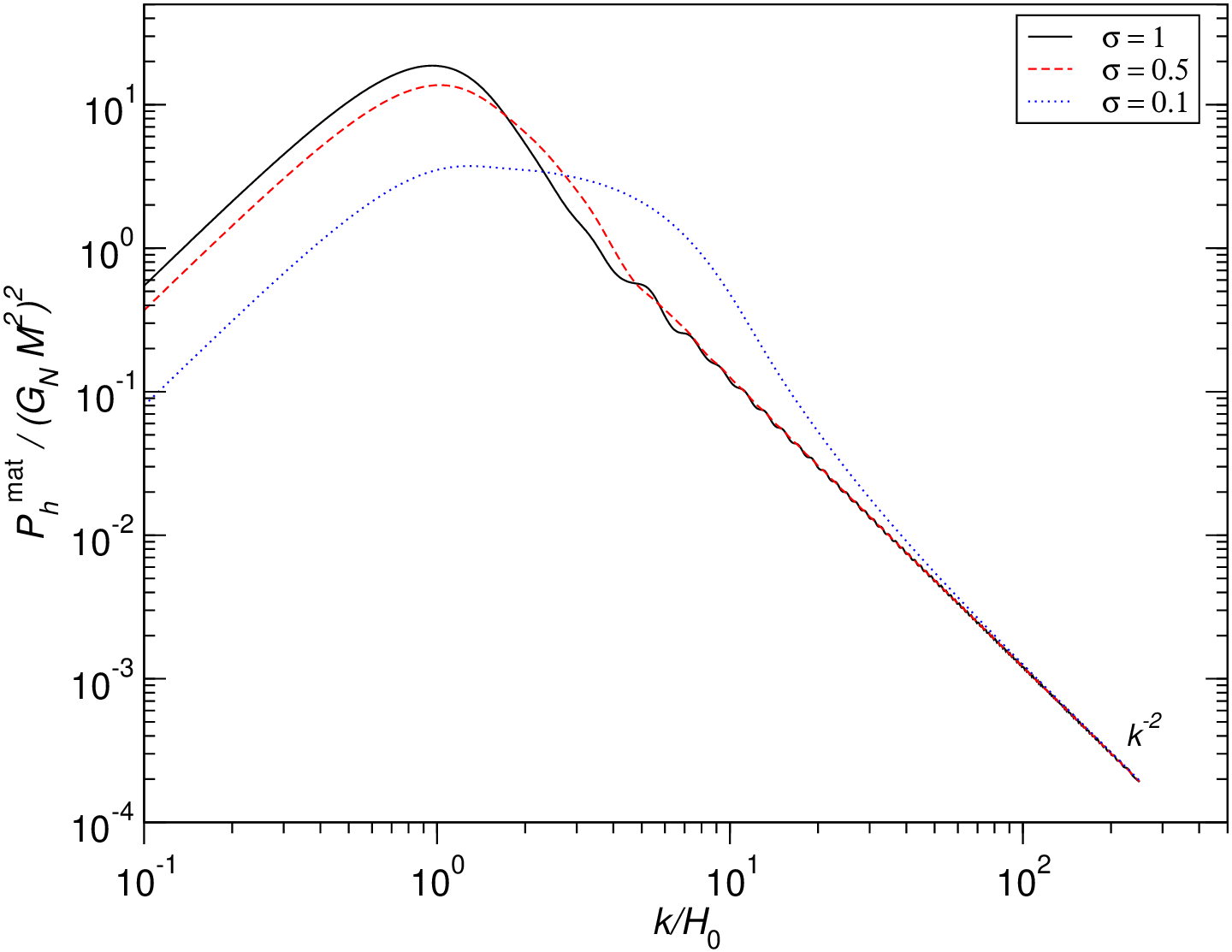}
  \includegraphics[width=\twofigw]{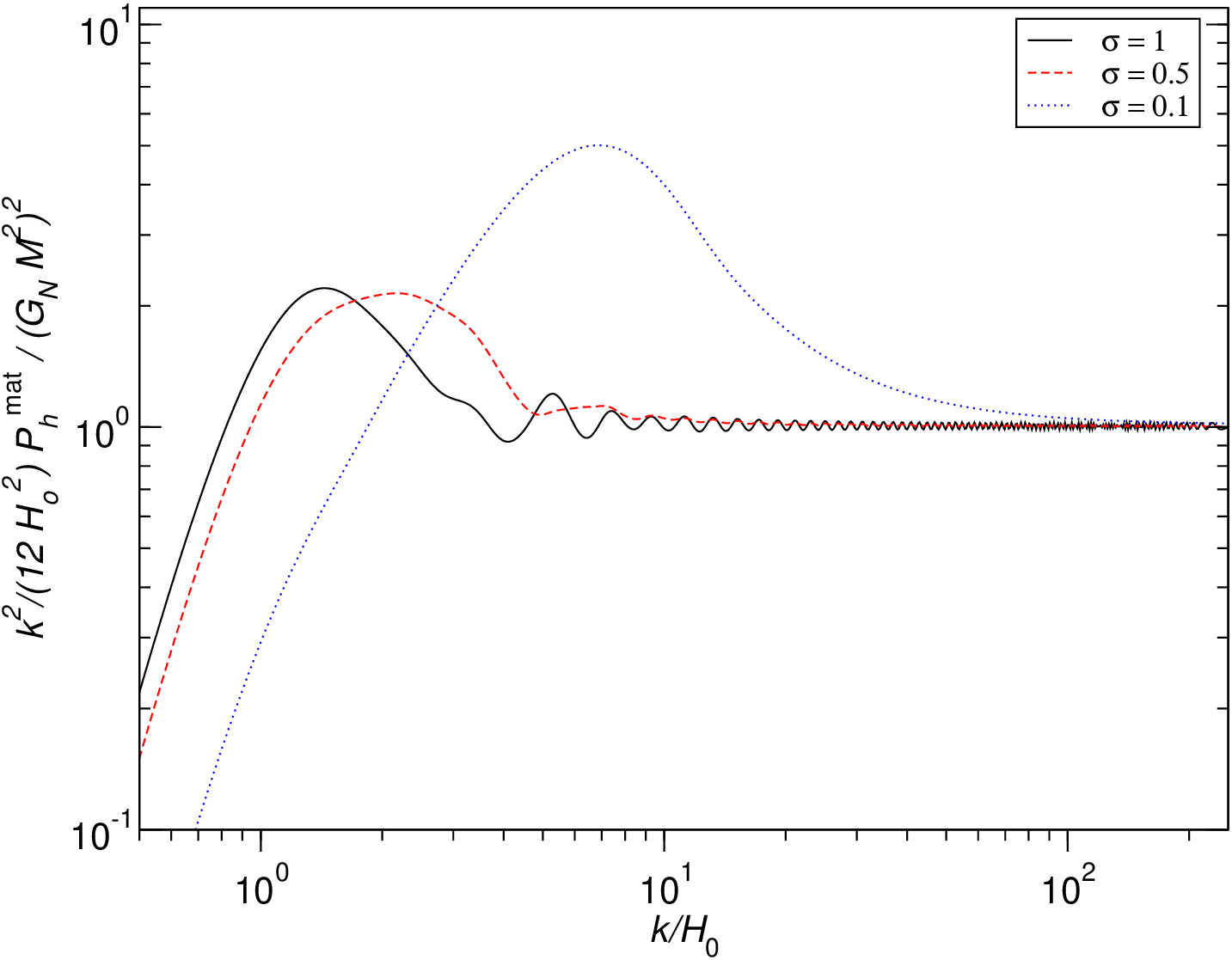}
\caption{Direct numerical evaluation of the normalised strain power
  spectrum $\calPh^\umat/(\GN M^2)^2$ (left panel) generated in the
  matter era by the smoothed singular source of
  equation~\eqref{eq:toysmooth} (with $\calUinf^\umat=1$), for
  different values of the smoothing width $\sigma$. The right panel
  shows the rescaled spectrum $k^2/(12 \calH_0^2)\calPh^\umat/(\GN
  M^2)^2$. The smoothing does not affect the behaviour at large
  wavenumbers $\calPh^\umat \propto 1/k^2$ (see also
  figure~\ref{fig:strainsingmat}).}
\label{fig:straingaussmat}
\end{center}
\end{figure}

We have represented in figure~\ref{fig:strainsingmat} the shape of the
equal-time matter era spectra coming from the singular correlator and
compared them to the ones associated with the extinct sources. In the
right panel of this figure, both $\OmegaGWmat$ and $k^2/(12\calH_0^2)
\calPh^\umat$ are represented. They strongly differ at all
wavenumbers. The perfectly coherent correlator of
equation~\eqref{eq:toycoherent} exhibits a high degree of symmetry and
decreases very slowly as $1/\sqrt{x}$ at fixed $x'$. But the main
reason for the appearance of the singular behaviour described above
lies in the fact that the function $\calC_k(y,y')$ is not square
integrable, and this statement depends not only on how the correlator
$\calU(x,x')$ behaves at large $(x,x')$ but also on how fast the scale
factor $a(\eta)$ grows. That is why the singular spectra associated
with equation~\eqref{eq:toycoherent} exhibit more pronounced
differences with respect to the extinct sources case in the matter era
than in the radiation era. Concerning the choice of a coherent
correlator, one could easily check that a perfectly \emph{incoherent}
correlator, varying as $\calU(x,x') = \calUinf \delta(x-x')/x$ would
induce an even more pronounced effect in the matter era, the strain
power spectrum decreasing only as $1/k$ at large wavenumbers (the
Dirac distribution makes it more singular). One can also check
that smoothing the transverse structure of the correlator with some
Gaussian function does not change the result. In
figure~\ref{fig:straingaussmat}, we have represented the matter era
strain power spectrum numerically computed from a smoothed correlator
varying as
\begin{equation}
\calU^\umat(x,x') = \dfrac{\calUinf^\umat}{\sqrt{x x'}}
\exp\left[-\dfrac{\left(\frac{x}{x'} - 1 \right)^2}{2 \sigma^2} \right],
\label{eq:toysmooth}
\end{equation}
and for various values of $\sigma$. The effect of having a strong
smoothing $\sigma \ll 1$ is to damp the oscillations visible in
figure~\ref{fig:strainsingmat}, add a new correlation scale in the
spectrum around $k \simeq 1/\sigma$, but the slow decay of
$\calPh^\umat \propto 1/k^2$ at large wavenumbers remains. In
conclusion, the simplest way to determine if any singular behaviour is
present is to search for singularities in the Fourier transform
$\calCFT_k(\gamma,\gamma')$. When $\calC_k(x,x')$ is non-square
integrable, poles are expected to show up at null ``frequencies''
$(\gamma,\gamma')$, but any other singularities elsewhere would
equally trigger new features in the spectrum and deviations from the
extinct sources case.

A note of caution is however in order. All along the paper we have
considered an instantaneous transition from the radiation to the
matter era and we have assumed that the anisotropic stress could take
a scaling form instantaneously at the transition. All realistic
scaling sources are expected to not being in ``scaling'' during the
transition and various distortions on the spectra should be expected
around the length scales associated $k=1/\etaeq$. For instance, it is
perfectly possible that the matter era power spectrum associated with
cosmic strings exhibit the $1/k^2$ decrease (see
figure~\ref{fig:strainsingmat}) only over an intermediate range of
wavenumbers above which it could be sensitive to the non-scaling
anisotropic stress at $k>1/\etaeq$. Only a full numerical simulation
of cosmic strings would allow us to determine its precise
shape~\cite{inprogress}.

\section{Conclusion}
\label{sec:conclusion}

Let us briefly recap our main results. We have derived the explicit
unequal-time and wavenumber dependence of the strain power spectrum
$\calPh(\eta_1,\eta_2,k)$ as well as the energy density parameter
$\OmegaGW(\eta_1,\eta_2,k)$ for scaling sources. For a wide class of
sources, extinct and smooth, having a holomorphic Fourier transform
$\calCFT_k(\gamma,\gamma')$, we have derived their complete analytical
forms given in equations~\eqref{eq:strainradinradholo},
\eqref{eq:omegaradinradholo}, \eqref{eq:strainradinmatholo},
\eqref{eq:omegaradinmatholo}, \eqref{eq:strainmatholo} and
\eqref{eq:omegamatholo}. However, realistic scaling sources are
expected to be ``constant'' on large scales before turning ``extinct''
on smaller scales. The spectra for constant sources have been derived
in section~\ref{sec:constant} and exhibit only small modulations. As
such, realistic sources may only be strongly oscillating at small
scales, in a regime which is notoriously difficult to compute, but on
immediate reach by GW direct detection experiments. Let us notice that
other cosmological sources, not necessarily scaling, have been shown
to produce oscillations~\cite{Kawasaki:2011vv, Jung:2018kde,
  Fumagalli:2020nvq} or time variation~\cite{Mukherjee:2019oma}. The
precise determination of the SGWB fine structure is therefore of
immediate interest for their disambiguation. In
section~\ref{sec:singular}, we have discussed a counter-example of
extinct sources that we refer to as a singular source. It mimics the
behaviour of long cosmic strings at small scales and the function
$\calCFT_k(\gamma,\gamma')$ is no longer holomorphic. This results in
various drastic changes in the oscillatory structure of both the
strain and energy density spectra that would allow its disambiguation
from extinct sources. Interestingly such a case provides an example
for which only the presence of interferences on top of the fine
structure would allow for a clear disambiguation between the
radiation-era generated spectra. In the matter era, we have found
strong changes, such as a very slow decay of $\calPh^\umat \propto
1/k^2$ (instead of the expected $1/k^4$) and a violation of the
relation $\OmegaGW \simeq k^2/(12 \calH^2) \calPh$ for all
wavenumbers.

These results have various implications. One is that there is no
reason for cosmological predictions to use $\OmegaGW(\eta_1,\eta_2,k)$
as a proxy, being the two-point correlation function of $h_{ij}'$ it
is not the quantity of interest for direct measurements which are
sensitive to correlations in the strain. As we have shown, both
quantities can be significantly different for the singular sources and
this could be a source of errors in the predictions. The only usage
of $\OmegaGW$ should be in measuring the overall gravitating effect of
gravitational waves, as it is done during BBN for
instance. Another implication concerns the waveform measurable by
direct detection experiments. Our results are given in spatial Fourier
space, with time-dependent terms. Taking the inverse spatial Fourier
transform of our formulas as well as the forward Fourier transform
with respect to the time $\eta$ would give a function of spatial
separation $\vect{x}$ and angular frequency $\omega$. The fine
structure in $k$ implies that the correlators have also some fine
structure in $\vect{x}$ and it would be interesting to determine how
the signal changes with respect to the separation between the
interferometers. Concerning the angular frequencies, at fixed
wavenumber $k$, only four are excited $\omega=\pm k$ and $\omega = \pm
2k$. This is expected, we consider correlators which are the square of
the strain, this one being a superimposition of free waves having
$\omega=k$ and $\omega=-k$. However, the amplitude of each of these
four oscillatory terms is peculiar to each type of source and its
experimental determination would be interesting. Concerning cosmic
strings, let us recap that most of the overall GW emission is expected
to come from cosmic string loops and not from long strings, at least
for Nambu-Goto strings. Moreover, even if the matter era spectrum
$\calPh \propto 1/k^2$ instead of $1/k^4$, it is perfectly possible
that this effect remain completely negligible because the long
strings contribution from the radiation era is also varying as $1/k^2$,
and, it could be the dominating part. However, this is of clear
interest for models in which cosmic strings are formed during
inflation and would enter scaling only in the matter
era~\cite{Yokoyama:1988zza, Kamada:2014qta, Ringeval:2015ywa}. In view
of our results, these models could be constrained by GW direct
detection experiments.

Finally, it would be interesting to search for a generalisation of the
case of extinct sources, out of the scaling hypothesis. For instance,
it should be possible to extend the results derived for holomorphic
anisotropic stresses to explicit time-dependent sources provided they
can be factorized with some ``scaling terms''. We let however these
investigations for a future work.

\section*{Acknowledgements}

This work is supported by the ``Fonds de la Recherche Scientifique -
FNRS'' under Grant $\mathrm{N^{\circ}T}.0198.19$ as well as by the
Wallonia-Brussels Federation Grant ARC
$\mathrm{N^{\circ}}19/24-103$.

\appendix
\section{Holomorphic correlators in Fourier space}
\label{sec:holomorphic}

In this appendix, we rigorously derive the value of the integrals
$\Icc$, $\Iss$, $\Ics$ and $\Isc$ presented in the
section~\ref{sec:convolution} when the Fourier transform of the
correlator $\calCFT_k(\gamma,\gamma')$ is a holomorphic function.

Let us explain the method with equation~\eqref{eq:Iccsinc}. Expanding
the sine cardinal functions into complex exponentials, we can rewrite
$\Icc$ as
\begin{equation}
\Icc = -\dfrac{1}{16 \pi^2} \iint_{-\infty}^{+\infty} \ud \gamma \ud
\gamma' \left[e^{\imath(1-\gamma) y_1} - e^{-\imath(1-\gamma)y_1}
  \right]\left[e^{\imath(1-\gamma') y_2} - e^{-\imath(1-\gamma')y_2}
  \right] \dfrac{\calCFT_k(\gamma,\gamma')}{(1-\gamma)(1-\gamma')}\,,
\label{eq:Iccexpanded}
\end{equation}
where the ``natural'' poles in $\gamma=1$ and $\gamma'=1$ coming from
GW propagation are now made explicit. Expanding all terms give
\begin{equation}
\begin{aligned}
  \Icc & = -\dfrac{1}{4} \Bigg\{
  e^{\imath(y_1+y_2)}\invfourierb{\dfrac{\calCFT_k(-\gamma,-\gamma')}{(1+\gamma)(1+\gamma')}}
  - e^{\imath(y_1-y_2)}
  \invfourierb{\dfrac{\calCFT_k(-\gamma,\gamma')}{(1+\gamma)(1-\gamma')}}
  \\ & - e^{-i(y_1-y_2)}
  \invfourierb{\dfrac{\calCFT_k(\gamma,-\gamma')}{(1-\gamma)(1+\gamma')}}
  + e^{-i(y_1+y_2)}
  \invfourierb{\dfrac{\calCFT_k(\gamma,\gamma')}{(1-\gamma)(1-\gamma')}} \Bigg\},
\end{aligned}
\label{eq:Iccinvfourier}
\end{equation}
where $\invfourier{}$ denotes the inverse Fourier transform, going from
$(\gamma,\gamma')$ to $(y_1,y_2)$. The expression of $\Icc$ is known
if one can evaluate these inverse Fourier transforms, and they are
trivial provided the function $\calCFT_k(\gamma,\gamma')$ is
holomorphic. Let us focus on
\begin{equation}
\invfourierb{\dfrac{\calCFT_k(-\gamma,-\gamma')}{(1+\gamma)(1+\gamma')}}
= \dfrac{1}{4 \pi^2} \int_{-\infty}^{+\infty} \ud
  \gamma' \dfrac{e^{\imath \gamma' y_2}}{1+\gamma'} \int_{-\infty}^{+\infty}
\ud \gamma  \dfrac{\calCFT_k(-\gamma,-\gamma')}{1+\gamma}e^{\imath \gamma y_1 }.
\label{eq:inversefourier}
\end{equation}
The simple pole at $\gamma=-1$ in the last integral requires an
integration contour to be chosen in the complex plane to determine its
Cauchy principal value. This one is depicted in
figure~\ref{fig:poleminusone}. After pushing the upper contour to
complex infinity and the smaller one towards the pole, one finds
\begin{equation}
\int_{-\infty}^{+\infty}
\ud \gamma  \dfrac{\calCFT_k(-\gamma,-\gamma')}{1+\gamma}e^{\imath
  \gamma y_1 } = \imath \pi e^{-\imath y_1} \calCFT_k(1,-\gamma').
\end{equation}
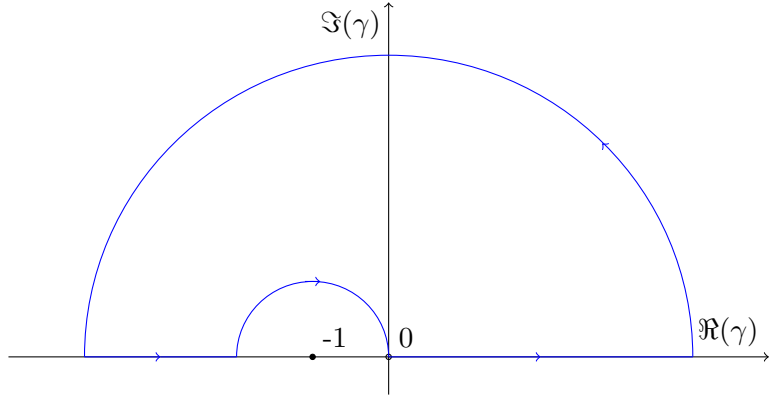
\begin{figure}
\begin{center}
  \begin{tikzpicture}[]
    \draw [->] (-5,0) -- (5,0) node [above left]  {$\Re(\gamma)$};
    \draw [->] (0,-0.5) -- (0,4.7) node [below left = -1pt] {$\Im(\gamma)$};
    
    \draw[] (0,0) circle (1pt) node [above right] {0};

    \draw[fill] (-1,0) circle (1pt) node [above right] {-1};
    
    \draw[blue] (-0,0) arc (0:180:1) ;
    \draw[blue,->] (-1.1,1) -- (-0.9,1);
    \draw[blue,->] (0,0) -- (2,0);
    \draw[blue] (2,0) -- (4,0);
    \draw[blue] (4,0) arc (0:180:4);
    \draw[blue,->] (-4,0) -- (-3,0);
    \draw[blue] (-3,0) -- (-2,0);
    \draw[blue,->] (2.82,2.82) -- (2.81,2.83);
  \end{tikzpicture}
\end{center}
\caption{Integration contour used to evaluate the inverse Fourier
  transform of equation~\eqref{eq:inversefourier}.}
\label{fig:poleminusone}
\end{figure}
Repeating the same procedure for the remaining integral in
equation~\eqref{eq:inversefourier}, in the complex
plane $[\Re(\gamma'),\Im(\gamma')]$, we get
\begin{equation}
\invfourierb{\dfrac{\calCFT_k(-\gamma,-\gamma')}{(1+\gamma)(1+\gamma')}}
= -\dfrac{1}{4} e^{-\imath(y_1+y_2)} \calCFT_k(1,1).
\end{equation}
The other inverse Fourier transforms appearing in
equation~\eqref{eq:Iccinvfourier} can be dealt in the same way. Notice
however that the poles are not at the exact same location, they are in
$\gamma=\pm 1$ and $\gamma'=\pm 1$. We obtain
\begin{equation}
\begin{aligned}
  \invfourierb{\dfrac{\calCFT_k(-\gamma,\gamma')}{(1+\gamma)(1-\gamma')}}
& = + \dfrac{1}{4} e^{-\imath(y_1-y_2)} \calCFT_k(1,1), \\ \invfourierb{\dfrac{\calCFT_k(\gamma,-\gamma')}{(1-\gamma)(1+\gamma')}}
  & = + \dfrac{1}{4} e^{\imath(y_1-y_2)} \calCFT_k(1,1), \\
  \invfourierb{\dfrac{\calCFT_k(\gamma,\gamma')}{(1-\gamma)(1-\gamma')}}
  & = - \dfrac{1}{4} e^{\imath(y_1+y_2)} \calCFT_k(1,1),
\end{aligned}
\end{equation}
from which equation~\eqref{eq:Iccinvfourier} gives
\begin{equation}
\Icc = \dfrac{1}{4} \calCFT_k(1,1).
\end{equation}

The other integrals $\Iss$, $\Ics$ and $\Isc$ can be explicitly
calculated with the same method. However, because they involve
functions of the form $\sinc^2[(1-\gamma)y_1/2]$, when doing an
expansion in terms of complex exponentials, constant terms appear and
one has to evaluate three new integrals. Two of them are
one-dimensional inverse Fourier transforms
\begin{equation}
  I_1 = \dfrac{1}{2}
  \invfourierb{\dfrac{\calCFT(\gamma,1)}{1-\gamma}}_{y_1 = 0}, \qquad
  I_1' = \dfrac{1}{2}
\invfourierb{\dfrac{\calCFT(1,\gamma')}{1-\gamma'}}_{y_2 = 0},
\end{equation}
and the last one is the two-dimensional inverse Fourier transform
\begin{equation}
I_2 = \invfourierb{\dfrac{\calCFT(\gamma,\gamma')}{(1-\gamma)(1-\gamma')}}_{(y_1,y_2)=(0,0)},
\end{equation}
all evaluated at the origin. To calculate their value one can make use
of the Dirichlet's theorem and evaluate the integrals at $y_1=0^\pm$
and $y_2=0^\pm$. Each sign requiring a different integration
contour. Taking the mean finally gives $I_1=I_1'=I_2=0$ which propagates to
$\Iss=\Ics=\Isc=0$ as stated in section~\ref{sec:convolution}.

\section{Spectra from extinct sources}
\label{sec:extinct}

As described in section~\ref{sec:matera}, the calculation of the
integral $\Imumat$ proceeds exactly as the one detailed for the
radiation era but starting from the matter era convolution kernels
given in equation~\eqref{eq:Ksmat}. From equation~\eqref{eq:Imumat},
using the definitions~\eqref{eq:calCmatdef} and \eqref{eq:calDEdef}, one
gets, for extinct sources in the matter era,
\begin{align}
  & \Imumat(x_1,x_2,k) = \dfrac{\calCFT_k^\umat +
  \calEFT_k^\umat}{8}
  \dfrac{1+(x_1+\xeq)(x_2+\xeq)}{(x_1+\xeq)(x_2+\xeq)} \cos(x_1-x_2)
  \nonumber \\ & + \dfrac{\calCFT_k^\umat +
  \calEFT_k^\umat}{8} \dfrac{x_1-x_2}{(x_1+\xeq)(x_2+\xeq)}
  \sin(x_1-x_2) \nonumber \\ & + \dfrac{(\calCFT_k^\umat -
  \calEFT_k^\umat)\left[1 - (x_1+\xeq)(x_2+\xeq) \right] - 2
  \calDFT_k^\umat(x_1+x_2+2\xeq)}{8(x_1+\xeq)(x_2+\xeq)} \cos(x_1+x_2
  - 2\xeq) \nonumber \\
  & + \dfrac{(\calCFT_k^\umat - \calEFT_k^\umat)(x_1+x_2+2\xeq) + 2
    \calDFT_k^\umat \left[1 - (x_1+\xeq)(x_2+\xeq)
      \right]}{8(x_1+\xeq)(x_2+\xeq)} \sin(x_1+x_2 - 2\xeq).
\label{eq:Imumatholo}
\end{align}
where we have used the abridged notation $\calCFT_k = \calCFT_k(1,1)$,
$\calDFT_k = \calDFT_k(1,1)$ and $\calEFT_k = \calEFT_k(1,1)$.
Plugging this expression into equation~\eqref{eq:strainmat} gives the
exact waveform of the unequal time strain power spectrum given in
equation~\eqref{eq:strainmatholo}.  The other integrals, $\Idmumat$ and
$\Ixmat$, entering the expression of $\OmegaGWmat$, can be immediately
obtained by using equation~\eqref{eq:partialx1x2}, i.e., by deriving
the above expression with respect to $x_1$ and $x_2$. After lengthy
algebra, one obtains

\begingroup
\allowdisplaybreaks
\begin{align}
 & \Idmumat(x_1,x_2,k) = \dfrac{\calCFT_k^\umat + \calEFT_k^\umat}{8
    (x_1+\xeq)^2(x_2+\xeq)^2} \left[1-x_1^2 -x_2^2 +x_1x_2(1+x_1 x_2)
    \right. \nonumber \\ & \left .+ (x_1+x_2) (2 x_1 x_2-1)\xeq +
    (x_1^2-1+4x_1x_2 +x_2^2)\xeq^2 +2(x_1+x_2) \xeq^3 + \xeq^4 \right]
  \cos(x_1-x_2) \nonumber \\ & + \dfrac{\calCFT_k^\umat +
    \calEFT_k^\umat}{8(x_1+\xeq)^2 (x_2+\xeq)^2}
  (x_1-x_2)\left[1+(x_1+\xeq)(x_2+\xeq) \right] \sin(x_1-x_2)
  \nonumber \\ & + \bigg\{\dfrac{\calCFT_k^\umat -
    \calEFT_k^\umat}{8(x_1+\xeq)^2 (x_2+\xeq)^2} \left[1-(x_1+\xeq)^2
    - (x_2+\xeq)^2 - (x_1+\xeq)(x_2+\xeq) \right. \nonumber
    \\ &\left. +(x_1+\xeq)^2(x_2+\xeq)^2 \right] - \dfrac{\calDFT_k^\umat}{4}
  \dfrac{(x_1+x_2 +
    2\xeq)\left[1-(x_1+\xeq)(x_2+\xeq)\right]}{(x_1+\xeq)^2
    (x_2+\xeq)^2} \bigg\} \nonumber \\ & \times \cos(x_1+x_2 - 2\xeq)
  \nonumber \\ & + \bigg\{\dfrac{\calCFT_k^\umat - \calEFT_k^\umat}{8}
  \dfrac{(x_1+x_2+2\xeq)\left[1-(x_1+\xeq)(x_2+\xeq)\right]}{(x_1+\xeq)^2
    (x_2+\xeq)^2} + \dfrac{\calDFT_k^\umat}{4(x_1+\xeq)^2
    (x_2+\xeq)^2} \nonumber \\ & \times \left[1 - x_1^2 - x_2^2
    -x_1x_2(1-x_1x_2) +(x_1+x_2)(2 x_1x_2-3) \xeq - (3 - x_1^2 - x_2^2
    - 4 x_1 x_2)\xeq^2 \right. \nonumber \\ & \left. + 2(x_1+x_2)
    \xeq^3 + \xeq^4 \right] \bigg\} \sin(x_1+x_2 - 2\xeq),
\label{eq:Idmumatholo}
\end{align}
\endgroup
and
\begingroup
\allowdisplaybreaks
\begin{align}
&  \Ixmat(x_1,x_2,k) = \dfrac{\calCFT_k^\umat+\calEFT_k^\umat}{8}
  \dfrac{(x_1-x_2)(x_1+\xeq)-1}{(x_1+\xeq)^2(x_2+\xeq)} \cos(x_1-x_2)
  \nonumber \\ & - \dfrac{\calCFT_k^\umat+\calEFT_k^\umat}{8}
  \dfrac{\xeq^3 +
    x_1^2(x_2+\xeq)+x_2(\xeq^2-1)+x_1(1+2x_2\xeq+2\xeq^2)}{(x_1+\xeq)^2(x_2+\xeq)}\sin(x_1-x_2)
  \nonumber \\ & + \bigg\{ \dfrac{\calCFT_k^\umat -
  \calEFT_k^\umat}{8}\dfrac{x_1^2-1+x_2 \xeq +2\xeq^2 +
    x_1(x_2+3\xeq)}{(x_1+\xeq)^2(x_2+\xeq)} +
  \dfrac{\calDFT_k^\umat}{4(x_1+\xeq)^2(x_2+\xeq)}
  \left[x_1+x_2 \right. \nonumber \\ & \left. +2\xeq-x_2\xeq^2 -
    \xeq^3 - x_1^2(x_2+\xeq) -2x_1 \xeq(x_2+\xeq)\right] \bigg\}
  \cos(x_1+x_2-2\xeq) \nonumber \\
  & + \bigg\{ - \dfrac{\calCFT_k^\umat -
    \calEFT_k^\umat}{8(x_1+\xeq)^2(x_2+\xeq)} \left[x_1+x_2+2\xeq-x_2
    \xeq^2 - \xeq^3 -x_1^2(x_2+\xeq) \right. \nonumber \\ &
    \left. -2x_1\xeq(x_2+\xeq)\right] + \dfrac{\calDFT_k^\umat}{4}
  \dfrac{x_1^2-1+x_2\xeq + 2\xeq^2
    +x_1(x_2+3\xeq)}{(x_1+\xeq)^2(x_2+\xeq)} \bigg\} \sin(x_1+x_2-2\xeq).  
\label{eq:Ixmatholo}
\end{align}
\endgroup
The last integral, $\Ixbar(x_1,x_2,k)$, defined by
equation~\eqref{eq:Ixbardef}, is obtained by complex conjugating the
Fourier transformed correlators while swapping $x_1$ and $x_2$. For
$\calU(x,x')$ symmetric, the $\calC(y,y')$, $\calD(y,y')$ and
$\calE(y,y')$ functions are even, and real, such that their Fourier
transform are also even and real. Therefore, it is enough to simply
swap $x_1$ and $x_2$ in the previous expression to obtains $\Ixbar$.
Plugging equations~\eqref{eq:Imumatholo} to \eqref{eq:Ixmatholo} into
the expression~\eqref{eq:omegamat} one gets for the energy density
parameter \begingroup \allowdisplaybreaks
\begin{align}
& \dfrac{\OmegaGWmat(\eta_1,\eta_2,k)}{\dfrac{4}{3} \left(\GN M^2 \right)^2}  = 
\left(\calCFT_k^\umat + \calEFT_k^\umat \right)  \left[\dfrac{1+ (x_1+\xeq)
    (x_2+\xeq)}{(x_1+\xeq)(x_2+\xeq)} \right. \nonumber \\ & \left. + \dfrac{k}{\calH_1} \dfrac{1-(x_1+\xeq)^2 + (x_1+\xeq)
    (x_2+\xeq)}{(x_1+\xeq)^2 (x_2+\xeq)} + \dfrac{k}{\calH_2} \dfrac{1+(x_1+\xeq)(x_2+\xeq)
    -(x_2+\xeq)^2}{(x_1+\xeq)(x_2+\xeq)^2} \right. \nonumber \\ & \left. + \dfrac{k^2}{\calH_1 \calH_2}\dfrac{1+ (x_1+\xeq)(x_2+\xeq)
      -(x_2+\xeq)^2 -(x_1+\xeq)^2 +(x_1+\xeq)^2 (x_2+\xeq)^2}{(x_1+\xeq)^2 (x_2+\xeq)^2}
  \right] \nonumber \\ & \times \cos(x_1 -
x_2) \nonumber \\ & + \left(\calCFT_k^\umat + \calEFT_k^\umat \right) \left[\dfrac{x_1 - x_2}{(x_1+\xeq)(x_2+\xeq)} +
  \dfrac{k}{\calH_1} \dfrac{x_1-x_2+(x_1+\xeq)^2
    (x_2+\xeq)}{(x_1+\xeq)^2 (x_2+\xeq)} \nonumber \right. \\ &
  \left. +
  \dfrac{k}{\calH_2} \dfrac{x_1-x_2 - (x_1+\xeq) (x_2+\xeq)^2}{(x_1+\xeq)(x_2+\xeq)^2}
  \right. \nonumber \\ &\left. +\dfrac{k^2}{\calH_1 \calH_2} \dfrac{x_1-x_2+(x_1+\xeq)^2 (x_2+\xeq)
    -(x_1+\xeq) (x_2+\xeq)^2}{(x_1+\xeq)^2 (x_2+\xeq)^2} \right]
\sin(x_1 -x_2) \nonumber \\ & +
\bigg\{\left(\calCFT_k^\umat -\calEFT_k^\umat\right) \left[-1 + \dfrac{1}{(x_1+\xeq)(x_2+\xeq)}\right] 
    - 2\calDFT_k^\umat \dfrac{x_1+x_2+2\xeq}{(x_1+\xeq)(x_2+\xeq)} 
    \nonumber \\ & + \dfrac{k}{\calH_1}\left[ \left(\calCFT_k^\umat -
    \calEFT_k^\umat \right)
    \dfrac{1-(x_1+\xeq)^2 - (x_1+\xeq)(x_2+\xeq)}{(x_1+\xeq)^2
      (x_2+\xeq)} \nonumber  \right. \\ & \left. -  2 \calDFT_k^\umat
    \dfrac{x_1 + x_2 +2\xeq -(x_1+\xeq)^2 (x_2+\xeq)}{(x_1+\xeq)^2 (x_2+\xeq)}
    \right] \nonumber \\ & +
    \dfrac{k}{\calH_2}\left[\left(\calCFT_k^\umat-\calEFT_k^\umat
      \right) \dfrac{1-(x_2+\xeq)^2 -(x_1+\xeq)
        (x_2+\xeq)}{(x_1+\xeq)(x_2+\xeq)^2} \nonumber \right. \\ &
      \left. - 2\calDFT_k^\umat \dfrac{x_1 +
        x_2 + 2\xeq -(x_1+\xeq) (x_2+\xeq)^2}{(x_1+\xeq)(x_2+\xeq)^2}  \right] \nonumber \\ & +
    \dfrac{k^2}{\calH_1 \calH_2}\left[- 2
    \calDFT_k^\umat\dfrac{x_1+x_2+2\xeq - (x_1+\xeq)^2(x_2+\xeq) -
      (x_1+\xeq)(x_2+\xeq)^2}{(x_1+\xeq)^2(x_2+\xeq)^2}
    \right. \nonumber \\ & \left.  + \dfrac{1-(x_1+\xeq)(x_2+\xeq) -
        (x_1+\xeq)^2 - (x_2+\xeq)^2 + (x_1+\xeq)^2
      (x_2+\xeq)^2}{(x_1+\xeq)^2 (x_2+\xeq)^2} \right. \nonumber
    \\ & \left. \times \left(\calCFT_k^\umat - \calEFT_k^\umat\right) \right]  \bigg\}
     \cos\left(x_1+x_2 - 2\xeq\right) \nonumber \\ & +
    \bigg\{ \left(\calCFT_k^\umat - \calEFT_k^\umat
    \right) \dfrac{x_1+x_2+2\xeq}{(x_1+\xeq)(x_2+\xeq)}  + 2
    \calDFT_k^\umat \dfrac{1 - (x_1+\xeq)
      (x_2+\xeq)}{(x_1+\xeq)(x_2+\xeq)} \nonumber \\ & +
    \dfrac{k}{\calH_1} \left[\left(\calCFT_k^\umat - \calEFT_k^\umat
      \right)  \dfrac{x_1+x_2 + 2\xeq -(x_1+\xeq)^2 (x_2+\xeq)}{(x_1+\xeq)^2 (x_2+\xeq)}
       \right. \nonumber \\ &
      \left. + 2\calDFT_k^\umat \dfrac{1-(x_1+\xeq)^2 - (x_1+\xeq)
        (x_2+\xeq)}{(x_1+\xeq)^2 (x_2+\xeq)}  \right] \nonumber \\ &
    + \dfrac{k}{\calH_2} \left[\left(\calCFT_k^\umat - \calEFT_k^\umat
      \right) \dfrac{x_1+x_2 + 2\xeq -(x_1+\xeq)
        (x_2+\xeq)^2}{(x_1+\xeq) (x_2+\xeq)^2} \right. \nonumber \\ &
      \left.  + 2 \calDFT_k^\umat
      \dfrac{1 - (x_1+\xeq)(x_2+\xeq)
        - (x_2+\xeq)^2}{(x_1+\xeq)(x_2+\xeq)^2} \right] \nonumber \\
    & + \dfrac{k^2}{\calH_1 \calH_2}\left[\left(\calCFT_k^\umat -
      \calEFT_k^\umat \right)  \dfrac{x_1+x_2+2\xeq - (x_1+\xeq)^2 (x_2+\xeq)
        -(x_1+\xeq)(x_2+\xeq)^2}{(x_1+\xeq)^2 (x_2+\xeq)^2} 
      \right. \nonumber \\ & \left. +
      2\calDFT_k^\umat \dfrac{1 - (x_1+\xeq)(x_2+\xeq) - (x_1+\xeq)^2
        -(x_2+\xeq)^2 + (x_1+\xeq)^2 (x_2+\xeq)^2}{ (x_1+\xeq)^2 (x_2+\xeq)^2}
       \right] \bigg\} \nonumber \\ & \times \sin\left(x_1 + x_2 - 2 \xeq \right).
\label{eq:omegamatholo}
\end{align}
\endgroup

\bibliographystyle{JHEP}

\bibliography{sgwb}

\end{document}